\documentclass[letterpaper,12pt]{article} 
\usepackage{amsthm}
\usepackage[]{algorithm2e}
\usepackage{amsmath}
\allowdisplaybreaks
\usepackage[usenames,dvipsnames]{color}
\usepackage[hmargin=0.7in,vmargin=1in]{geometry}
\usepackage[comma,longnamesfirst]{natbib}
\usepackage{amsfonts}
\usepackage{float}
\usepackage{amsmath, amssymb}
\usepackage{graphicx}
\usepackage{epstopdf}
\usepackage{caption}
\usepackage{blindtext}
\usepackage[utf8]{inputenc}
\usepackage{setspace}
\usepackage{pdflscape}
\usepackage{numprint} 
\usepackage{setspace}
\newtheorem{theorem}{Theorem}

\newtheorem{corollary}{Corollary}

\usepackage{fontenc}
\usepackage{epsfig}
\usepackage{rotating}
\usepackage{hyperref}
\usepackage{multirow}
\usepackage{bm}
\usepackage{tikz}
\def\checkmark{\tikz\fill[scale=0.4](0,.35) -- (.25,0) -- (1,.7) -- (.25,.15) -- cycle;}

\epsfverbosetrue
\usepackage{enumitem}
\newlist{props}{enumerate}{1}
\setlist[props,1]{
	label={\arabic*.},
	leftmargin=*,
	align=left,
	labelsep=5pt,
}

\makeatletter
\def\@seccntformat#1{\@ifundefined{#1@cntformat}%
   {\csname the#1\endcsname\quad}  % default
   {\csname #1@cntformat\endcsname}% enable individual control
}
\let\oldappendix\appendix %% save current definition of \appendix
\renewcommand\appendix{%
    \oldappendix
    \newcommand{\section@cntformat}{\appendixname~\thesection\quad}
}
\makeatother

\usepackage{listings}
\usepackage{color} %red, green, blue, yellow, cyan, magenta, black, white
\definecolor{mygreen}{RGB}{28,172,0} % color values Red, Green, Blue
\definecolor{mylilas}{RGB}{170,55,241}
\usepackage[mathlines,displaymath]{lineno}
\DeclareMathOperator{\diag}{diag}

\def \calL {\mathcal L}

\def \avec {\text{\boldmath$a$}}

\def \dvec {\text{\boldmath$d$}}

\def \hvec {\text{\boldmath$h$}}

\def \uvec {\text{\boldmath$u$}}    
\def \vvec {\text{\boldmath$v$}}    
\def \wvec {\text{\boldmath$w$}}    
\def \xvec {\text{\boldmath$x$}}    
\def \yvec {\text{\boldmath$y$}}    
\def \zvec {\text{\boldmath$z$}}    
\def \nuvec {\text{\boldmath$\nu$}}

\def \alphavec        {\text{\boldmath$\alpha$}}
\def \betavec         {\text{\boldmath$\beta$}}
\def \gammavec        {\text{\boldmath$\gamma$}}

\def \epsilonvec      {\text{\boldmath$\epsilon$}}
\def \varepsilonvec   {\text{\boldmath$\varepsilon$}}
\def \zetavec         {\text{\boldmath$\zeta$}}
\def \etavec          {\text{\boldmath$\eta$}}
\def \thetavec        {\text{\boldmath$\theta$}}
\def \varthetavec     {\text{\boldmath$\vartheta$}}

\def \lambdavec       {\text{\boldmath$\lambda$}}
\def \muvec           {\text{\boldmath$\mu$}}
\def \nuvec           {\text{\boldmath$\nu$}}
\def \xivec           {\text{\boldmath$\xi$}}

\def \rhovec          {\text{\boldmath$\rho$}}

\def \tauvec          {\text{\boldmath$\tau$}}

\def \psivec          {\text{\boldmath$\psi$}}
\def \chivec          {\text{\boldmath$\chi$}}
\def \omegavec        {\text{\boldmath$\omega$}}

\usepackage{titlesec}
\titlespacing*\section{0pt}{0pt plus 4pt minus 2pt}{0pt plus 2pt minus 2pt}
\titlespacing*\subsection{0pt}{0pt plus 4pt minus 2pt}{0pt plus 2pt minus 2pt}
\titlespacing*\subsubsection{0pt}{0pt plus 4pt minus 2pt}{0pt plus 2pt minus 2pt}
\titlespacing*\paragraph{0pt}{5pt plus 4pt minus 2pt}{7pt plus 2pt minus 2pt}

\begin{document}

%\tableofcontents
\pagestyle{empty}
%this file produces the title page for the paper
\begin{titlepage}
\title{Fast and Accurate Variational Inference for Models with Many Latent Variables}
\author{Rub\'{e}n Loaiza-Maya, Michael Stanley Smith, David J. Nott \& Peter J. Danaher}
\date{First Version: May 2020\\ This Version: \today}
\maketitle
\vspace{3cm}
\noindent {\small 
	Rub\'{e}n Loaiza-Maya is a Lecturer at the Department of Econometrics and Business Statistics,
	Monash University; Michael Stanley Smith is Chair of Management (Econometrics)
	at Melbourne Business School, University of Melbourne; David J. Nott is 
	Associate Professor of Statistics, National University of Singapore; and 
	Peter J. Danaher is Professor of Marketing and Econometrics, Monash University.  
	Correspondence should be directed to Michael Smith at
	{\tt mike.smith@mbs.edu}.}\\

\vspace{1cm}
\noindent {\small
	{\bf Acknowledgements}: the authors thank the Editor, Associate Editor and two referees for 
	their encouragement and comments that helped improve the paper.  
	They also thank Wharton Customer Analytics for providing the consumer panel data used in the second example. 
	Rub\'{e}n Loaiza-Maya is an associate investigator with the Australian Centre of Excellence for Mathematical and Statistical Frontiers.
	David Nott is affiliated with the Operations Research and Analytics Research cluster at the National University of Singapore. }

\newpage
\begin{center}
%{\LARGE Models with Many Latent Variables: Efficient Variational Inference and Econometric Applications}\\
{\LARGE Fast and Accurate Variational Inference for Models with Many Latent Variables}\\
%{\LARGE Efficient variational inference for models with many latent variables}\\
\vspace{15pt}
{\Large Abstract}
\end{center}
\vspace{-10pt}
Models with a large number of latent
variables are often used to fully utilize the information
in big or complex data. However, they can be difficult
to estimate using standard approaches, and
variational inference methods are a popular alternative. Key to the success of these is the selection of 
an approximation to the target density that is accurate, tractable
and fast to calibrate using optimization methods. Most existing choices 
can be inaccurate or slow to calibrate when
there are many latent variables. Here, we propose a 
family of tractable variational approximations that are more accurate
and faster to calibrate for this case.
It combines a parsimonious parametric approximation for the parameter
posterior, with
the exact conditional posterior of the latent variables. 
We derive a simplified expression for the re-parameterization gradient of the variational lower bound,
which is the main ingredient of efficient optimization 
algorithms used to implement variational estimation. To do so only requires
the ability to 
generate exactly or approximately
from the conditional posterior of the latent variables, rather than to compute its
density.
We illustrate using two complex contemporary econometric examples. The first is a
nonlinear multivariate state space model for U.S. macroeconomic variables. 
The second
is a random coefficients tobit model applied to two million sales by 20,000 individuals
in a large consumer panel from a marketing study. In both cases, we
show that our approximating family is considerably more accurate than mean
field or structured Gaussian approximations, and faster than Markov chain Monte
Carlo. Last, we show how
to implement data sub-sampling in variational inference for our
approximation, which can
lead to a further reduction in computation time. MATLAB code implementing the method for our
examples is included in supplementary material.
\vspace{1cm}

\noindent 
{\bf Key Words}: Latent variable models; Time-varying VAR with stochastic volatility; Large consumer panels; Sub-sampling variational inference; Stochastic gradient ascent; Variational Bayes.

\end{titlepage}

\newpage
\pagestyle{plain}
\newpage
\doublespacing
\setlength{\abovedisplayskip}{0.15cm}
\setlength{\belowdisplayskip}{0.15cm}
\section{Introduction}\label{sec:intro}
Models with large numbers of latent variables\footnote{In the 
machine learning literature latent variables are sometimes called ``local variables''
and model parameters are sometimes called ``global variables''; for example, see~\cite{hoffman2013}.} are
increasingly popular for capturing
nuances in large or complex datasets. Examples
include topic models~\citep{hoffman2013}, state space models~\citep{durbin2012time},
choice models~\citep{train2009} and mixed models~\citep{gelman2006}, among others. 
In some cases it is possible to integrate out
the latent variables, 
although frequently it is not.
Instead,
Bayesian estimation usually
proceeds by considering the joint posterior distribution of the model parameters and latent
variables, often called an ``augmented posterior''.
Conventional Markov chain Monte Carlo (MCMC) methods for 
evaluating the augmented posterior are
computationally burdensome, and 
optimization-based variational inference \citep{ormerod2010explaining,blei2017} 
is a scalable alternative. However,
existing variational methods make strong independence or other parametric
assumptions which often result in a poor approximation to the 
augmented posterior. In this paper we  
suggest a general variational inference method which provides a more accurate approximation to the augmented posterior, 
while being
scalable to the case where there are a 
large number of latent variables and global model parameters.

Key to effective variational inference (also called ``variational Bayes'') is the selection of a 
suitable variational
approximation (VA). We suggest a VA which uses an arbitrary parametric approximation
%which uses a copula model defined through a parametric transformation~\citep{Smith2020} 
for 
the model parameters, along with the exact conditional posterior distribution for the latent
variables.
We show that this approximation is tractable, and derive a method for estimating the re-parameterization
gradient of the variational lower bound~\citep{Kingma+w13,rezende+mw14}, which is the main input to efficient stochastic gradient ascent (SGA)
methods widely used for calibrating VAs~\citep{bottou10,hoffman2013,salimans2013fixed}. 
Importantly, our method
does not require evaluation of the conditional posterior
density of the latent variables, or its derivative, which are often
unavailable or intractable. Instead, only a draw from the conditional 
posterior (either exactly or approximately) is needed, which is typically straightforward using a variety of well-explored methods for most models.
%including sub-sampling 
%approaches suitable for very large numbers of latent variables~\citep{quiroz+kvt19}. 
Thus, our approach 
can combine the
strengths of recent developments in MCMC and variational inference.

In some applications poor parameter uncertainty quantification
may not harm predictive inference~\citep{wang+b19}.
However, this is not the case for many latent variable models, including
the two examples considered here.
%For example, in panel datasets where it is desired
%to perform within subject prediction for individuals with sparse existing data, 
%quantification of uncertainty in individual specific latent variables is often crucial. 
Although marginal posterior distributions of global model parameters are often
well approximated as Gaussian for large data sets~\citep{titsias2014doubly,archer2015,kucukelbir2016automatic,ong2017gaussian}, 
observation-specific latent variables about which there is little information may exhibit highly non-Gaussian 
marginal posteriors. 
Furthermore, when the number of observation-specific latent variables grows with the sample size, poor
uncertainty quantification for the latent variables can cause poor inference for the 
global model parameters, including inaccurate point estimates.
A review of the existing literature on variational 
inference for complex 
latent variable models is given later in Section~\ref{sec:litreview}.

To illustrate the advantages of our approach we employ it to estimate
two contemporary and complex
econometric models. The first is a time-varying parameter vector 
autoregression with stochastic volatility (TVP-VAR-SV) proposed by~\cite{huber2020}. This is
a nonlinear state space model, for which mean field approximations are known 
to be poor~\citep{wang2004}, and Gaussian VAs with parsimonious structured covariance matrices are the most popular choice~\citep{quiroz+nk18}. 
Because this is a setting
where the exact posterior can be computed using existing
MCMC methods, the accuracy of our VA can be assessed. 
We find that our method provides
an approximation that is close to the exact posterior and is 
much more accurate than even a well-structured Gaussian VA. 

The second example is a random coefficient tobit (i.e. censored regression) model applied to a large consumer 
panel originating from~\citep{danaher2020}. Random 
coefficient models are widely used in marketing 
to capture consumer-level heterogeneity~\citep{allenby1998}
and in these models estimates
of the random coefficients are central as they
are employed to tailor
advertising and promotions at the consumer level.
%~\citep{rossi1996,danaher2015}.
It is impractical
to employ MCMC methods to estimate such models for large datasets, so that recent marketing
studies employ variational
inference instead~\citep{ansari2018,danaher2020}. In their original study~\cite{danaher2020} employ a structured Gaussian VA, and 
we show that this is less accurate than our
proposed VA for a small sample of 1,000 consumers, which is a case where the exact posterior can 
be computed. However, evaluation of the
exact posterior is impractical for a large sample of 20,000 individuals,
 yet we show that our approach is readily applied to this 
case and more effective than the Gaussian VA. 
%Interestingly, when the trade-off 
%between per-iteration computation time and speed of convergence in the variational optimization
%is considered, our proposed approximations can be calibrated accurately in
%20\% to 50\% of the time taken by the Gaussian alternatives in these examples.

We also show how our proposed VA is suitable 
for sub-sampling variational inference for augmented posteriors that can be factorized. 
This is common in practice, and includes the random coefficient tobit model considered here.
Application of sub-sampling further reduces the computational burden of calibrating
our proposed VA for this example.

In the next section we introduce variational inference, outline our new methodology, 
and discuss related existing methods. 
Section~\ref{sec:eg1} applies our approach to the TVP-VAR-SV model, and Section~\ref{sec:eg2} to the mixed effects tobit model. Section~\ref{sec:subsampling} outlines sub-sampling
variational inference for our approximation, while Section~\ref{sec:discuss} concludes. Appendices~\ref{app:proof} and~\ref{app:gradients} provide 
a proof and implementation details for our method and Appendix~\ref{app:tvpreg} gives key derivations
for the TVP-VAR-SV model. An extensive Online Appendix
provides additional details and results for the two applications.
\section{Methodology}\label{sec:vb}
We first provide a short overview of variational
inference, and then outline our new family of VAs for 
models with latent variables, along with how they can be used to provide  variational inference.

\subsection{Variational Inference}\label{ssec:VI}
We consider Bayesian inference with data $\yvec$ having density $p(\yvec|\psivec)$,
where in our paper, $\psivec$ contains the model parameters
augmented with a potentially large number of latent variables. 
Assuming a prior density $p(\psivec)$,
Bayesian inference is based on the density $p(\psivec|\yvec)\propto p(\psivec)p(\yvec|\psivec)$, which   
in our paper is the augmented posterior.
We will consider variational inference methods, 
in which a member $q_\lambda(\psivec)$ of some parametric family of densities is used to 
approximate the target $p(\psivec|\yvec)$, where $\lambdavec\in \Lambda$ is a vector of variational parameters.
%For example, for the Gaussian family
%$\lambdavec$ would consist of the distinct elements of the mean vector and covariance matrix.  
Approximate Bayesian inference is then formulated as an optimization
problem, where a measure of divergence between $q_\lambda(\psivec)$ and $p(\psivec|\yvec)$ 
is minimized with respect to $\lambdavec$. The  
Kullback-Leibler divergence is typically used, and it is straightforward 
to show (see, for example, \citet{ormerod2010explaining}) that this 
is equivalent to maximizing the variational lower bound (also called
the evidence lower bound, or ``ELBO'') given by
\[
\calL(\lambdavec)=\int \log \frac{p(\psivec)p(\yvec|\psivec)}{q_\lambda(\psivec)} q_\lambda(\psivec) d\psivec\,.
\]
%
%
%%Writing $p(\yvec)=\int p(\psivec)p(\yvec|\psivec) d\psivec$ for the marginal likelihood, and 
%\begin{align*}
%\text{KL}(q_\lambda(\psivec)||p(\psivec|\yvec) ) & = \int \log \frac{q_\lambda(\psivec)}{p(\psivec|\yvec)} q_\lambda(\psivec) \,d\psivec\,,
%\end{align*}
%is typically used, and we employ it here.
%If $p(\yvec)=\int p(\psivec)p(\yvec|\psivec) d\psivec$ denotes the marginal likelihood, then it is easily shown
%(see, for example, \citet{ormerod2010explaining}) that
%\begin{align}
%\text{KL}(q_\lambda(\psivec) ||p(\psivec|\yvec) ) & = \log p(\yvec)-\int \log \frac{p(\psivec)p(\yvec|\psivec)}{q_\lambda(\psivec)} q_\lambda(\psivec) d\psivec \nonumber \\
%& = \log p(\yvec)-\calL(\lambdavec), \label{kldexpression}
%\end{align}
%where $\calL(\lambdavec)$ is called the variational
% lower bound.\footnote{This is also widely
%	called the Evidence Lower Bound (ELBO).} Because $\log p(\yvec)$ does not depend on $\lambdavec$, minimization of the Kullback-Leibler
%divergence above with respect to $\lambdavec$ is equivalent to maximizing the variational lower bound 
%$\calL(\lambdavec)$. 

The lower bound takes the form of an intractable integral, so it seems challenging to optimize. However, notice
that 
it can be written as an expectation with respect to $q_\lambda$ as 
\begin{equation}
\calL(\lambdavec)=E_{q_\lambda}\left[\log g(\psivec) - \log q_\lambda(\psivec)\right]\,,
\label{eq:lowerbound}
\end{equation}
where $g(\psivec)=p(\psivec)p(\yvec|\psivec)$. This expression
allows easy application of stochastic gradient ascent (SGA) methods \citep{bottou10}.  
In SGA, given an initial
value $\lambdavec^{(1)}$, $\lambdavec$ is updated recursively as
\begin{align*}
\lambdavec^{(i+1)} & = \lambdavec^{(i)}+\rhovec_i \circ \widehat{\nabla_\lambda \calL(\lambdavec^{(i)})},
\;\mbox{ for } i=1,2,\ldots\,,
\end{align*}
where $\rhovec_i=(\rho_{i1},\dots, \rho_{im})^\top$ is a vector of step sizes, `$\circ$' denotes the element-wise product of two vectors, and $\widehat{\nabla_\lambda \calL(\lambdavec^{(i)})}$ is an unbiased estimate of the gradient of $\calL(\lambdavec)$ evaluated at $\lambdavec=\lambdavec^{(i)}$.  For appropriate step size
choices this will converge to a local optimum of $\calL(\lambdavec)$.
Adaptive step size choices are often used in practice, and we use
the popular automatic ADADELTA method of \cite{zeiler2012adadelta}.  

To implement SGA, unbiased estimates of the gradient of the lower bound are the key requirement.
These can be obtained directly by differentiating~\eqref{eq:lowerbound} with respect
to $\lambdavec$, and evaluating the expectation
in a Monte Carlo fashion 
by simulating from $q_\lambda$.
%; an example of this approach is given in Section~\ref{sec:tsapp}. 
However, variance reduction methods for
the gradient estimation are often also important for fast convergence and stability.  One of the most useful is the ``re-parametrization
trick'' \citep{Kingma+w13}. In this approach, 
%which is used in Section~\ref{sec:skcopva}, 
it is assumed that an iterate $\psivec$ can be generated from $q_\lambda$ by first drawing $\varepsilonvec$ from a density $f_\varepsilon$ which does not depend on $\lambdavec$, and then transforming $\varepsilonvec$ by a deterministic function $\psivec=h(\varepsilonvec,\lambdavec)$.  From~\eqref{eq:lowerbound}, the lower bound can be written as the following
expectation with respect to $f_\varepsilon$:
\begin{align}
\calL(\lambdavec) & = E_{f_\varepsilon}\left[\log g(h(\varepsilonvec,\lambdavec))-\log q_\lambda(h(\varepsilonvec,\lambdavec))\right]\,. \label{lbdrepar}
\end{align}
Differentiating under the integral
sign in (\ref{lbdrepar}) gives the ``re-parameterization gradient''
\begin{equation}
\nabla_\lambda \calL(\lambdavec) = E_{f_\varepsilon}\left[\nabla_\lambda \left\{ \log g\left(h(\varepsilonvec,\lambdavec)\right)-\log q_\lambda\left(h(\varepsilonvec,\lambdavec)\right)\right\}\right]\,, \label{lbdgradexpr}
\end{equation}
and approximating the expectation at~(\ref{lbdgradexpr}) by
one or more draws from $f_\varepsilon$ gives an unbiased
estimate of $\nabla_\lambda \calL(\lambdavec)$. 
An intuitive reason for the success of the re-parameterization trick is that it allows gradient information from the log-posterior to be used, by moving the variational parameters
inside $g(\psivec)$ in (\ref{lbdrepar}). For a well-chosen VA, only one draw from $f_\varepsilon$ is 
typically sufficient for the SGA to converge reliably.
% \cite{xu2018} show how the trick
%reduces the variance of the gradient estimates when $q_\lambda$ is a Gaussian with diagonal 
%covariance matrix (i.e. a
%mean field Gaussian approximation). 
We employ the 
re-parameterization trick throughout, and specify $h$
in Sections~\ref{sec:gapprox} and~\ref{sec:gcopapprox}. 

\subsection{Variational approximations for models with latent variables}\label{ssec:VAlatent}
In this paper we consider the case where
$\psivec^\top=(\thetavec^\top,\zvec^\top)$ is a parameter
vector $\thetavec=(\theta_1,\ldots,\theta_m)^\top$ augmented with additional latent
variables $\zvec$. 
Examples include state space models where $\zvec$ are latent states~\citep{durbin2012time}, 
discrete choice models where $\zvec$ are latent utilities~\citep{train2009} and
mixed models where $\zvec$ are the realizations of random coefficients~\citep{gelman2006}. 
The prior $p(\psivec)=p(\zvec|\thetavec)p(\thetavec)$, where $p(\zvec|\thetavec)$
is known from the model specification, and we approximate
the augmented posterior density 
$p(\psivec|\yvec)=p(\thetavec,\zvec|\yvec)\propto p(\yvec|\thetavec,\zvec)p(\zvec|\thetavec)p(\thetavec)\equiv g(\thetavec,\zvec)$ with VAs 
of the form
\begin{equation}\label{EQ:approximation}
q_\lambda\left(\bm{\theta},\zvec\right)= p\left(\zvec|\bm{y},\bm{\theta}\right)q_\lambda^0\left(\bm{\theta}\right)\,.
\end{equation}
As discussed below, it must be feasible to generate directly or approximately
from $p(\zvec|\bm{y},\bm{\theta})$, although it is unnecessary to
evaluate this density or its derivatives. 
The density $q^0_\lambda$ is chosen to be analytically tractable and from which it is convenient to sample, 
and we outline two effective choices in Section~\ref{ssec:copula}.
Calibration of $q_\lambda$ at~\eqref{EQ:approximation}
has the potential of being much more efficient than 
exact sampling using MCMC, because at each VB step we draw the vector $\bm{\theta}$ jointly from $q^0_\lambda$, and do not require partitioning of $\thetavec$ or Metropolis-Hastings steps
as in many MCMC schemes.

A key reason why~\eqref{EQ:approximation} is an attractive choice can be seen
by evaluating the lower bound~\eqref{eq:lowerbound} as
\begin{eqnarray}
\calL(\lambdavec) &=& E_{q_\lambda}\left[\log g\left(\bm{\theta},\zvec\right) -\log q_\lambda\left(\bm{\theta,\zvec}\right)\right]\nonumber\\
&= & E_{q_\lambda}\left[\log p\left(\bm{y}|\zvec,\bm{\theta}\right)+\log p\left(\zvec|\bm{\theta}\right)+\log p\left(\bm{\theta}\right)-\log q^0_\lambda\left(\bm{\theta}\right)-\log p\left(\zvec|\bm{y},\bm{\theta}\right)\right]\,.
\label{eq:lower2}
\end{eqnarray}
From Bayes theorem $p(\zvec|\bm{y},\bm{\theta})=
p(\bm{y}|\zvec,\bm{\theta})p(\zvec|\bm{\theta})/p(\bm{y}|\bm{\theta})$, substituting 
this into~\eqref{eq:lower2} gives
\begin{equation}
\calL(\lambdavec) = E_{q_\lambda}\left[\log p(\bm{y}|\bm{\theta})+\log p(\bm{\theta})-\log q^0_\lambda\left(\bm{\theta}\right)\right] = \calL^0(\lambdavec)\,.
\end{equation}
Here, $\calL^0(\lambdavec)$ is the variational lower bound arising from approximating the posterior $p(\thetavec|\yvec)$ directly by the VA $q^0_\lambda$. 
Thus, maximizing $\calL(\lambdavec)$ using SGA
methods is equivalent to maximizing $\calL^0(\lambdavec)$ for the posterior of $\thetavec$
with $\zvec$ marginalized out exactly, yet avoids the computation
of the (often intractable) density $p(\thetavec|\yvec)$ and its derivative.

A second major advantage of the VA at~\eqref{EQ:approximation} is that the gradient
of the lower bound at~\eqref{lbdgradexpr} has a simplified expression when using the
re-parameterization trick, as summarized in the theorem below.

\begin{theorem}[Re-parameterization Gradient]\label{thm:gradient}
Let $\varepsilonvec=((\varepsilonvec^0)^\top,\zvec^\top)^\top$ have the product density 
$f_{\varepsilon}(\varepsilonvec)=f_{\varepsilon^0}(\varepsilonvec^0)p(\zvec|h^0(\varepsilonvec^0,\lambdavec),\yvec)$, where $f_{\varepsilon^0}$ does not depend on $\lambdavec$, such that there
exists a (vector-valued) transformation $h$ from $\varepsilonvec$ to the 
augmented parameter
space given by $\psivec=h(\varepsilonvec,\lambdavec)=(h^0(\varepsilonvec^0,\lambdavec)^\top,\zvec^\top)^\top$, with $\thetavec=h^0(\varepsilonvec^0,\lambdavec)$.
Then the re-parametrization gradient used to implement SGA is 
\begin{equation}
\nabla_\lambda\calL(\lambdavec) =E_{f_\varepsilon}\left[\frac{\partial\bm{\theta}}{\partial\bm{\lambda}}^\top\left(\nabla_\theta\log g\left(\bm{\theta},\zvec\right) -\nabla_\theta\log q^0_\lambda(\bm{\theta})\right)\right]\,.
\label{eq:thm1}
\end{equation}
\end{theorem} 
\noindent {\em Proof}: See Appendix~\ref{app:proof}.

In~\eqref{eq:thm1}, the term $\nabla_\theta \log p(\zvec|\bm{\theta},\bm{y})$ is not needed, nor are derivatives with respect to $\zvec$, greatly simplifying calculation of the re-parameterization gradient. 
Instead, only
a draw from the conditional posterior $p(\zvec|\thetavec,\yvec)$ is required.
There is a large literature on drawing either exactly, 
or approximately, from this distribution for a wide range of
latent variable models using filtering, particle, MCMC, Hamiltonian Monte Carlo or 
other methods.
In the applications considered here, 
we use either a single, or a small number, of sweeps from a 
Gibbs sampler initialized at the draw from the previous SGA step, which proves
effective and simple to implement.

A third reason why~\eqref{EQ:approximation} is an attractive choice is because reducing the 
approximation error due to $\zvec$ also improves calibration of the marginal VA of $\thetavec$, 
as summarized in the corollary below.
\begin{corollary}\label{thm:betterapprox}
	Consider the VA at~\eqref{EQ:approximation} and a second VA with the same
	approximating family  
	for the marginal of $\thetavec$ with density 
	\begin{equation}
	\widetilde{q}_{\widetilde{\lambda}}(\thetavec,\zvec)=q^0_{\lambda_a}(\thetavec) q_{\lambda_b}(\zvec|\thetavec)\,,
	\label{eq:VA2}
	\end{equation}
	and variational parameters $\widetilde{\lambdavec}=(\lambdavec_a,\lambdavec_b)$. 
	Let $\lambdavec^\star$ and $\widetilde{\lambdavec}^\star=(\lambdavec_a^\star,\lambdavec_b^\star)$ be
	the values of the variational parameters that maximize the lower bound
	for the VAs at~\eqref{EQ:approximation} and~\eqref{eq:VA2}, respectively. Then our proposed
	VA is a more accurate approximation for the global parameters, in that it has lower Kullback-Leibler divergence
	$$\text{KL}(q_{\lambda^\star}^0(\thetavec) || p(\thetavec|\yvec)) \leq \text{KL}(q_{\lambda_a^\star}^0(\thetavec)||p(\thetavec|\yvec))\,.$$
\end{corollary}
\noindent {\em Proof}: See Appendix~\ref{app:proof}.

Algorithm~\ref{alg:VB} calibrates our proposed VA to the augmented posterior using SGA with the 
re-parameterization trick and the ADADELTA learning rate.

\begin{algorithm}[h]
	\hrulefill\\
	Initialize $\lambdavec^{(1)}$
	%$\lambdavec^{(1)}=\left(\gammavec^{(1)},\muvec_\vartheta^{(1)},
	%\mbox{vech}(B^{(1)}),\dvec^{(1)}\right)$, $\zvec^{(0)}$ 
	and set $s=1$
	\begin{itemize}
		\item[(a)] Generate $\thetavec^{(s)} \sim q^0_{\lambda^{(s)}}(\thetavec)$ using its re-parameterized representation
		%			$\zetavec^{(s)}\sim N(0,I_k)$ and $\epsilonvec^{(s)} \sim N(0,I_m)$  
		%			\item[(b)] Compute $\varthetavec^{(s)}=\muvec_\vartheta^{(s)}+B^{(s)}\zetavec^{(s)}+D^{(s)}\epsilonvec^{(s)}$
		%			\item[(c)] Compute $\bm{\theta}^{(s)} = (t_{\gamma_1^{(s)}}^{-1}(\vartheta_1^{(s)}),\ldots,t_{\gamma_m^{(s)}}^{-1}(\vartheta_m^{(s)}))^\top$ using the closed form inverse YJ transform
		\item[(b)] Generate $\zvec^{(s)} \sim p(\zvec|\thetavec^{(s)},\yvec)$ (either exactly or approximately)
		\item[(c)] Compute $\widehat{\nabla_{\lambda}\mathcal{L}\left(\bm{\lambda}^{(s)}\right)}= \left.\frac{\partial\bm{\theta}}{\partial\bm{\lambda}}^\top\right|_{\bm{\lambda}=\bm{\lambda}^{(s)}}\times\left[\nabla_\theta\log g\left(\bm{\theta}^{(s)},\zvec^{(s)}\right) -\nabla_\theta\log q^0_{\lambda^{(s)}}(\bm{\theta}^{(s)})\right]$ 
		\item[(d)] Compute step size $\rhovec^{(s)}$ using the ADADELTA method.
		\item[(e)] Set  $\bm\lambda^{(s+1)}=\bm\lambda^{(s)}+\rhovec^{(s)}\circ
		\widehat{\nabla_{\lambda}\mathcal{L}\left(\bm{\lambda}^{(s)}\right)} $
		\item[(f)] Set $s = s+1$
	\end{itemize}
	If stopping rule not satisfied go to step (a)\\ \vspace{5pt}
	\hrule
	\vspace{5pt}
	\caption{SGA Algorithm to calibrate our proposed variational approximation}
	\label{alg:VB}
\end{algorithm}

\subsection{Marginal approximation}\label{ssec:copula}
The final ingredient of our VA at~\eqref{EQ:approximation} is 
$q_\lambda^0(\thetavec)$, along with a matching re-parameterization transformation $\thetavec=h^0(\varepsilonvec^0,\lambdavec)$.  
An advantage of our approach is that any existing variational family
can be used for $q^0_\lambda$. Popular candidates include mean field
Gaussian or elliptical distribution approximations, although we stress this is not the same as employing
such approximations for the entire augmented posterior $p(\thetavec,\zvec|\yvec)$.
Below we discuss two flexible choices for $q^0_\lambda$ that are generic in
that they do not exploit any model-specific properties of the posterior distributions. They are more accurate than mean field approximations, have
generative representation allowing easy computation of the
gradient at~\eqref{lbdgradexpr}, and involve computations that 
increase only linearly with the dimension of $\thetavec$.

\subsubsection{Gaussian approximation}\label{sec:gapprox}
Gaussian VAs are popular, but can be computationally burdensome or inaccurate
when an unrestricted covariance matrix is employed and the dimension $m$ is high.
\cite{miller2017} and~\cite{ong2017gaussian} use a factor covariance structure to reduce the number of variational parameters.
Here, $q^0_\lambda(\thetavec)=\phi_m(\thetavec;\muvec_\theta,B_\theta B_\theta^\top+D_\theta^2)$, where $B_\theta$ is an $(m\times k)$ matrix with $k<<m$ and 
zero upper triangular elements, 
$D_\theta=\mbox{diag}(\dvec_\theta)$ is a diagonal matrix, $\phi_m(\cdot;\avec,A)$ is the density of a $m$-dimensional $N(\avec,A)$ distribution,
and $\lambdavec=(\muvec_\theta^\top,\mbox{vech}(B_\theta)^\top,\dvec_\theta^\top)^\top$.  
The ``vech'' operator is the half-vectorization of a rectangular matrix (i.e.
the vectorization of the non-zero elements). 
The distribution has generative representation
$\thetavec=\muvec_\theta+B_\theta \zetavec_1 + \dvec_\theta \circ \zetavec_2$,  
where $\varepsilonvec^0=(\zetavec_1^\top,\zetavec_2^\top)^\top\sim N(\bm{0},I_{k+m})$,
which
defines the re-parameterization transformation $h^0$ and is used 
at step~(a) of Algorithm~\ref{alg:VB}. \cite{ong2017gaussian} derive
analytical expressions for $\nabla_{\theta}\log q_\lambda^0(\thetavec)$ and 
$\frac{\partial \thetavec}{\partial \lambdavec}^\top$ that can be used 
for fast evaluation of the unbiased 
estimate of the re-parameterization gradient 
at step~(c) of Algorithm~\ref{alg:VB}. When $k=0$ this approximation reduces to a product
of independent Gaussians (i.e. a mean field approximation), 
while these authors found that setting $k=5$ balanced accuracy and computations
for their examples.

\subsubsection{Gaussian copula approximation}\label{sec:gcopapprox}
Gaussian copula VAs are more flexible because they can accommodate different
marginal distributions for each element of $\thetavec$. \cite{han2016} suggest such an approximation,
but with nonparametric or mixture density margins, which are difficult to calibrate quickly. 
Instead, we follow
\cite{Smith2020} and employ a Gaussian copula VA constructed using an element-wise transformation\footnote{This is not to be confused with
	the transformation $h^0$ associated with the re-parameterization trick.}
of $\thetavec$ and a factor decomposition of the copula correlation matrix. 
We provide an outline here, but refer to the work of these authors for details. 

Let $t_{\gamma}:\mathbb{R} \rightarrow \mathbb{R}$ be a family of one-to-one transformations 
with  parameter vector $\gammavec$. If a parameter $\theta_i$ is constrained 
we first transform it to the real 
line; for example, with a scale or variance parameter we set $\theta_i$ to its
logarithm.
To construct $q^0_\lambda$,
we transform each parameter as $\vartheta_i=t_{\gamma_i}(\theta_i)$ and 
 adopt the Gaussian distribution  $\varthetavec=(\vartheta_1,\ldots,\vartheta_m)^\top \sim 
 N(\muvec_\vartheta,\Sigma_\vartheta)$. The density $q^0_\lambda$
 can be recovered by computing the Jacobian of the transformation, so that
 \begin{equation}
 q^0_\lambda(\thetavec)=\phi_m(\varthetavec;\muvec_\vartheta,\Sigma_\vartheta)\prod_{i=1}^m t_{\gamma_i}'(\theta_i)\,,
 \label{eq:q}
 \end{equation} 
 where the variational parameters are
 $\lambdavec^\top=(\gammavec_1^\top,\ldots,\gammavec_m^\top,\muvec_\vartheta^\top,
 \mbox{vech}(\Sigma_\vartheta)^\top)$ and
 $t_{\gamma_i}'(\theta_i)=\frac{d\vartheta_i}{d\theta_i}$.  
 Let
 $\muvec_{\vartheta}=(\mu_{\vartheta,1},\ldots,\mu_{\vartheta,m})^\top$
 and $\sigma_{\vartheta,i}^2$ be the $i$th leading diagonal element of $\Sigma_{\vartheta}$,
 then the marginal densities of the approximation are 
 \begin{equation}
 q^0_{\lambda_i}(\theta_i)=\phi_1(\vartheta_i,\mu_{\vartheta,i},\sigma_{\vartheta,i}^2)t_{\gamma_i}'(\theta_i)\,, \mbox{ for }i=1,\ldots,m\,,
 \label{eq:qi}
 \end{equation}
 with $\lambdavec_i^\top=(\gammavec_i^\top,\mu_{\vartheta,i},\sigma_{\vartheta,i}^2)$ 
 a sub-vector of $\lambdavec^\top$.
It is straightforward to show
that the distribution with density at~\eqref{eq:q} has a Gaussian copula. \cite{Smith2020} point
out that while $q^0_\lambda$ has a copula representation, it is more computationally 
efficient to utilize that at~\eqref{eq:q}, and we do so here.

To allow for large $m$, we adopt a factor 
structure for $\Sigma_\vartheta$ as above. 
Let $B_\vartheta$ be an $(m\times k)$ matrix with zeros in the upper triangle, $D_\vartheta=\mbox{diag}(\bm{d}_\vartheta)$ a diagonal matrix, then we assume that
$\Sigma_\vartheta = B_\vartheta B_\vartheta^\top+D_\vartheta^2$.
Thus, the variational parameters are $\lambdavec^\top=(\gammavec_1^\top,\ldots,\gammavec_m^\top,\muvec_\vartheta^\top,
\mbox{vech}(B_\vartheta)^\top,\dvec_\vartheta^\top)$. 
We note that this  
copula is equivalent to the Gaussian factor copula used to model data~\citep{oh2017}, although here it is a VA.
The distribution has generative representation
$\thetavec=h^0(\varepsilonvec^0,\lambdavec)=(t_{\gamma_1}^{-1}(\vartheta_1),\ldots,
t_{\gamma_m}^{-1}(\vartheta_m))^\top$, where
$\varthetavec=\muvec_\vartheta+B_\vartheta \zetavec_1 + \dvec_\vartheta \circ \zetavec_2$,  
and $\varepsilonvec^0=(\zetavec_1^\top,\zetavec_2^\top)^\top\sim N(\bm{0},I_{k+m})$.
Closed form expressions for $\frac{\partial\bm{\theta}}{\partial\bm{\lambda}}$
and $\nabla_\theta\log q^0_\lambda(\bm{\theta})$ 
required 
to compute~\eqref{eq:thm1} can be derived as in Appendix~\ref{app:gradients}.

We employ a transformation suggested by
\cite{yeojohnson2000} that has proven successful in transforming data to near normality. This extends the Box-Cox transformation to the entire real line
and, for $0<\gamma<2$, is given by
\[
t_\gamma(\theta)=
\left\{\begin{array}{cl}
-\frac{(-\theta+1)^{2-\gamma}-1}{2-\gamma} &\mbox{if }\theta<0\\
\frac{(\theta+1)^\gamma -1}{\gamma} &\mbox{if }\theta\geq 0\,.
\end{array} \right.
\]
When implementing SGA $t_\gamma$ is not evaluated,
but its (closed form) inverse $t_\gamma^{-1}(\vartheta)$
%\[
%t_\gamma^{-1}(\vartheta)=
%\left\{\begin{array}{cl}
%1-\left(1-\vartheta(2-\gamma)\right)^{1/(2-\gamma)}&\mbox{if }\vartheta<0 \\
%(1+\vartheta\gamma)^{1/\gamma}-1 &\mbox{if }\vartheta \geq 0\,,
%\end{array} \right.
%\] 
is repeatedly. With this transformation, the margins at~\eqref{eq:qi} are 
flexible parametric densities that can exhibit skew and/or excess kurtosis, as illustrated in~\cite{Smith2020}. Other
transformations may also be used for $t_\gamma$, producing alternative margins.

\subsection{Discussion of alternative variational approximations}\label{sec:litreview}
The alternative of applying generic approximations to the augmented posterior
$p(\thetavec,\zvec|\yvec)$, or the marginal posterior of the latent variables $p(\zvec|\yvec)$,
has been considered previously. For example,~\cite{braun2010} do so for the posterior of a multinomial logistic regression
augmented with random coefficient realizations, \cite{hui2017} for latent variables
in a generalized linear model, \cite{LoaizaSmith2018} for the augmented
posterior of a discrete-margined copula model, and \cite{archer2015} for the augmented posterior of a state space
model, among others. Such approaches require a much larger number of variational parameters, 
resulting in three drawbacks compared to adopting the VA at~\eqref{EQ:approximation}. First, additional error is introduced into the variational estimate of $p(\thetavec|\yvec)$ through imprecision in the approximation of the posterior of $\zvec$.
Second, there is 
an increased computational burden at each step of the SGA algorithm because 
the re-parameterization gradient %for such an approximation
is a much larger vector than that at~\eqref{eq:thm1}. Last, the SGA algorithm typically requires more steps, because additional
noise is introduced into the Monte Carlo estimate of the gradient. 

A seminal paper on variational inference for latent variable models is \cite{hoffman2013}. These authors 
consider mean field approximations in models with global model parameters and latent variables (which are called ``local variables''), 
and describe how sub-sampling methods can be used in the variational optimization. Their method
requires the model to have a conjugate exponential family structure. 
\cite{hoffman+b15} consider structured stochastic variational inference methods where
dependence between global parameters and local variables can be accommodated.  
They consider the possibility of using the exact conditional posterior for the latent variables as part
of the approximation, similar to our approach, but 
unlike our method theirs also requires conjugacy. 
\cite{tan+n14} consider a stochastic variational inference implementation of non-conjugate variational message passing
using sub-sampling which is useful for generalized linear mixed models, and some diagnostics for prior-data conflict 
checking.
\cite{tan18} consider affine transformations for 
re-parametrizations of latent variable models which improve accuracy, and 
\cite{nolan+mw19} consider efficient variational inference in models with multi-level
random effects structures.  

\cite{tran2017variational} consider variational inference in latent variable models when only an unbiased estimate of 
the likelihood is available. 
Their approach requires tuning log-likelihood estimates to achieve constant variance, and
\cite{gunawan+tk17} overcome this disadvantage using sub-sampling methods and re-parametrization gradient estimates
%implemented for large panel data models 
based on Fisher's identity.
Their method is related to ours, but our work differs from theirs in two main ways.
First, their applications to latent variable models are limited to random effects models for
panel data, and we extend to more general latent variable models such as state space models.
Second, \cite{gunawan+tk17} use importance sampling for integrating out any latent variables in gradient estimation, whereas here we use MCMC methods. The use of importance sampling limits applications to models with low-dimensional latent variables, unlike with our approach. For example, the panel data application in~\cite{gunawan+tk17} uses a logistic model with a scalar intercept random effect, whereas later we consider a tobit model with a larger random effect vector. \cite{tomasetti2019} consider a VA similar to~\eqref{EQ:approximation}, but
for a partition of the parameter vector, not for the augmented posterior of a
latent variable model. They 
do not use re-parameterization gradients, nor simulate from the conditional posterior, but instead use an importance sampler for sequential posterior
inference. 

There have been a number of general efforts to combine variational inference and MCMC. These include
\cite{salimans2015markov}, \cite{domke17}, \cite{li+tl17}, \cite{zhang2018ergodic} and \cite{ye+bdh20} 
among others, although the methods do not focus specifically on 
computation for latent variable models. The method of \cite{ruiz+t19}
does consider such models. The VA they consider
corresponds to a parametrized approximation to an initial value of an MCMC algorithm which is run for a fixed
small number of iterations. 
For the case of latent variable models, scalable amortized variational inference methods can be 
used in learning the variational parameters for the latent variables.     \cite{hoffman17} considers
SGA algorithms in which 
gradient estimates based on short MCMC runs starting from samples
from a VA are considered.  The approach is used for
marginal maximum likelihood computations in deep Gaussian latent variable models.

\section{Example: TVP-VAR-SV model for macro variables}\label{sec:eg1}
To illustrate our method, we use 
it to estimate a time-varying parameter vector autoregression with 
stochastic volatility. This complex time series model is a leading model for 
forecasting macroeconomic variables~\citep{clark2015,carriero2019} and we employ a  variant
suggested by~\cite{huber2020} where a horseshoe prior is 
used to regularize the time-varying parameters. 
This is a challenging example for two reasons. First, the TVP-VAR-SV model is an example of a
nonlinear state space model, for which mean field approximations of the augmented posterior
are known to be poor~\citep{wang2004,karl2016}. To address this, for state space
models
previous authors have used structured Gaussian or other parametric
VAs to the augmented posterior or marginal posterior of the latent states~\citep{ghahramani2000,daunizeau2009,archer2015,naesseth2017}. 
In contrast, our approach has a smaller number of variational parameters,
involves less computation and provides greater accuracy than these approaches.
%than even well-structured parsimonious Gaussian
%approximations to the augmented posterior. 
The second challenge is that horseshoe regularization produces posterior densities that are funnel-shaped and difficult to approximate.
To solve this problem we adopt the re-parameterization of~\cite{ingraham2017}.
 
\subsection{The model}\label{Sec:UCSV}
\subsubsection{Specification}
Let $\{\yvec_t\}_{t=1}^T$ be a time series of $N$ macroeconomic variables with $\yvec_t=(y_{1,t},y_{2,t},\ldots,y_{N,t})^\top$.
We employ the following conditionally Gaussian TVP-VAR-SV model:
\begin{eqnarray}
\yvec_t &=  &\betavec_{0,t}+\sum_{s=1}^p B_{s,t}\yvec_{t-s}+L_t\epsilonvec_t \,, \nonumber\\
\betavec_t &= &\betavec_{t-1} + \wvec_t \,, \nonumber  \\
h_{i,t}& = & \bar{h}_i+\rho^h_i (h_{i,t-1}-\bar{h}_i)+e_{i,t}\,,\mbox{ for }
i=1,\ldots,N\,. \label{eq:tvpvarsv1}
\end{eqnarray}
Here, 
%$\hvec_t=(h_{1,t},\ldots,h_{N,t})^\top$, 
$\epsilonvec_t=(\epsilon_{1,t},\ldots,\epsilon_{N,t})^\top\sim N(\bm{0},H_t)$, $H_t=\mbox{diag}(e^{h_{1,t}},\ldots,e^{h_{N,t}})$ is a diagonal matrix,
$L_t$ is a lower triangular matrix with unit-valued leading diagonal (i.e. a ``unitriangular'' matrix), $\betavec_{0,t}$ is an intercept vector and $B_{1,t},\ldots,B_{p,t}$ 
are $(N\times N)$ autoregressive parameter matrices. The intercept and autoregressive 
parameters
follow a random walk, with the $(pN^2+N)$-dimensional vector $\betavec_t^\top\equiv
(\betavec_{0,t}^\top,\mbox{vec}(B_{1,t})^\top,\ldots,\mbox{vec}(B_{p,t})^\top)$ and $\wvec_t\sim N(0,V)$ with $V=\mbox{diag}(v_1,\ldots,v_K)$. The $N(N-1)/2$ free elements of $L_t$ are 
assumed to follow independent random walks.
The terms
$h_{i,1},\ldots,h_{i,T}$ are the logarithm of the volatilities for the $i$th variable, and these follow a stationary
first order autoregression with mean $\bar{h}_i$, autoregressive parameter $-1<\rho^h_i<1$ 
and independent disturbances $e_{i,t}\sim N(0,\sigma_i^2)$. 
%Models of the form at~\eqref{eq:tvpvarsv1} are popular for the
%probabilistic forecasting of macroeconomic variables; see~\cite{clark2015,carriero2019,huber2020} and references
%therein.

\subsubsection{Horseshoe regularization}
Likelihood-based estimation
using the parameterization in~\eqref{eq:tvpvarsv1} is difficult, and following~\cite{huber2020} and others,
it is convenient to both transform the system and adopt a non-centered parameterization. As outlined in Appendix~\ref{app:tvp1}, this
results in $i=1,\ldots,N$ unrelated regressions
\begin{equation}
y_{i,t} = \bm{x}_{i,t}^\top \alphavec_i + \epsilon_{i,t}\,,\label{eq:tvpreg}
\end{equation}
where $\alphavec_i$ and $\xvec_{i,t}$ are vectors of dimension $J_i=2(pN+i)$.
Each equation
can then be estimated
separately. The vector $\xvec_{i,t}$ is a function
of both the observed times series values and time-varying latent variables
$\widetilde \etavec_{i,t}$ specified at~\eqref{eq:apptvp1} in Appendix~\ref{app:tvp1}. \cite{huber2020}~compare different priors
for the regularization of $\alphavec_i=(\alpha_{i,1},\ldots,\alpha_{i,J_i})^\top$, including
the horseshoe prior of~\cite{carvalho2010} which we also use here. 
For each element $\alpha_{i,j}$, the horseshoe is specified by adopting the hyper-priors
\[
\alpha_{i,j}|\xi_i,\chi_{i,j}~\sim N(0,\xi_i\chi_{i,j}),\ \  \chi_{i,j}|\nu_{i,j}\sim \mathcal{G}^{-1}\left(\frac{1}{2},\frac{1}{\nu_{i,j}}\right),\ \  \xi_i|\kappa_{i}\sim \mathcal{G}^{-1}\left(\frac{1}{2},\frac{1}{\kappa_{i}}\right),\ \  \nu_{i,1},\dots,\nu_{i,J_i},\kappa_{i}\sim \mathcal{G}^{-1}\left(\frac{1}{2},1\right),
\]
where $\mathcal{G}^{-1}(a,b)$ denotes an inverse gamma distribution
with parameters $a,b$. This provides both global and local shrinkage
of the coefficients in each vector $\alphavec_i$. 

\subsubsection{The augmented posterior}
For the $i$th equation, let   $\chivec_i=(\chi_{i,1},\ldots,\chi_{i,J_i})^\top$ and
$\nuvec_i=(\nu_1,\ldots,\nu_{J_i})^\top$. Then 
the parameter vector is $\bm{\theta}_i=\left(\bm{\alpha}_i^\top,\bm{\chi}_i^\top,\xi_i,\bm{\nu}_i^\top,\kappa_i,\bar{h}_i,\rho^h_i,\sigma_i^2\right)^\top$, which has a total of $3J_i+5$ elements. 
The latent variables 
consist of the $T$-dimensional vector of log-volatilities $\hvec_i=(h_{i,1},\ldots,h_{i,T})^\top$
and the $(TJ_i/2)$-dimensional vector $\widetilde \etavec_i=(\widetilde \etavec_{i,1}^\top,\ldots,\widetilde \etavec_{i,T}^\top)^\top$. 
While the likelihood is intractable because the latent variables cannot be 
integrated out analytically, the augmented posterior can be expressed in closed form.
Let $\yvec_{(i)}
\equiv (y_{i,1},\ldots,y_{i,T})^\top$ be the observations on
the $i$th macroeconomic variable, and $\yvec_{(\backslash i)}$ be the observations
on the other $N-1$ macroeconomic variables, then the augmented posterior is
\begin{eqnarray}
\lefteqn{p(\thetavec_i,\hvec_i,\widetilde \etavec_i|\yvec) \propto p(\yvec_{(i)}|\thetavec_i,\hvec_i,\widetilde \etavec_i,\yvec_{(\backslash i)})p(\hvec_i,\widetilde \etavec_i|\thetavec_i)p(\thetavec_i)} \nonumber\\
&& =
\prod_{t=1}^{T}\left\{\phi_{1}\left(y_{i,t};\xvec_{i,t}^\top
%\left(\sqrt{\xi_i}\left(\tauvec_i \circ \sqrt{\chivec_i}\right)\right)
\alphavec_i
,e^{h_{i,t}}\right)\right\}
\phi_{J_i/2}\left( \widetilde{\etavec}_{i,1}; \bm{0},I\right)
\phi_1\left(h_{i,1};\bar{h}_i,\frac{\sigma_i^2}{1-(\rho^h_i)^2}\right) \nonumber \\
&& \times 
\prod_{t=2}^T \left\{\phi_{J_i/2}\left( \widetilde{\bm{\eta}}_{i,t}; \widetilde{\bm{\eta}}_{i,t-1},I\right) \phi_1\left(h_{i,t};\bar{h}_i+\rho^h_i(h_{i,t-1}-\bar{h}_i),\sigma_i^2\right)\right\}
p(\thetavec_i)\,,\label{eq:eg1augpost}
\end{eqnarray}
where  observe that $p(\hvec_i,\widetilde \etavec_i|\thetavec_i)=p(\hvec_i|\thetavec_i)p(\widetilde \etavec_i|\thetavec_i)$. The remaining priors are as specified in~\cite{huber2020}, with
$\bar{h}_i\sim N(0,10^2)$, the beta prior $\frac{\rho^h_i+1}{2}\sim\mathcal{B}(25,5)$ and $\sigma_i^2\sim\mathcal{G}\left(\frac{1}{2},\frac{1}{2}\right)$.
These authors
provide an MCMC scheme to compute the
augmented posterior above. 

\subsection{Variational approximation}
The posterior densities of parameters that are regularized
using the horseshoe prior have pathological funnel-shaped geometries, making them difficult
to approximate~\citep{betancourt2015,ghosh2019}.
\cite{ingraham2017} suggest a ``non-centered'' re-parameterization
of $\alphavec_i$ that simplifies the posterior densities so that they are 
easier to approximate, 
where each element $\alpha_{i,j}=\tau_{i,j}\sqrt{\xi_i\chi_{i,j}}$ with  
$\tau_{i,j}\sim N(0,1)$. This re-parameterization is unnecessary when
using MCMC,
although doing so may affect sampling efficiency.
Let $\tauvec_{i}=(\tau_{i,1},\ldots,\tau_{i,J_i})^\top$, so that
$\alphavec_i=\sqrt{\xi_i}(\tauvec_i \circ \sqrt{\chivec_i})$, then the
model parameters that
we employ for variational inference
are $\bm{\theta}_i=\left(\bm{\tau}_i^\top,\bm{\chi}_i^\top,\xi_i,\bm{\nu}_i^\top,\kappa_i,\bar{h}_i,\rho^h_i,\sigma_i^2\right)^\top$ which remains of length
$3J_i+5$. Here,
 $\chivec_{i}$, $\xi_i$, $\nuvec_{i}$, $\sigma^2_i$
and $\kappa_i$ are all transformed to their logarithms, and $\rho^h_i$  to $\Phi_1^{-1}\left(\frac{\rho^h_i+1}{2}\right)$,
so that all parameters are unconstrained on the real line. 

Let $\zvec\equiv(\hvec_i,\widetilde \etavec_i)^\top$, then 
we approximate the augmented posterior at~\eqref{eq:eg1augpost}
using our suggested VA at~\eqref{EQ:approximation} with $q^0_\lambda$ the Gaussian factor model at Section~\ref{sec:gapprox} with $k=5$ factors.
The variational parameters
$\lambdavec$ are calibrated 
using Algorithm~\ref{alg:VB}.\footnote{We apply Algorithm~\ref{alg:VB} to estimate
	 each equation separately, so that the parameter vector is $\thetavec_i$ in the variational
	 approximation at~\eqref{EQ:approximation}. However, we continue to denote the latent variable
	 vector as $\zvec$ and variational parameters as $\lambdavec$ to be consistent with the notation
	 in Section~\ref{ssec:VAlatent}.}
At step~(b), drawing directly from $p(\zvec|\thetavec_i^{(s)},\yvec)$ is infeasible.
Therefore, we draw $\zvec$ by generating between one and five sweeps
from the densities $p(\hvec_i|\bm{y},\bm{\theta}_i^{(s)},\widetilde{\etavec}_i^{(s-1)})$ and $p(\widetilde{\etavec}_i|\bm{y},\bm{\theta}_i^{(s)},\bm{h}_i^{(s)})$ using standard 
state space methods detailed
in Appendix~\ref{app:tvp2} and initialized at the value of $\zvec$ from the previous
SGA step.
While this provides only an approximate draw from 
$p(\zvec|\thetavec_i^{(s)},\yvec)$, it is fast and we find it provides accurate estimates as documented below, although 
other approaches to generating
a draw may also be used.
When computing the re-parameterization gradient at~\eqref{eq:thm1},
$\nabla_{\theta_i} \log g(\thetavec_i,\zvec)$ is 
available in closed form; see the Online Appendix.

The MCMC estimator of~\cite{huber2020} is used with their recommended number of draws (15,000 burn-in and 15,000 Monte Carlo sample) to compute the ``exact'' posterior, and we employ it to
judge the accuracy of our VA. 
For comparison we also calibrate the following Gaussian VA
\begin{equation}\label{Eq:UCSV_VA_competing}
q_\lambda\left(\bm{\theta}_i,\bm{z}\right) = \phi_{K_i}\left(\left(\bm{\theta}_i^\top,\widetilde \etavec_i^\top\right)^\top;\bm{\mu}_{G,\theta},\Sigma_G\right)\phi_{T}\left(\bm{h}_i;\bm{\mu}_{G,h},C_{G,h}C_{G,h}^\top\right)\,,
\end{equation}
where $K_i=J_i(3+T/2)+5$,
$C_{G,h}$ is a band one lower triangular Cholesky factor, and
a factor decomposition with five factors is used for $\Sigma_G$. The banded structure
of the covariance matrix $C_{G,h}C_{G,h}^\top$ can better capture the Markov
dependence in $\hvec_i$ than a factor decomposition; see~\cite{quiroz+nk18} for a
similar structured Gaussian VA.

\subsection{Empirical results}
The TVP-VAR-SV with $p=2$ lags is used to model the same data and 
eight macroeconomic variables used by~\cite{huber2020} that these authors label
the ``medium case''. The data are quarterly observations 
from 1980:Q3 to 2017:Q4, so that  $T=150$. In the eight equations, the
number of parameters $|\thetavec_i|$ is between 107 and 149, while the 
number of latent variables $|\zvec|$ is between 2700 and 3750.
The purpose of this empirical analysis is to demonstrate the high degree
of accuracy of our proposed VA.
\subsubsection{Estimates}
\begin{figure}[H]
 	\caption{Comparison of posterior mean estimates of the latent volatilities
 	in the TVP-VAR-SV example}
	\begin{center}
		\includegraphics[width=0.95\textwidth]{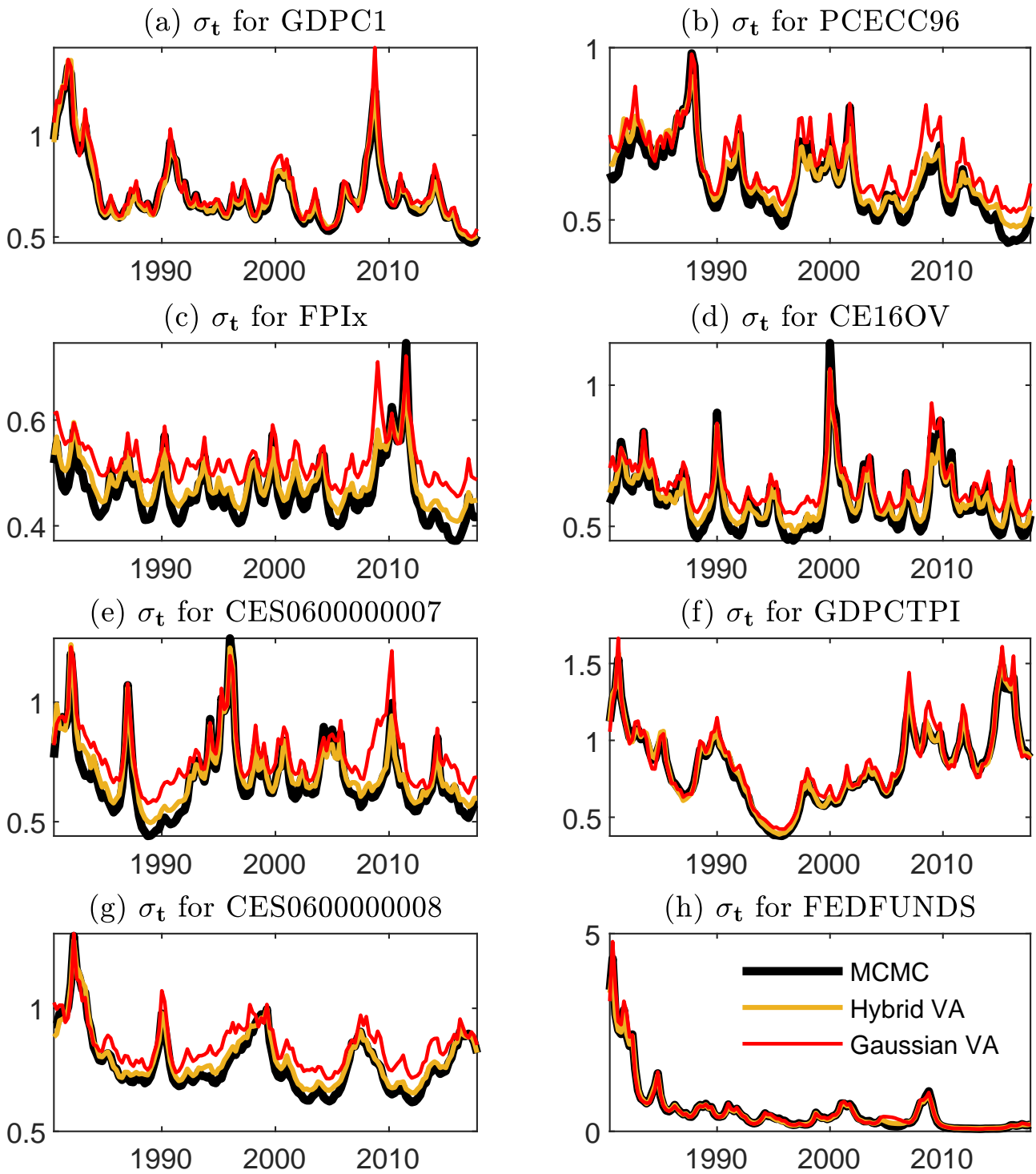}
	\end{center}
	Each panel plots the disturbance standard deviations $\sigma_{i,t}=\exp(h_{i,t}/2)$, for $t=1,\ldots,T$, for a different macroeconomic variable $i=1,\ldots,8$. The eight variables are defined in 
	\cite{huber2020}. The exact posterior mean (computed by MCMC) is plotted in black, the mean of the hybrid VA at~\eqref{EQ:approximation} in yellow, and the mean of the structured Gaussian VA in red.
	\label{fig:States_posteriors}
\end{figure}

\begin{figure}[H]	
 \caption{Parameter posterior density estimates for the Real GDP equation 
 in the TVP-VAR-SV example}
\begin{center}
\includegraphics[width=0.7\textwidth]{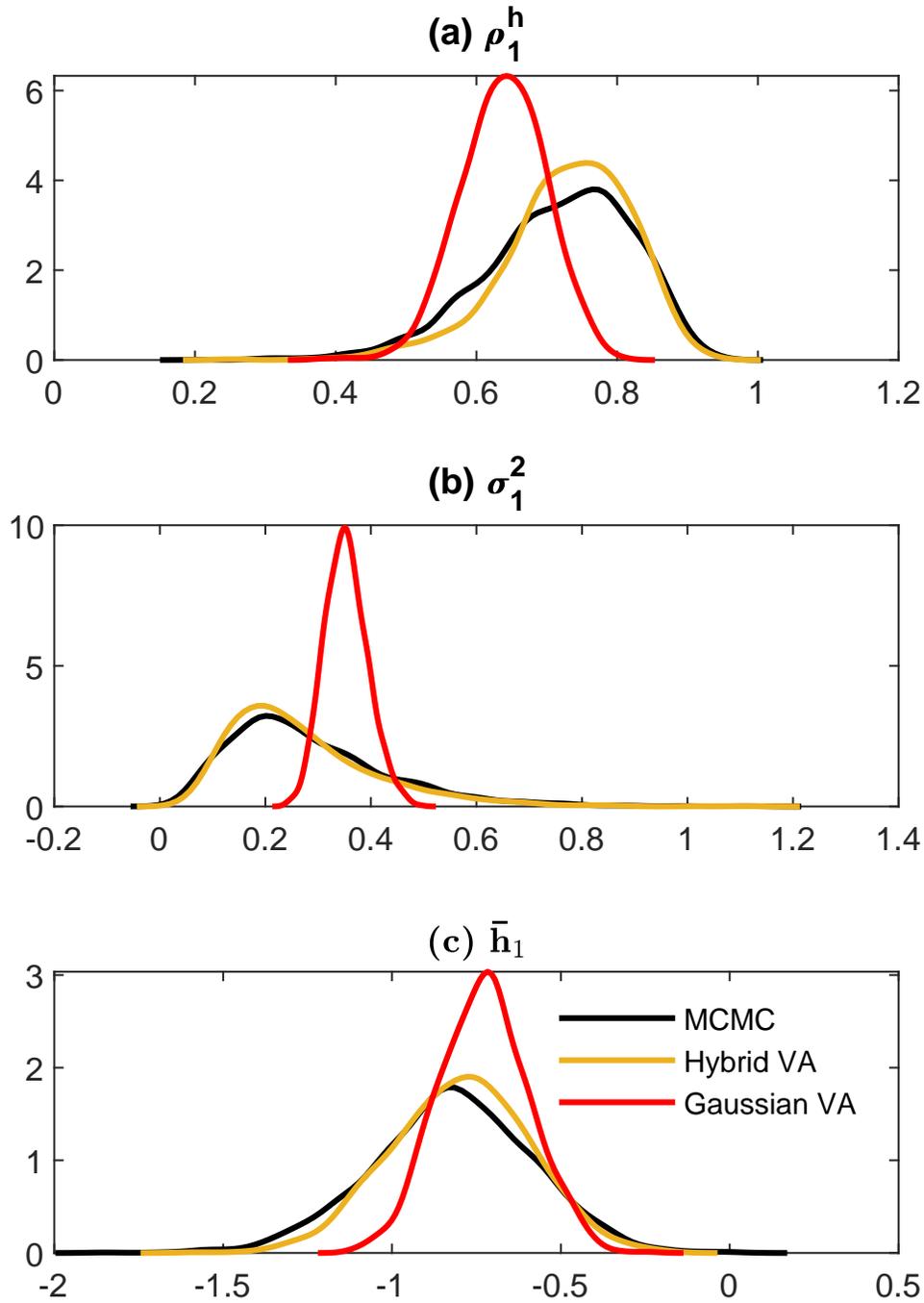}
\end{center}
Marginal posterior density estimates of $\rho_1^h,\sigma_1^2,\bar{h}_1$ 
for the Real GDP equation of the
TVP-VAR-SV model. Exact posterior estimates (computed by MCMC) are plotted in black, 
the hybrid VA at~\eqref{EQ:approximation} in yellow,
and the structured Gaussian VA in red. Results for the other seven equations can be found in the Online Appendix.\label{fig:Parameter_posteriors}
\end{figure}

Figure~\ref{fig:States_posteriors} plots the exact posterior mean of 
the time-varying standard deviations $\exp(h_{i,t}/2)$ for each of the 
eight series, along with the mean of $\exp(h_{i,t}/2)$ for both calibrated VAs.
Our proposed VA (labeled ``Hybrid VA'') is more accurate than the structured Gaussian VA. Figure~\ref{fig:Parameter_posteriors} plots the exact posterior 
densities of 
$(\rho_i^h,\bar{h}_i,\sigma^2_i)$, along with those of the two VAs, for the variable Real GDP (equation $i=1$). 
Again, the densities from the structured Gaussian VA are
poor, whereas those from our method are accurate. The Online Appendix
provides the equivalent densities for the other seven variables, and 
very similar results are observed. Note that an accurate approximation for
$\hvec_i$ is necessary to also obtain an accurate approximation of $(\rho_i^h,\bar{h}_i,\sigma^2_i)$,
and vice-versa.

\begin{figure}[H]	
	\caption{Posterior estimates of time-varying autoregressive coefficients in the TVP-VAR-SV example}
	\begin{center}
		\includegraphics[scale=0.7]{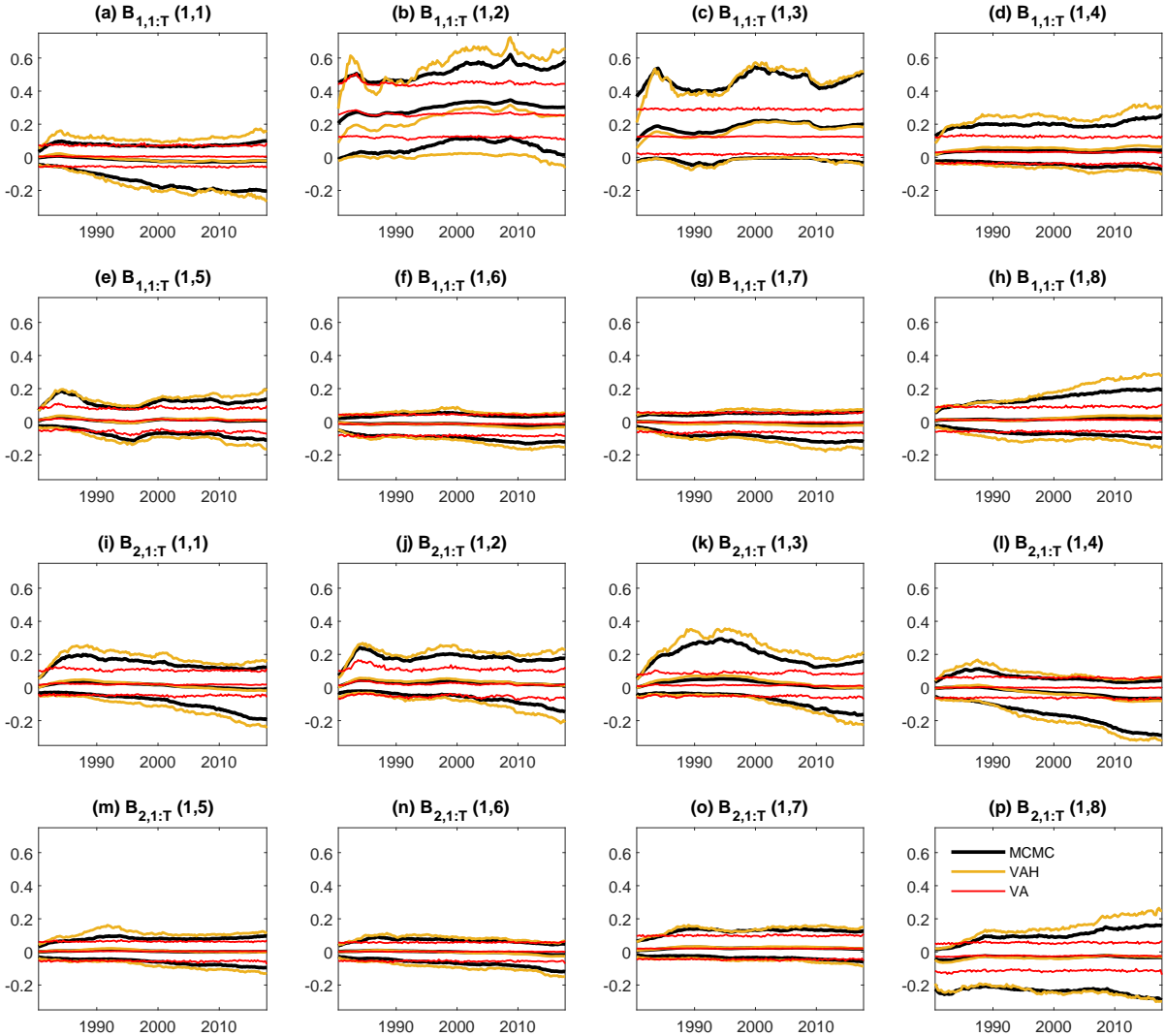}
	\end{center}
	Each panel plots the posterior mean and 90\% intervals of 
	an element in a time-varying autoregressive coefficient matrix $B_{s,t}$ against time $t=1,\ldots,T$.  The top eight panels are the 
	elements in the first row of $B_{1,t}$, and the bottom eight panels
	are the elements in the first row of $B_{2,t}$; these are the coefficients
	for the Real GDP equation. Exact posterior estimates (computed by MCMC) are plotted in black, those for the hybrid VA at~\eqref{EQ:approximation} in yellow,
	and that for the structured Gaussian VA in red. Equivalent plots for all other autoregressive coefficients can be found in the Online Appendix.\label{fig:Parameter_posteriors_betas}
\end{figure}

Figure~\ref{fig:Parameter_posteriors_betas}
plots the exact posterior of a subset of the time-varying
autoregressive coefficients in black. These are the first rows of $B_{1,t}$ and $B_{2,t}$
which correspond to the first equation (for Real GDP). Both the posterior
mean and the upper/lower 90\% intervals are plotted. Some coefficients
(e.g. those in panels~(f,g,m,n,o)) are heavily regularized to zero, whereas others
vary substantially over time (e.g. those in panels~(b,c,i,j,k,p)). The equivalent
mean and intervals are also plotted for the structured Gaussian VA (in red) and
our hybrid VA (in yellow). The latter is much more accurate than the former. 
This posterior output for the VAs was computed using Monte Carlo draws of $(\thetavec_i,\zvec)$
from the calibrated VAs, and from which draws of $B_{1,t}$ and $B_{2,t}$ were evaluated using the relationships in Appendix~\ref{app:tvpreg}. Estimates of
the time-varying lower unitriangular matrix $L_t$ (see the Online Appendix)
also show that our hybrid VA is accurate, and the structured Gaussian 
VA much less so.

\begin{figure}[H]
 	 	\caption{Accuracy of latent volatility estimates from the hybrid VA in the TVP-VAR-SV example}
	\begin{center}
		\includegraphics[width=0.45\textwidth]{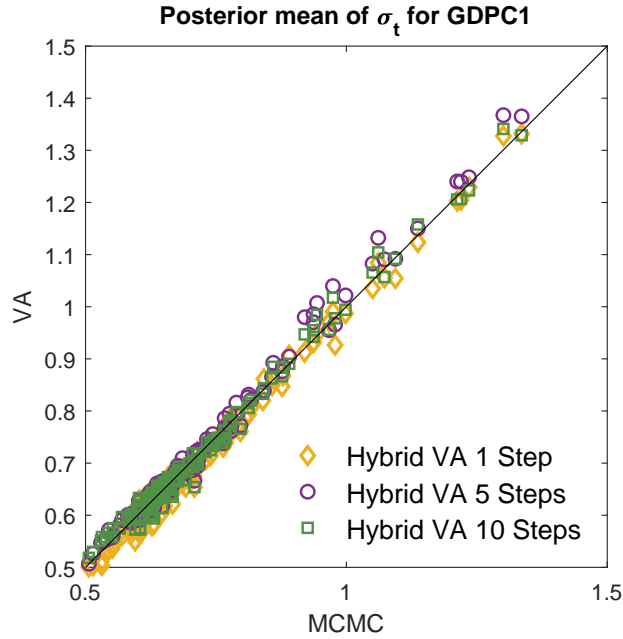}
	\end{center}
	Results are given for the latent standard deviations of Real GDP,  $\sigma_{1,t}=\exp(h_{1,t}/2)$, for $t=1,\ldots,T$. 
	The three scatter plots are of the exact posterior means (computed using MCMC) against the means of the hybrid VA at~\eqref{EQ:approximation} computed
	using Algorithm~\ref{alg:VB} with 1, 5 and 10 sweeps
	of a Gibbs sampler at step~(b) of the algorithm. 
	\label{fig:States_Compare_Gibbs}
\end{figure}

Last, we explore
the impact of drawing $\zvec$
using multiple sweeps of a Gibbs sampler at step~(b) of Algorithm~\ref{alg:VB}. Figure~\ref{fig:States_Compare_Gibbs} plots the means of the VAs of
$\exp{(h_{1,t}/2)}$ against their exact posterior means when 1, 5 and 10 sweeps
of a Gibbs sampler are used 
at step~(b). While a greater number of sweeps results in increased
accuracy, it is only a minor improvement. Similar results are observed
for the other seven equations.

\subsubsection{Accuracy and calibration speed}
It is typical to judge the accuracy and calibration speed of different
VAs using the 
lower bound ${\cal L}$. However, this is unavailable for the VA at~\eqref{EQ:approximation} because
the density $p(\zvec|\thetavec_i,\yvec)$ is intractable. Therefore,
we instead compute the accuracy of the one-step-ahead 
posterior predictive densities
estimated using each VA. Let $\zvec_t\equiv (\widetilde \etavec_{i,t},h_{i,t})$
and $\yvec_{1:t}=(\yvec_1^\top,\ldots,\yvec_t^\top)^\top$, then the predictive density
is computed for the TVP-VAR-SV model as
\begin{eqnarray*}
	p_{t+1|t}(y_{i,t+1}|\thetavec_i,\zvec_{t})&\equiv &\int \int p(y_{i,t+1}|\widetilde{\etavec}_{i,t+1},h_{i,t+1},\thetavec_i,\yvec_{1:t})p(\widetilde{\etavec}_{i,t+1}|\widetilde{\etavec}_{i,t})p(h_{i,t+1}|h_{i,t},\thetavec_i) \mbox{d}\widetilde{\etavec}_{i,t+1}\mbox{d}h_{i,t+1}\\
	&= &\int \phi_1\left(y_{i,t+1};\,\widetilde\xvec_{i,t+1}^\top \etavec_{i,0}+\widetilde \xvec_{i,t+1}^\top \mbox{diag}(\sqrt{\vvec_i})\widetilde\etavec_{i,t},\,\widetilde \xvec_{i,t+1}^\top \mbox{diag}(\vvec_i)\widetilde \xvec_{i,t+1}+e^{h_{i,t+1}}\right)\times\\
	& & \ \ \ \ \phi_1\left(h_{i,t+1};\bar{h}_i+\rho_i^h(h_{i,t}-\bar{h}_i),\sigma^2_i\right)\mbox{d}h_{i,t+1}\,.
\end{eqnarray*}
%\begin{eqnarray*}
%p_{t+1|t}(y_{t+1}|\thetavec,\zvec_t) &= &\int \int 
%p(y_{t+1}|\mu_{t+1},\eta_{t+1},\thetavec)p(\mu_{t+1},\eta_{t+1}|\mu_t,
%\eta_{t},\thetavec) \mbox{d}\mu_{t+1}\mbox{d}\eta_{t+1}\\
% &= &\int \phi_1(y_{t+1};\bar{\mu}+\rho_\mu(\mu_t-\bar{\mu}),\sigma^2_\mu+e^{\eta_{t+1}})
% \phi_1(\eta_{t+1};\bar{\eta}+\rho_\eta(\eta_t-\bar{\eta}),\sigma^2_\eta)\mbox{d}\eta_{t+1}\,,
%\end{eqnarray*}
Here, the integral in $\widetilde{\etavec}_{i,t+1}$ is evaluated analytically 
by recognising a Gaussian density as in Appendix~\ref{app:tvp1}, and the integral
in $h_{i,t+1}$ is computed numerically. The accuracy of these predictive densities
is measured using their average Kullback-Leibler divergence
over time when computed from the VA and also from the exact posterior as follows.
Let $\bar{\thetavec}_i=E(\thetavec_i|\yvec)$ and $\bar{\zvec}_{t}=E(\zvec_{t}|\yvec)$
be the exact posterior means (computed using MCMC), and 
$\widetilde{\thetavec}_i,\widetilde{\zvec}_t$ be the means from the VA $q_\lambda$.
The average Kullback-Leibler divergence is then
$\overline{\mbox{KL}}(\lambdavec)=\frac{1}{T}\sum_{t=1}^T\mbox{KL}_{t+1|t}(\lambdavec)$, where
\[
\mbox{KL}_{t+1|t}(\lambdavec)=\int p_{t+1|t}(y_{i,t+1}|\widetilde{\thetavec}_i,\widetilde{\zvec}_{t})
\log\left(\frac{p_{t+1|t}(y_{i,t+1}|\widetilde{\thetavec}_i,\widetilde{\zvec}_{t})}{p_{t+1|t}(y_{i,t+1}|\bar{\thetavec}_i,\bar{\zvec}_{t})}\right)\mbox{d}y_{i,t+1}\,,
\] 
is computed by numerical integration. Lower values of $\overline{\mbox{KL}}(\lambdavec)$ suggest the VA $q_\lambdavec$ has increased accuracy, with 
$\overline{\mbox{KL}}(\lambdavec)=0$ when the VA is exact (i.e. when $q_\lambdavec(\thetavec_i,\zvec)=p(\thetavec_i,\zvec|\yvec)$).

\begin{figure}[H]
	\caption{Average KL divergence between variational and exact posterior predictive densities for Real GDP in the TVP-VAR-SV example}
	\begin{center}
		\includegraphics[scale=0.8]{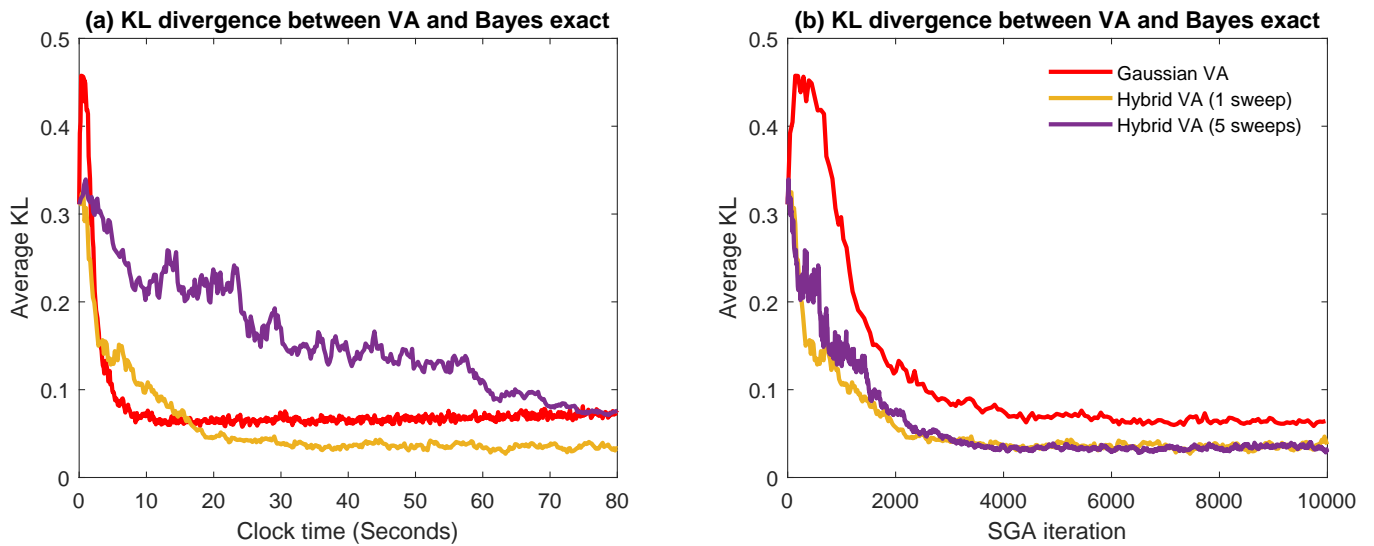}
	\end{center}
	\label{fig:KLcompare}
\end{figure}

We compute this diagnostic for the variable Real GDP (i.e. equation $i=1$) although note that
results for the other seven variables are similar. The minimum value of $\overline{\mbox{KL}}(\lambdavec)$
was 0.0578 and 0.0282 for the Gaussian and our hybrid VA, respectively,
so that 
our VA is substantially more accurate than the benchmark by this
metric,
consistent with the other empirical results given above. Last, 
Figure~\ref{fig:KLcompare} plots $\overline{\mbox{KL}}(\lambdavec)$ for our hybrid VA with 1 and 5 sweeps
of the sampler at step~(b) of Algorithm~\ref{alg:VB}, and also the structured Gaussian VA. Panel~(a) plots the divergence against walk clock time, and panel~(b) against step number of the SGA, with
all methods coded in MATLAB and run on a standard laptop.
A single sweep
of the sampler appears sufficient to calibrate the posterior predictive densities well by this metric. Calibration 
of the VA takes about 30s for one sweep and 180s for five sweeps, whereas in comparison the MCMC sampler took 246s using the same computing environment.
%Using an absolute change in $\overline{\mbox{KL}}(\lambdavec)$ of
%0.0001 as a stopping rule, coded in MATLAB and run
%on a standard laptop,
%the Gaussian VA took 4960 steps in 8.17s, whereas the Hybrid 
%VA took 1403 steps in 1.73s. It is often observed in 
%variational inference that more accurate 
%VAs can be faster to calibrate using optimization methods.

%Last, while unreported here, we also used a Gaussian copula VA~\citep{Smith2020} with the same covariance matrix structure outlined in~\eqref{Eq:UCSV_VA_competing}, but found accounting for marginal asymmetry in this way provided only a very
%minor improvement in VA accuracy.`
\section{Example: Mixed Tobit Model for Disaggregate Sales}\label{sec:eg2}
The second example applies our variational estimator to a mixed effects tobit model for $T=100$ weekly sales by members of a large consumer panel of U.S. customers from the field of marketing.
The panel originates from~\cite{danaher2020}, who
create a rich set of covariates 
that they model with both  
fixed and individual-level random coefficients. The authors 
point out that exact Bayesian estimation using MCMC is computationally
infeasible for large numbers of customers from this panel, and use variational inference with a structured Gaussian VA as in~\cite{ong2017gaussian} to fit models for 4,000 individuals.
Using this panel, we first show that for a small sample of 1,000
individuals (for which  MCMC can be used), our proposed VA is very close to the exact posterior 
and also more accurate than either the Gaussian VA used
by~\cite{danaher2020} or a mean field Gaussian VA. 
We then apply our approach to a large sample of 20,000 customers (so that the number
of weekly sales observations is two million) and show it 
improves upon both Gaussian VAs. 

%model for one million weekly sales amounts for a panel of 10,000 U.S. customers over $T=100$ weeks. The 
%panel originates from~\cite{danaher2020} and these authors 
%create a rich set of covariates 
%that they model with both  
%fixed and individual-level random coefficients. The authors 
%point out that exact Bayesian estimation using MCMC is 
%not computationally
%viable for the full dataset, and use variational inference with a structured Gaussian VA as in~\cite{ong2017gaussian}. We show here that for a sub-sample of 100
%individuals, our proposed VA is very close to the exact posterior
%and more accurate than either the Gaussian VA used
%by~\cite{danaher2020} or a mean field Gaussian VA. 
%We then apply our approach to the full dataset and show it 
%improves upon both Gaussian VAs. 

\subsection{The model}
The response $y_{i,t}$ for individual
$i = \{1,\dots,N\}$ at week $t = \{1,\dots,T\}$ in a tobit model is an observation of
a latent variable $y^\star_{i,t}$ censored at zero, so that 
$y_{i,t}=y^{*}_{i,t}$ if $ y^{*}_{i,t}>0$, and $y_{i,t}=0$ if $y^{*}_{i,t}\leq 0$.
The latent response follows a Gaussian mixed effects model
\[
y^{*}_{i,t} = \bm{x}_{i,t}^\top\bm{\beta}+\wvec_{i,t}^\top\bm{\alpha}_i+\sigma\epsilon_{i,t}\,,\;\;\epsilon_{i,t}\sim N(0,1)\,,\;\; \alphavec_i\sim N(\bm{0},V_\alpha)\,,
\]
where $\xvec_{i,t}$ is a $(p\times 1)$ vector of fixed effect covariates, 
$\wvec_{i,t}$ is a $(r\times 1)$ vector of random effect covariates that 
is a sub-vector of $\xvec_{i,t}$.
Bayesian analysis of panel data in marketing is popular
because estimates of random coefficient values are often
a key output~\citep{allenby1998,manchanda2004response}, as is the case
here with $\alphavec=(\alphavec_1^\top,\ldots,\alphavec_N^\top)^\top$.
A Bayesian analysis requires specification of priors. Here we set $V_\alpha=LL^\top+\Omega$, with $L$ a $r\times k_\alpha$ factor loading matrix (with zeros in the upper triangle and positive leading diagonal elements for identification) and the diagonal matrix $\Omega=\mbox{diag(\omegavec)}$, and then adopt 
independent uninformative 
priors for $\mbox{vech}(L),\omegavec,\betavec,\sigma^2$ as 
outlined in the Online Appendix.

\subsection{Estimation}
Let $\yvec$ be the vector of the $n=NT$ values
of $y_{i,t}$, and $\yvec^\star_{{\tiny U}}$ be the vector of values of $y^\star_{i,t}$ which are unobserved (i.e. those where $y_{i,t}=0$), and set
$\eta_{i,t}=\bm{x}_{i,t}^\top\bm{\beta}+\wvec_{i,t}^\top\bm{\alpha}_i$.
The likelihood $p(\thetavec|\yvec)$ is intractable, 
so the focus is often on the posterior augmented with $\alphavec$, which 
has density
\begin{equation}
p(\bm{\alpha},\bm{\theta}|\bm{y}) \propto \prod_{\left\{i,t|y_{i,t}=0\right\}}\Phi_1\left(0;\eta_{i,t},\sigma^2\right) \prod_{\left\{i,t|y_{i,t}>0\right\}}\phi_1(y_{i,t};\eta_{i,t},\sigma^2)\prod_{i=1}^N\phi_{r}\left(\bm{\alpha}_i;\bm{0},V_\alpha\right)
p(\thetavec)\,.\label{eq:tobitlike}
\end{equation}
In many studies, including ours, weekly individual-level sales
values $y_{i,t}$ are mostly zero, so that the first product at~\eqref{eq:tobitlike} is over more terms than the second. For $r$ and/or $N$ large,
integration over $\alphavec$ to obtain the likelihood is 
computationally difficult, and MCMC methods that simulate
$\alphavec$ are popular. 

It is often simpler and  
faster to consider the posterior augmented with both $\alphavec$
and $\yvec^\star_{\tiny U}$. This can be derived by first noting that
$p(\bm{y}^*_{\tiny U}|\bm{y},\bm{\alpha},\bm{\theta})=
\prod_{\{i,t|y_{i,t}=0\}}p(y_{i,t}^*|y_{i,t},\bm{\alpha},\bm{\theta})$, with
\begin{equation*}
p\left(y_{i,t}^*|y_{i,t},\bm{\alpha},\bm{\theta}\right)= \left\{\begin{array}{lr}
\frac{\phi_1\left(y^{*}_{i,t};\eta_{i,t},\sigma^2\right)}{\Phi_1\left(0;\eta_{i,t},\sigma^2\right)}I\left(y^{*}_{i,t}\le 0\right) & \text{for } y^{*}_{i,t}< 0\\
I\left[y^{*}_{i,t}=y_{i,t}\right]& \text{for } y^{*}_{i,t}\geq 0\,.
\end{array}\right.
\end{equation*}
Then the augmented posterior is 
\begin{align}
p(\bm{y}^*_{\tiny U},\bm{\alpha},\bm{\theta}|\bm{y}) = & 
p(\bm{y}^*_{\tiny U}|\bm{y},\bm{\alpha},\bm{\theta})p(\alphavec,\thetavec|\yvec) \nonumber \\
%\propto 
%\, p(\bm{y}^*|\bm{y},\bm{\alpha},\bm{\theta})
%p(\bm{y}|\bm{\alpha},\bm{\theta})
%p(\bm{\alpha}|\thetavec)
%p(\bm{\theta})\nonumber\\
\propto&\left[\prod_{i=1,t=1}^{N,T}\phi_1\left(y^{*}_{i,t};\eta_{i,t},\sigma^2\right)\right]\left[\prod_{i=1}^N\phi_{r}\left(\bm{\alpha}_i;\bm{0},V_\alpha\right)\right]
p\left(\sigma^2\right)p\left(\bm{\beta}\right)p\left(\mbox{vech}(L)\right)p\left(\bm{\omega}\right)\,,\label{eq:tobitaugpost}
\end{align}
which was computed by substituting in the expressions above. Using~\eqref{eq:tobitaugpost} a Gibbs sampler that generates from 
both $\yvec^\star_{\tiny U}$ and $\alphavec$
and evaluates the
augmented posterior exactly is given in the Online Appendix. 
However, even MCMC methods applied to either~\eqref{eq:tobitlike} or~\eqref{eq:tobitaugpost} become impractical for larger values of $r$ and/or $N$, and~\cite{danaher2020} propose using variational inference instead.

The model is estimated using our approach with 
$\zvec=\left((\yvec^\star_{\tiny U})^\top,\alphavec^\top\right)^\top$ and 
parameter vector
$\thetavec=(\betavec^\top,\mbox{vech}(L)^\top,\omegavec^\top,\sigma^2)^\top$, where some elements are transformed to the real line
for ready application of the VA as outlined in the Online Appendix.
Thus, the VA is to the augmented posterior at~\eqref{eq:tobitaugpost},
where the Gaussian copula at Section~\ref{sec:gcopapprox} with $k=10$ factors
is used for $q^0_\lambda$.
When implementing Algorithm~\ref{alg:VB}, at step~(b)
 $\zvec$ is generated from 
the conditionals $p(\alphavec|(\yvec^\star_{\tiny U})^{(s-1)},\thetavec^{(s)},\yvec)$ and $p(\yvec^\star_{\tiny U}|\alphavec^{(s)},\thetavec^{(s)}),\yvec)$ given in the Online
Appendix, and initialized at the iterate from the previous SGA step. This forms a Gibbs sampler to produce a draw from $p(\zvec|\thetavec^{(s)},\yvec)$, and five sweeps are used to produce our results.

Two Gaussian VAs to~\eqref{eq:tobitlike} are included as benchmarks, which 
are strong contenders because $\yvec_{\tiny U}^*$ (but not $\alphavec$) is integrated out
from the posterior exactly. The
first VA is that proposed by~\cite{danaher2020} with 
density 
$q^{G}_\lambda(\thetavec,\alphavec)=q^{G,a}_{\lambda^a}(\thetavec)
\prod_{i=1}^N
q^{G,i}_{\lambda^i}(\alphavec)$. Here, $q^{G,a}_{\lambda^a}$ and $q^{G,i}_{\lambda^i}$ are the densities of Gaussian factor VAs as in~\cite{ong2017gaussian}, and $\lambdavec^\top=\left((\lambdavec^a)^\top,(\lambdavec^1)^\top,
\ldots,(\lambdavec^N)^\top\right)$. 
One factor is used for $q^{G,i}_{\lambda^i}$, while
10 factors are used for $q^{G,a}_{\lambda^a}$ to produce a
richer approximation. The second benchmark is a Gaussian
mean field VA (i.e. where the VA density is a product of univariate Gaussian densities), which is a popular choice in practice for models with many
parameters.
Both benchmarks are calibrated using SGA with the re-parameterization trick.
% as discussed in~\cite{ong2017gaussian}.

\subsection{Empricial results: small data}
To assess the accuracy of our proposed VA we first consider weekly sales by 
a sample of $N=1000$ individuals, for which the exact posterior
can be calculated in reasonable time (27 hours) using MCMC.
There are
$p=32$ covariates which are described in the Online Appendix, and 
include 
measurements of an individual's exposure to advertisements (ads)
in three media (email, catalogs and paid search) to their in-store and online purchases for three retailer-brands (``B1'', ``B2'' and ``B3'') in the clothing category. The response $y_{i,t}$ 
is the logarithm of in-store spend (plus one) for retailer-brand B1 (called
the ``focal brand'' here) by individual $i$ in week $t$.
We employ random coefficients at the individual consumer level 
for the three media ad exposure variables of the focal brand and the intercept, so that $r=4$. 
 
\begin{figure}[H]
		\caption{Posterior distribution of the random effect variance $V_\alpha$ for the small tobit example.}
	\begin{center}
		\includegraphics[scale =0.5]{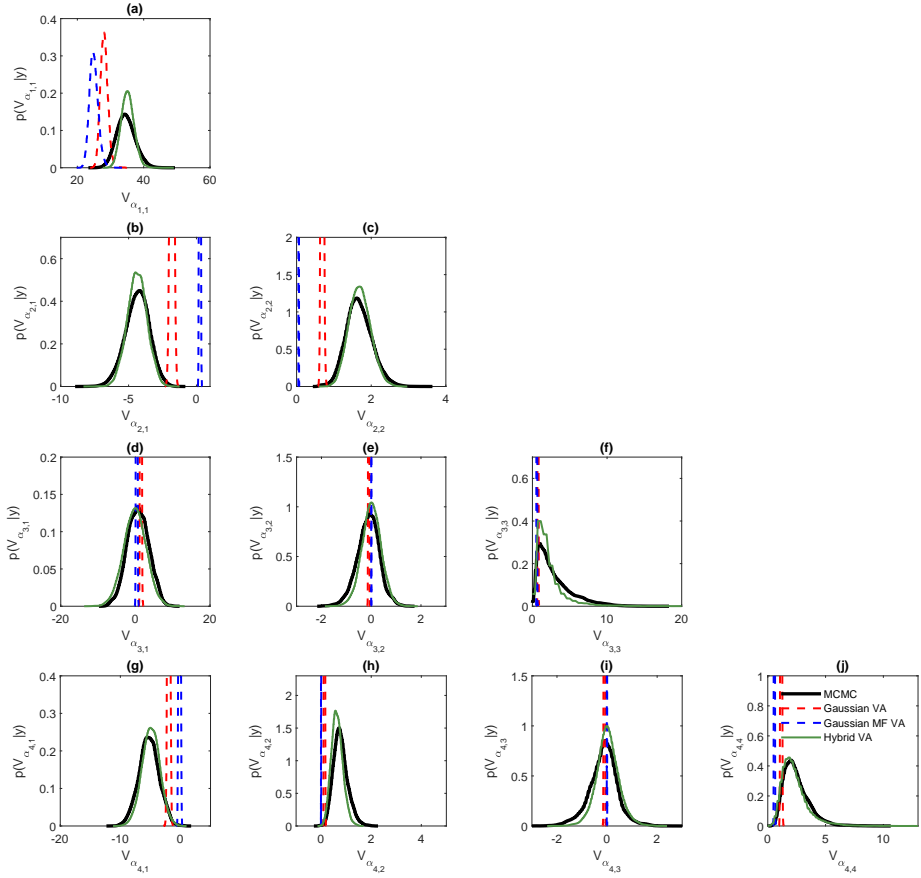}
	\end{center}
The panels give the distributions for the lower triangular elements in the 
$(4 \times 4)$ matrix $V_\alpha$. The black line depicts the exact posteriors computed using MCMC.
The blue dashed line depicts the Gaussian Mean Field VA, the red dashed line the Gaussian VA, and our proposed estimator as a green thin line. In a number of panels (e.g. panel~i) the two Gaussian VA estimates are 
so similar they are indistinguishable (i.e. the lines sit on top
of one another).
	\label{fig:postV}
\end{figure}

%%In this small data case, the exact
%%posterior can be computed
%%using a (slow) Gibbs sampler that generates from both $\yvec^\star$ and 
%%$\alphavec$; see the Online Appendix. 
%Figure~\ref{fig:poteriormoments} plots the exact posterior mean and standard deviation of $\thetavec$ (computed using MCMC)
%against the mean and standard deviation of the three calibrated VAs. 
%The mean field VA is poor, greatly understating the posterior
%variance for all parameters. 
%The Gaussian VA is better, but our proposed VA (labeled `hybrid')
%is closest to the true posterior. In particular, the posterior mean is well-approximated, although there are a few parameters with understated variance,
%which is common in variational inference in general. 
Figure~\ref{fig:postV} presents
the posteriors of the elements of $V_\alpha$, where there is a striking
difference between the three VAs. Both Gaussian VAs under-estimate the level
of posterior variance, and are incorrectly located 
for some elements.
In contrast, our proposed hybrid VA estimates the posteriors
very well-- not only getting the correct location and variance, but also the skewness for all elements of $V_\alpha$. Part~B.4 of the Online Appendix shows
that the hybrid VA also approximates
the posterior of $\thetavec$ more accurately
than the two Gaussian VAs.

%The hybrid VA results took XXX h to compute, compared to 4.19 hours for MCMC. 

\begin{figure}[H]
	\caption{Accuracy of the random coefficient estimates
		for the tobit small data example.} 
	\begin{center}
		\includegraphics[width=0.8\textwidth]{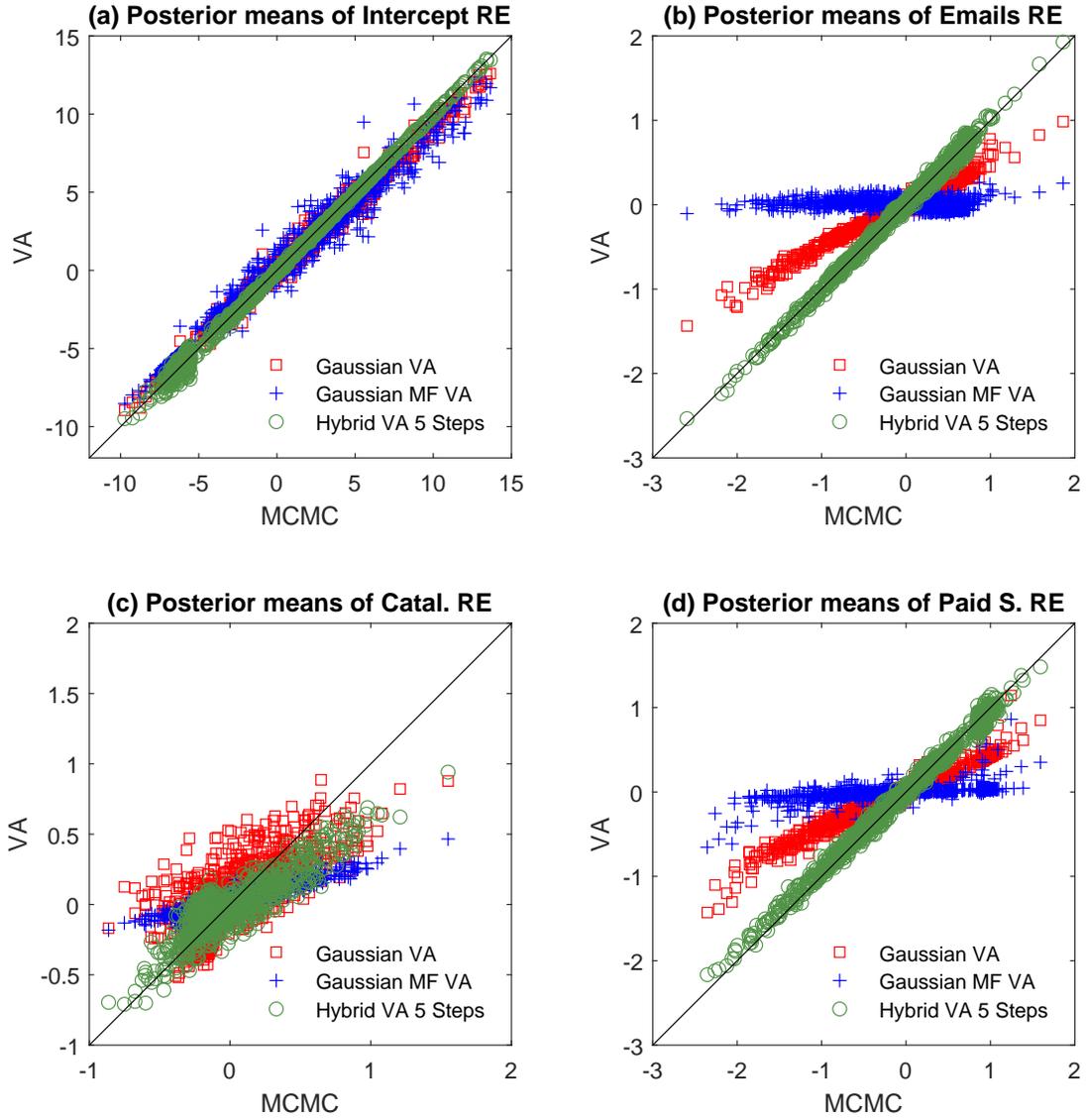}
	\end{center}
	Scatter-plots of VB mean estimates of the $Nr=4000$ random coefficients $\alphavec$ against their true posterior means computed
	using MCMC. Panels~(a) to~(d) correspond the four random coefficients. 
	Accurate estimates have scatters on the 45 degree line. 
	Results are given for the Gaussian mean field approximation (blue scatter),
	the Gaussian approximation with factor structure (red scatter), and our proposed VA using 5 sweeps
	of a Gibbs sampler at step~(b) of Algorithm~1 (green scatter).
	\label{fig:RE_posteriors_gibbs_tobit_VBG}
\end{figure}

\begin{figure}[H]
	\caption{Effect of more sweeps 
		for the small tobit example.} 
	\begin{center}
		\includegraphics[width=0.8\textwidth]{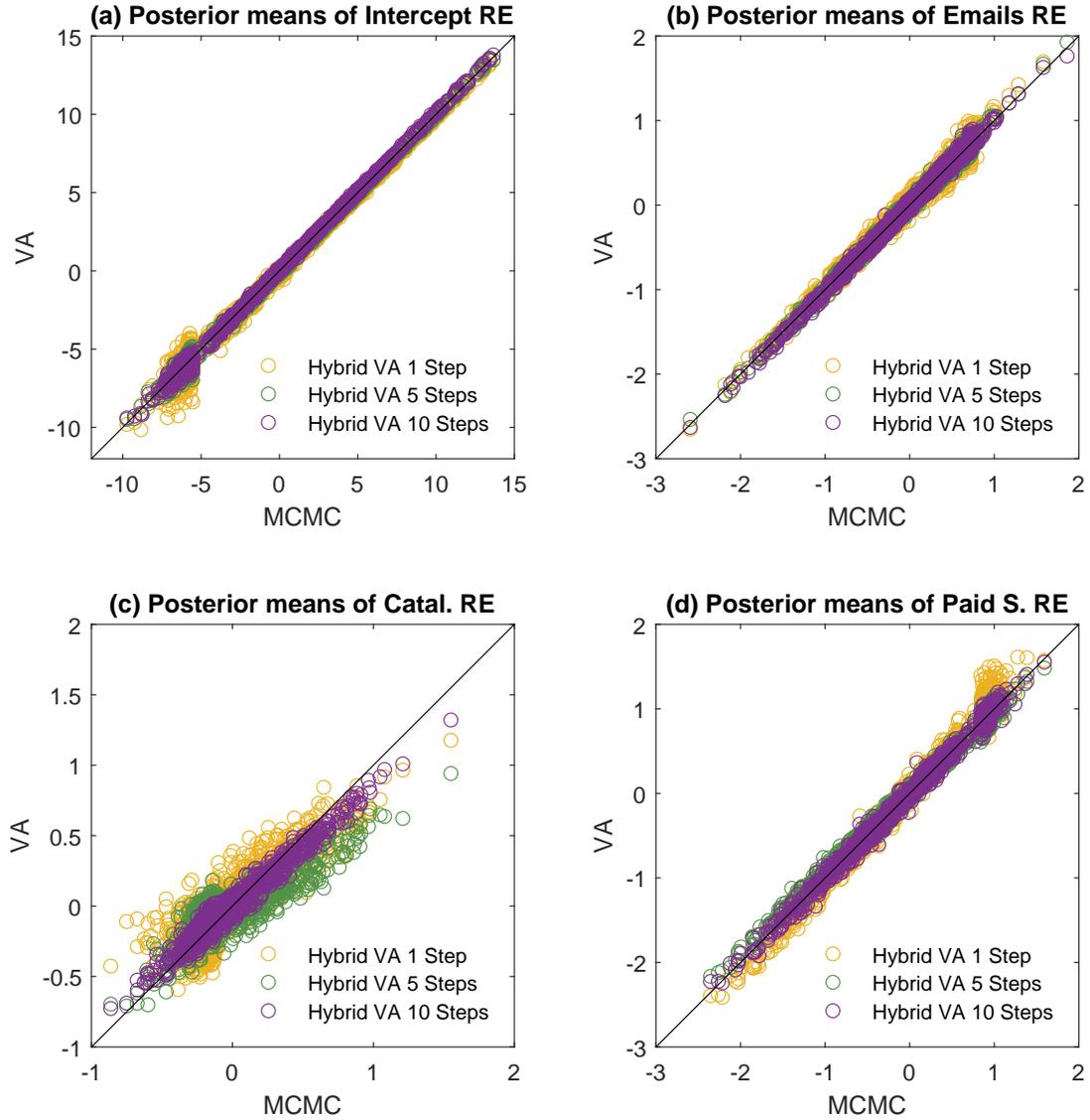}
	\end{center}
Scatter-plots of VB mean estimates of the $Nr=4000$ random coefficients $\alphavec$ against their true posterior means computed
using MCMC. Panels~(a) to~(d) correspond the four random coefficients. Accurate estimates have scatters on the 45 degree line.
Results are given for our proposed VA using 1, 5 and 10 sweeps
of a Gibbs sampler at step~(b) of Algorithm~1.
	\label{fig:RE_posteriors_gibbs_tobit}
\end{figure}

A key output of the study by~\cite{danaher2020} are estimates of the random coefficient values $\alphavec$. To judge
their accuracy, Figure~\ref{fig:RE_posteriors_gibbs_tobit_VBG}
plots the $rN=4,000$ point estimates against their exact posterior means
using the three different VAs. Scatters more closely aligned 
along the 45 degree line indicate increased accuracy, and the hybrid VA is much
more accurate than both Gaussian VAs. Figure~\ref{fig:RE_posteriors_gibbs_tobit} produces the same plot 
for posterior of $\alphavec$, but for the hybrid VA when 1, 5 and 10 sweeps of the Gibbs sampler are used
at step~(b) of Algorithm~\ref{alg:VB}. Increasing the number of 
sweeps improves the estimation accuracy of $\alphavec$, although
a reasonable degree is obtained with only five sweeps. Additional
results for this example are in the Online Appendix.

\subsection{Empirical results: large data}\label{sec:tobitlarge}
\subsubsection{Estimates and inference}
We now apply our method to the large dataset using the 
same response but for $N=20,000$ individuals
and the same $r=10$ random coefficients considered by~\cite{danaher2020}, for which exact
posterior inference cannot be easily computed.
Table~\ref{tab:tobitbeta} reports the estimates of
$\betavec$ and $\sigma$ for the three VAs. The results are broadly consistent,
although the choice of VA affects some key parameters. 
For example, the coefficient
of ``Log Price'' (a key effect in the study) is $-0.271$ with 
a 95\% posterior interval of $(-0.361,-0.178)$ 
using our VA, compared to $-0.169$ with a 95\% posterior
interval of $(-0.333, -0.004)$ using the Gaussian factor VA. 
Given the retail category is
off-the-peg clothing, a sizable 
negative coefficient is to be expected.
Similarly, the coefficient of ``B1 Catalog'' is twice as large
in our analysis at $1.335$, compared to $0.751$ using the Gaussian VA,
suggesting that focal brand advertising through catalogs
is more effective. 
 
Table~\ref{tab:REvariance} reports the 
estimate of $V_\alpha$. In~\cite{danaher2020} the correlations were estimated as close to zero,
whereas with the larger sample and our  VA we find many to be non-zero. In particular,
the correlations between the intercept and the three focal brand advertising variables are all negative ($-0.62, -0.468, -0.248$) suggesting that individuals that are heavier buyers of the focal brand (i.e. with a larger intercept
random effect value) are less 
affected by advertising; a marketing insight not made previously.
A key objective of the original study was to measure
the individual-level heterogeneity in the effects of focal and cross brand advertising, and we 
assess this here as a function of 
$V_\alpha$ as follows. For individual $i$, let $\wvec_{i,t}^{B1}$
be the three focal brand advertising spend covariates,
and $\wvec_{i,t}^{B23}$
be the six other brand advertising spend covariates. Further, let $V_\alpha^{B1}$ and $V_\alpha^{B23}$ 
be the corresponding sub-matrices of $V_\alpha$. Then 
for each VA we compute
the distribution of the following three measures of heterogeneity:
\begin{itemize}
	\item[] Total Heterogeneity: $\mbox{TH}(V_\alpha)=
	%\frac{1}{Nr}\sum_{i,t} \mbox{Var}(\wvec_{i,t}^\top \alphavec_i)=
	\frac{1}{Nr}\sum_{i,t}\wvec_{i,t}'V_\alpha \wvec_{i,t}$.
	\item[] Focal Brand Ad Heterogeneity: $\mbox{FBH}(V_\alpha)=
	%\frac{1}{Nr}\sum_{i,t} \mbox{Var}((\wvec_{i,t}^{B1})^\top \alphavec_i^{B1})=
	\frac{1}{Nr}\sum_{i,t}(\wvec_{i,t}^{B1})^\top V_\alpha^{B1} \wvec_{i,t}^{B1}$.
	\item[] Cross Brand Ad Heterogeneity: $\mbox{CBH}(V_\alpha)=
	%\frac{1}{Nr}\sum_{i,t} \mbox{Var}((\wvec_{i,t}^{B23})^\top \alphavec_i^{B23})=
	\frac{1}{Nr}\sum_{i,t}(\wvec_{i,t}^{B23})^\top V_\alpha^{B23} \wvec_{i,t}^{B23}$.
\end{itemize}
Table~\ref{tab:TH} reports their estimates, and total heterogeneity is similar for both dependent VAs.
However,
the hybrid VA estimates a substantially higher level of 
heterogeneity in advertising effectiveness, particularly for 
cross-brand advertising.

\subsubsection{Calibration speed}
\begin{figure}[H]
	\caption{Comparison of calibration speeds for the large data tobit example.}
	\begin{center}
		\includegraphics[width=1\textwidth]{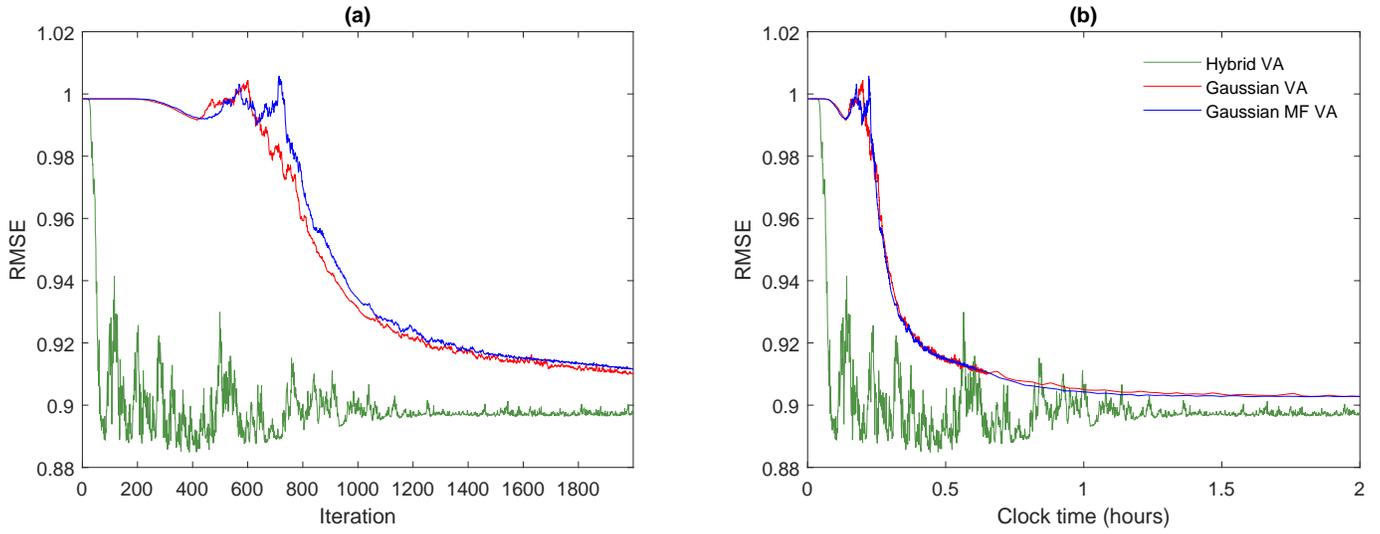}
	\end{center}
Calibration accuracy is measured by the $\mbox{RMSE}(\alphavec,\thetavec)$. This is plotted
against both (a)~step number, and (b)~ clock time, for the Gaussian factor (red),
mean field Gaussian (blue), and our proposed hybrid (green) variational approximations.
	\label{fig:RMSE_trace}
\end{figure}

We measure calibration speed using the point 
predictions $\hat{y}_{i,t}=E(y_{i,t}|\alphavec_i,\thetavec)$,
which can be calculated analytically from the tobit model.
From these we can 
compute
the root mean squared error  
$\mbox{RMSE}(\alphavec,\thetavec)=(\frac{1}{NT}(\sum_{i,t}(y_{i,t}-\hat{y}_{i,t})^2))^{1/2}$ for the values of $\alphavec,\thetavec$ obtained during the SGA. 
Figure~\ref{fig:RMSE_trace} plots the RMSE against both step number and clock time for all
three VAs. By this metric, convergence of the SGA is several times faster for our proposed
VA than for the two Gaussian benchmarks. 

The estimated computation time for the MCMC sampler 
is between 13.6 and 54.4 days, depending on the number
of draws required, as discussed in the Online Appendix (Part~B.4). In contrast, 15,000 steps of Algorithm~\ref{alg:VB} were sufficient to calibrate our hybrid VA in 15.8 hours
using 5 sweeps at step~(b),
although less sweeps or
sub-sampling decreases computation time further as discussed below.
\section{Sub-sampling variational inference}\label{sec:subsampling}
Sub-sampling methods that use only mini-batches of data at each step of an SGA algorithm are common in machine learning for large data sets and have two main 
advantages. The first is that the need to read the whole data set into computer memory is avoided, which is important in some applications. A second advantage is that a favourable trade-off between computation time per iteration and gradient variance can be made, 
reducing overall computation time to reach convergence.  

\subsection{Fast sub-sampling for models with latent variables}
Many models with latent variables have an augmented posterior density that
factors as 
\[
p(\thetavec,\zvec|\yvec)\propto g(\thetavec,\zvec) = \prod_{i=1}^n p(\yvec_i,\zvec_i|\thetavec) p(\thetavec)\,,
\]
where $\zvec^\top=(\zvec_1^\top,\ldots,\zvec_n^\top)$ and 
$\yvec^\top=(\yvec_1^\top,\ldots,\yvec_n^\top)$.
For these models the VA at~\eqref{EQ:approximation} provides a convenient 
framework to implement sub-sampling. 
%Following~\cite{gunawan+tk17}, 
Let
$\widehat{G}(\thetavec,\zvec,\uvec)$ be an unbiased estimator of 
$G(\thetavec,\zvec) \equiv \nabla_\theta \log g(\thetavec,\zvec)$, where $\uvec\sim f_u$ are a set of random variables determining a sub-sampling mechanism. 
%For example, $\uvec$ may be
%uniforms that define a random sample without replacement from the indices $\{1,2,\ldots,n\}$. 
Then the re-parameterization gradient at Theorem~\ref{thm:gradient} 
can be re-written as the expectation
\begin{equation*}
	\nabla_\lambda\calL(\lambdavec) =E_{f_{\varepsilon,u}}\left(\frac{\partial\bm{\theta}}{\partial\bm{\lambda}}^\top\left[ \widehat G\left(\bm{\theta},\zvec,\uvec\right) -\nabla_\theta\log q^0_\lambda(\bm{\theta})\right]\right)\,,
\end{equation*}
where $f_{\varepsilon,u}(\varepsilonvec,\uvec)=f_\varepsilon(\varepsilonvec)f_u(\uvec)$.
An estimator with one draw
$\varepsilonvec^{(s)}=(\varepsilonvec^{0,(s)},\zvec^{(s)})\sim f_\varepsilon$
(and thus also $\thetavec^{(s)}=h^0(\varepsilonvec^{0,(s)},\lambdavec)$) and 
one draw $\uvec^{(s)}\sim f_u$, is simply
\begin{equation}
	\widehat{\nabla_\lambda\calL(\lambdavec)} =\frac{\partial\bm{\theta}^{(s)}}{\partial\bm{\lambda}}^\top
	\left[\widehat G\left(\bm{\theta}^{(s)},\zvec^{(s)},\uvec^{(s)}\right) -\nabla_\theta\log q^0_\lambda(\bm{\theta}^{(s)})\right]
	\label{eq:ssrpg}\,.
\end{equation}

The key to successful sub-sampling within variational inference is
the choice of $\widehat G$.
\cite{gunawan+tk17} propose an unbiased estimator for models without 
latent variables based on a Taylor series
expansion around $\thetavec=\bar{\thetavec}$ with coefficients that only 
need to be computed once. However, this is not applicable to the VA at~\eqref{EQ:approximation} because these coefficients would be functions of
$\zvec$ and have to be re-computed every step of the SGA, negating
the computational improvements. For random coefficient models
\cite{gunawan+tk17}  also suggest
integrating out
$\zvec$ using importance sampling, but this is also prohibitively slow when the number and complexity of the random coefficients is high. We use a more
conventional sub-sampling approach that is much faster here.
Let $k_i(\thetavec,\zvec_i)=\nabla_\theta \log p(\yvec_i,\zvec_i|\thetavec)$, so that
$G(\thetavec,\zvec)=\sum_{i=1}^n k_i(\thetavec,\zvec_i)+\nabla_\theta \log p(\thetavec)$. Consider a sub-sample $S(\uvec)\subset \{1,2,\ldots,n\}$ of size
$|S(\uvec)|$ with $\uvec\sim f_u$ (e.g. uniform random variables 
that correspond to a simple random sample without replacement), then we use
the unbiased estimator of $G(\thetavec,\zvec)$ below:
\[\widehat{G}(\thetavec,\zvec,\uvec)=\frac{n}{|S(\uvec)|}\sum_{j\in S(\uvec)}
k_j(\thetavec,\zvec_j) + \nabla_\theta \log p(\thetavec)\,.
\] 
Employing~\eqref{eq:ssrpg} with this approximation produces two immediate computational
savings in Algorithm~\ref{alg:VB}: (i) only $\zvec_i$ for $i\in S(\uvec)$ have to be generated at 
step~(b),  and (ii)~the evaluation of  $\widehat{G}(\thetavec,\zvec,\uvec)$ is much faster
than $G$ at step~(c). The trade-off is that more steps of the SGA may be necessary 
as the gradient approximation at~\eqref{eq:ssrpg}
is likely to have higher variance.

\subsection{Example: mixed tobit model for disaggregate sales}
To illustrate the potential of sub-sampling, we
employ it to compute variational inference for the tobit model with the augmented 
posterior at~\eqref{eq:tobitaugpost} factored across individuals $N$. We consider
the large dataset in Section~\ref{sec:tobitlarge} and sample at random without replacement 5\%, 10\%, 25\% and 50\% of the individuals at each step of the SGA. Table~\ref{tab:tobitss} reports the speed (time per SGA step) and the 
RMSE metric averaged over the last half of a 10,000 step run of Algorithm~\ref{alg:VB}. It does
so for sub-samples of different size, as well as for 1 to 5 sweeps of the Gibbs sampler at step~(b) of the 
algorithm. By the RMSE metric, sub-samples of size 25\% give comparable estimates
to the full sample, even when only one sweep is used at step~(b), resulting in a substantial 
decrease in computation time. 
\section{Discussion}\label{sec:discuss}
Variational inference has great potential for big models; especially those that employ a large number of 
latent variables. However, popular
mean field and structured Gaussian VAs for the augmented posterior can lack accuracy, and we present a more accurate VA for this case. 
It combines a parametric approximation to the marginal
 posterior of the global parameters with the exact conditional posterior of the latent variables. This removes the approximation 
error due to the latent variables, also
improving calibration of the VA of the global parameter posterior as in Corollary~\ref{thm:betterapprox}. 
Our VA admits an efficient representation of the re-parameterization gradient in Theorem~\ref{thm:gradient}
that allows the SGA algorithm to be implemented 
without requiring computation of $p(\zvec|\thetavec,\yvec)$
or its gradient. However, our approach requires a draw from this distribution
for the estimate of $\nabla_\lambda {\cal L}(\lambdavec)$ to be unbiased. For 
more complex latent variable models this may be difficult or slow. This is the case
in our two applications, although we 
find that even an approximate draw using a few sweeps of simple Gibbs samplers  
produces much more accurate VAs than Gaussian approximations to the augmented
posterior. One major advantage of our method is that it provides a flexible framework
for incorporating MCMC within variational inference.
  For augmented posteriors that can be factorized---a common case for latent variable models---our VA is well-suited to sub-sampling variational inference. The two applications demonstrate the generality and accuracy of our method. In particular, the 
tobit example shows how our approach can be applied to large data sets, and
how sub-sampling is easily implemented.

We finish with some comments on using our approach in practice. First, 
in the computation of Bayesian inference,
the parameters and latent variables in a model are often 
treated similarly. In such cases, 
we recommend selecting $\zvec$ based on the viability 
of sampling efficiently from their conditional posterior at step~(b) of Algorithm~\ref{alg:VB}. 
Second, while any tractable parametric density
can be used for $q_\lambda^0$, either
a Gaussian or Gaussian copula with a factor correlation structure is
a good default choice. \cite{ong2017gaussian} and~\cite{Smith2020} show they are scalable and
well-suited to the re-parameterization trick, improving
the performance of SGA optimization. The efficient
selection of the number of factors $k$ is an open problem, but these authors
found that between three and ten factors worked well in a wide range of applications.
In our empirical work we show that our choices for $k$ produce VAs that are fast to calibrate and accurate. 
Finally, we list here the circumstances where our approach has potential to provide much 
faster Bayesian inference than MCMC, while retaining a high degree of accuracy. They include whenever: (i) generation from $p(\thetavec|\zvec,y)$ in MCMC is either slow or difficult; (ii) generating a draw from the 
conditional posterior at step~(b) is fast and accurate; and, (iii) the augmented posterior is factorizable so as to allow
sub-sampling.

%Finally, we point out some interesting extensions to our method.  
%A second extension of our methodology is to use new MCMC coupling methods \citep{jacob+oa20} at 
%step~(d) to compute exactly unbiased gradient estimates. 
%Such coupling
%methods can be employed when there is no convenient blocked Gibbs sampler available, and alternative methods like
%Hamiltonian Monte Carlo can be used for the MCMC implementation \citep{heng+j19}. Combining coupling with 
%sub-sampling may be particularly attractive, because the sampling then only needs to occur in a lower-dimensional space if all the latent
%variables do not need to be considered at once. Implementing such a scheme would involve a careful balance between
%reducing gradient variance, which
%requires sufficient large sub-samples, and reducing coupling times, which would require smaller sub-samples and hence
%sampling the conditional posterior for a lower-dimensional subset of the latent variables.  Investigating the
%practicality of these ideas is left to future work.
\newpage
\onehalfspacing
\newpage
\appendix
\section{Proofs}\label{app:proof}
\subsection{Proof of Theorem~\ref{thm:gradient}}
First, because $\calL(\lambdavec) =\calL^0(\lambdavec)=E_{q_\lambda}\left[\log p(\bm{y}|\bm{\theta})+\log p(\bm{\theta})-\log q^0_\lambda\left(\bm{\theta}\right)\right]$, the re-parameterization
gradient of $\calL$ is the same as that of $\calL^0$, so that
\begin{align}\label{Eq:proof}
\nabla_\lambda\calL(\lambdavec) =& E_{f_{\varepsilon^0}}\left\{\frac{\partial\bm{\theta}}{\partial\bm{\lambda}}^\top\left[\nabla_\theta\log p\left(\bm{\theta}\right)+\nabla_\theta\log p\left(\bm{y}|\bm{\theta}\right) -\nabla_\theta\log q^0_\lambda\left(\bm{\theta}\right)\right]\right\}\,.
\end{align}
Here, the
random vector $\varepsilonvec^0$ has density $f_{\varepsilonvec^0}$
that does not depend on $\lambdavec$, and $h^0$ is the
one-to-one vector-valued re-parameterization transformation from $\varepsilonvec^0$
to the parameter vector, such that $\thetavec=h^0(\varepsilonvec^0,\lambdavec)$.

Next, note that Fisher's identity gives (see, for example Equation~(4) of~\cite{poyiadjis2011})
\begin{equation*}
\nabla_\theta\log p\left(\bm{y}|\bm{\theta}\right) =\int\nabla_\theta\left[\log p\left(\bm{y}|\zvec,\bm{\theta}\right)p\left(\zvec|\bm{\theta}\right)\right]p\left(\zvec|\bm{\theta},\bm{y}\right)d\zvec\,.
\end{equation*}
Substituting this expression into Equation (\ref{Eq:proof}), and
writing $E_{f_\varepsilon}\left(.\right)$ for expectation with respect to $f_{\varepsilon}(\varepsilonvec)=f_{\varepsilon^0}\left(\varepsilonvec^0\right)p\left(\zvec|\bm{\theta},\bm{y}\right)$, 
and because $g(\bm{\theta},\zvec)=p(\yvec|\zvec,\thetavec)p(\zvec|\thetavec)p(\thetavec)$, we get
\begin{align*}
\nabla_\lambda\calL(\lambdavec) =& E_{f_\varepsilon}\left\{\frac{\partial\bm{\theta}}{\partial\bm{\lambda}}^\top\left[\nabla_\theta\log p\left(\bm{\theta}\right)+\nabla_\theta\log p\left(\zvec|\bm{\theta}\right)+\nabla_\theta\log p\left(\bm{y}|\zvec,\bm{\theta}\right) -\nabla_\theta\log q^0_\lambda\left(\bm{\theta}\right)\right]\right\}\\
=&E_{f_\varepsilon}\left\{\frac{\partial\bm{\theta}}{\partial\bm{\lambda}}^\top\left[\nabla_\theta\log g\left(\bm{\theta},\zvec\right) -\nabla_\theta\log q^0_\lambda\left(\bm{\theta}\right)\right]\right\}\,,
\end{align*}
which is the required result.

\subsection{Proof of Corollary~\ref{thm:betterapprox}}
For any approximating density $q(\thetavec,\zvec)=q(\thetavec)q(\zvec|\thetavec)$ of
$p(\thetavec,\zvec|\yvec)$,
the Kullback-Leibler divergence is
\begin{align}
	\text{KL}\left(q(\thetavec,\zvec)|| p(\thetavec,\zvec|\yvec)\right) & = \text{KL}\left(q(\thetavec)||p(\thetavec|\yvec)\right)+
	\int \text{KL}\left(q(\zvec|\thetavec)|| p(\zvec|\thetavec,\yvec)\right) q(\thetavec)\,d\thetavec.  \label{kld-identity}
\end{align}
For the VA at~\eqref{EQ:approximation}, the second term on the right-hand side of~\eqref{kld-identity} is 
always zero because $q(\zvec|\thetavec)=p(\zvec|\thetavec,\yvec)$. 
Therefore, minimizing the divergence
between $q(\thetavec,\zvec)$ and $p(\thetavec,\zvec|\yvec)$ for $q(\thetavec,\zvec)$ of the form at~\eqref{EQ:approximation} is equivalent to minimizing the KL divergence 
between $q^0_{\lambda}(\thetavec)$ and $p(\thetavec|\yvec)$. For any other VA of the form
at~\eqref{eq:VA2}, 
the approximation to the marginal posterior distribution of $\thetavec$
cannot improve on the KL-optimal approximation within the family $q_\lambda^0(\thetavec)$ for approximation~\eqref{EQ:approximation}.
%the second term on the right-hand side of~\eqref{kld-identity} is 
%a non-negative function of $\thetavec$. Therefore, minimizing the divergence between 
%\eqref{eq:VA2} and $p(\thetavec,\zvec|\yvec)$ does not minimize the KL divergence
%between $q^0_{\lambda}(\thetavec)$ and $p(\thetavec|\yvec)$.
This proves the result.
%
%\subsubsection{Ruben's notes}
%We have two approximating families: $q_\lambda\left(\bm{\theta},\bm{z}\right)=p(\bm{z}|\bm{y},\bm{\theta})q^0_\lambda\left(\bm{\theta}\right)$, and $\tilde{q}_{\tilde{\lambda}}\left(\bm{\theta},\bm{z}\right)=q_{\tilde{\lambda}_b}(\bm{z}|\bm{\theta})q^0_\lambda\left(\bm{\theta}\right)$ where $\tilde{\bm\lambda} = \left(\tilde{\bm\lambda}_b^\top,\bm\lambda^\top\right)^\top$. The variational parameters have corresponding support $\bm{\lambda}\in\Lambda$ and $\tilde{\bm\lambda}_b\in\tilde{\Lambda}_b$.
%It follows from David's result that
%\begin{align*}
%	\bm{\lambda}^\star &= \argmin_{\bm{\lambda}\in\Lambda}\text{KL}\left(p(\bm{z}|\bm{y},\bm{\theta})q^0_\lambda\left(\bm{\theta}\right)|| p(\thetavec,\zvec|\yvec)\right)\\
%	&  = \argmin_{\bm{\lambda}\in\Lambda}\text{KL}\left(q^0_\lambda(\thetavec)|| p(\thetavec|\yvec)\right)
%\end{align*}
%The optimization process implies by definition that 
%$$\text{KL}\left(q_{\lambda^\star}^0(\thetavec)|| p(\thetavec|\yvec)\right)\le \text{KL}\left(q_\lambda^0(\thetavec)|| p(\thetavec|\yvec)\right) \ \ \ \ \ \forall \ \ \ \bm{\lambda}\in\Lambda$$
%which also includes 
%$$\text{KL}\left(q_{\lambda^\star}^0(\thetavec)|| p(\thetavec|\yvec)\right)\le \text{KL}\left(q_{\lambda_a^*}(\thetavec)|| p(\thetavec|\yvec)\right)$$
%for 
%\begin{align*}
%\{\bm{\lambda_a}^\star,\bm{\tilde{\lambda}_b}^\star\} &= \argmin_{\{\bm{\lambda}\in\Lambda,\tilde{\lambda}_b\in\tilde{\Lambda}_b\}}\text{KL}\left(q_{\tilde{\lambda}_b}(\bm{z}|\bm{\theta})q^0_\lambda\left(\bm{\theta}\right)|| p(\thetavec,\zvec|\yvec)\right).
%\end{align*}

\section{Terms Required in Theorem~\ref{thm:gradient}}\label{app:gradients}
Here we derive closed form expressions for two terms required 
to compute the re-parameterization gradient at~\eqref{eq:thm1}
for the factor Gaussian copula for $q_\lambda^0$ in Section~\ref{sec:gcopapprox}.
The necessary results for the factor Gaussian for $q_\lambda^0$ in Section~\ref{sec:gapprox} are found in~\cite{ong2016variational}. 
The third term, $\nabla_\theta \log g(\thetavec,\zvec)$,
is computed from the augmented likelihood, and is therefore model specific.

\subsection{Computation of $\nabla_\theta\log q^0_\lambda(\thetavec)$}
From~\eqref{eq:q} and the decomposition 
$\Sigma_\vartheta=B_\vartheta B_\vartheta^\top+D_\vartheta^2$, we get the gradient
$$\nabla_\theta\log q^0_\lambda(\bm{\theta}) = -\left[\frac{\partial \varthetavec}{\partial\bm{\theta}}\right]^\top (B_\vartheta B_\vartheta^\top+D_\vartheta^2)^{-1}\left(\varthetavec-\bm{\mu}_{\vartheta}\right)+\left[\frac{\partial }{\partial\bm{\theta}}\sum_{i=1}^m \log t_{\gamma_i}'(\theta_i)\right]^\top\,,$$
where the diagonal matrix $\frac{\partial \varthetavec}{\partial\bm{\theta}} = \text{diag}\left(t_{\gamma_1}'(\theta_1),\dots,t_{\gamma_m}'(\theta_m)\right)$.  For the Yeo-Johnson transformation, the derivative
\[
t_\gamma'(\theta)=
\left\{\begin{array}{cl}
(-\theta+1)^{1-\gamma} &\mbox{if }\theta<0\\
(\theta+1)^{\gamma-1} &\mbox{if }\theta\geq 0\,,
\end{array} \right.
\]
so that $\frac{\partial }{\partial\bm{\theta}}\sum_{i=1}^m \log t_{\gamma_i}'(\theta_i) = \left(\frac{\gamma_1-1}{|\theta_1|+1},\dots,\frac{\gamma_m-1}{|\theta_m|+1}\right)$. For large
$m$ the inverse of the $(m\times m)$ matrix $(B_\vartheta B_\vartheta^\top+D_\vartheta^2)$ (or the solutions of linear systems in this matrix) can be computed efficiently 
using the Woodbury formula
\[
(BB^\top+D^2)^{-1}=D^{-2}-D^{-2}B(I+B^\top D^{-2}B)^{-1}B^\top D^{-2}\,,
\]
when
$(I+B^\top D^{-2}B)$ is a $(k\times k)$ matrix with $k<<m$.
 
\subsection{Computation of $\frac{\partial \thetavec}{\partial\lambdavec}^\top$}
The variational parameter vector is 
$\lambdavec=(\muvec_\vartheta^\top,\mbox{vech}(B_\vartheta)^\top,\dvec_\vartheta^\top,\gammavec^\top)^\top$, 
so that we compute the derivative of $\thetavec$ with respect to each of these four parameter vectors. To compute
these first note that 
from the copula model and re-parameterization trick we can write
$$\bm{\theta}=t^{-1}_{\bm{\gamma}}\left(\bm{\varthetavec}\right)=t^{-1}_{\bm{\gamma}}\left(\muvec_\vartheta+B_\vartheta\zetavec+D_\vartheta\epsilonvec\right)\,,$$
where we denote $t^{-1}_{\gamma}\left(\bm{\varthetavec}\right)\equiv
(t^{-1}_{\gamma_1}(\vartheta_1),\ldots,t^{-1}_{\gamma_m}(\vartheta_m))^\top$. Then by 
repeated application of the chain rule
\begin{eqnarray*}
\mbox{(i)}\;\frac{\partial\bm{\theta}}{\partial\muvec_\vartheta} &=& \frac{\partial\bm{\theta}}{\partial\varthetavec}\frac{\partial\varthetavec}{\partial\muvec_\vartheta} = \frac{\partial\bm{\theta}}{\partial\varthetavec}\,,\;\;\;\;\;\;\mbox{(ii)}\;
\frac{\partial\bm{\theta}}{\partial B_\vartheta} = \frac{\partial\bm{\theta}}{\partial\varthetavec}\frac{\partial\varthetavec}{\partial B_\vartheta} = \frac{\partial\bm{\theta}}{\partial\varthetavec}\left(\zetavec^\top\otimes I_m\right)\,,\\
\mbox{(iii)}\;\frac{\partial\bm{\theta}}{\partial\bm{d}} &=& \frac{\partial\bm{\theta}}{\partial\varthetavec}\frac{\partial\varthetavec}{\partial\bm{d}} = \frac{\partial\bm{\theta}}{\partial\varthetavec}\text{diag}(\epsilonvec)\,,\;\;\mbox{(iv)}\;
\frac{\partial\bm{\theta}}{\partial\bm{\gamma}} = -\frac{\partial\bm{\theta}}{\partial\varthetavec}\frac{\partial\varthetavec}{\partial\bm{\gamma}} \,,
\end{eqnarray*}
with the diagonal matrices  
$\frac{\partial\bm{\theta}}{\partial\varthetavec} =
\text{diag}\left(1/t_{\gamma_1}'(\theta_1),\dots,1/t_{\gamma_m}'(\theta_m)\right)$, 
 $\frac{\partial\varthetavec}{\partial\bm{\gamma}} = \text{diag}\left(\frac{\partial}{\partial \gamma_1}t_{\gamma_1}(\theta_1),\dots,\frac{\partial}{\partial \gamma_m}t_{\gamma_m}(\theta_m)\right)$ and $\text{diag}(\epsilonvec)$, where
\[
\frac{\partial}{\partial \gamma}t_{\gamma}(\theta)=
\left\{\begin{array}{cl}
\frac{\left(2-\gamma\right)(1-\theta)^{2-\gamma}\ln\left(1-\theta\right)-(1-\theta)^{2-\gamma}+1}{\left(2-\gamma\right)^2}&\mbox{if }\theta<0\\
\frac{\gamma\left(1+\theta\right)^\gamma\ln\left(\theta+1\right)-\left(1+\theta\right)^\gamma+1}{\gamma^2} &\mbox{if }\theta\geq 0\,.
\end{array} \right.
\]
Notice that evaluation of the derivatives at (i)--(iv) above
only involves sparse matrix computations, which can be employed 
for larger values of $m$ in practice. Last, to obtain
$\frac{\partial\bm{\theta}}{\partial \mbox{\footnotesize vech}(B_\vartheta)}$
simply extract the corresponding
elements from $\frac{\partial\bm{\theta}}{\partial B_\vartheta}$.

%\textbf{Computation of} $\nabla_{\lambda}\calL(\lambdavec)$\\
%Finally we can construct
%\begin{align*}
%\nabla_{\muvec_\vartheta}\calL(\lambdavec) =&E_{f_\varepsilon}\left(\frac{\partial\bm{\theta}}{\partial\muvec_\vartheta}^\top\left[\nabla_\theta\log g\left(\bm{\theta},\zvec\right) -\nabla_\theta\log q^0_\lambda(\bm{\theta})\right]\right)\\
%\nabla_{B}\calL(\lambdavec) =&E_{f_\varepsilon}\left(\frac{\partial\bm{\theta}}{\partial B}^\top\left[\nabla_\theta\log g\left(\bm{\theta},\zvec\right) -\nabla_\theta\log q^0_\lambda(\bm{\theta})\right]\right)\\
%\nabla_{\bm{d}}\calL(\lambdavec) =&E_{f_\varepsilon}\left(\frac{\partial\bm{\theta}}{\partial\bm{d}}^\top\left[\nabla_\theta\log g\left(\bm{\theta},\zvec\right) -\nabla_\theta\log q^0_\lambda(\bm{\theta})\right]\right)\\
%\nabla_{\bm{\gamma}}\calL(\lambdavec) =&E_{f_\varepsilon}\left(\frac{\partial\bm{\theta}}{\partial\bm{\gamma}}^\top\left[\nabla_\theta\log g\left(\bm{\theta},\zvec\right) -\nabla_\theta\log q^0_\lambda(\bm{\theta})\right]\right)\\
%\end{align*}
%and 
%$$\nabla_{\lambda}\calL(\lambdavec) = \left(\nabla_{\muvec_\vartheta}\calL(\lambdavec)^\top,\nabla_{\text{vech}(B)}\calL(\lambdavec) ^\top,\nabla_{\bm{d}}\calL(\lambdavec)^\top,\nabla_{\bm{\gamma}}\calL(\lambdavec)^\top\right)^\top$$
%Extract from $\nabla_{B}\calL(\lambdavec)$ the corresponding elements to  $\nabla_{\text{vech}(B)}\calL(\lambdavec)$.
%\newpage

\section{TVP-VAR-SV model}\label{app:tvpreg}
In this appendix we provide further details on the TVP-VAR-SV model in 
Section~\ref{sec:eg1}.
\subsection{Representation and parameterization}\label{app:tvp1}
Following~\cite{huber2020} and references therein,
 the response equation of the TVP-VAR-SV model at~\eqref{eq:tvpvarsv1} 
 can be rewritten as the set of $N$ regressions 
at~\eqref{eq:tvpreg} as follows. First, pre-multiply~\eqref{eq:tvpvarsv1} 
by $L_t^{-1}$, so that 
\begin{eqnarray*}
	L_t^{-1}\yvec_t &= &L_t^{-1}\betavec_{0,t}+\sum_{s=1}^p L_t^{-1}B_{s,t}\yvec_{t-s}+\epsilonvec_t \\
	&= &\gammavec_{0,t}+\sum_{s=1}^p\Gamma_{s,t}\yvec_{t,s}+\epsilonvec_t\,.
\end{eqnarray*}
Notice that the parameters of the TVP-VAR are easily recovered by setting
$\betavec_{0,t}=L_t\gammavec_{0,t}$ and $B_{s,t}=L_t\Gamma_{s,t}$. Because $L_t$ is a lower unitriangular matrix, then 
$L_t^{-1}=\{l_{i,j,t}\}$ is also a lower unitriangular matrix with $l_{i,i,t}=1$ and $l_{i,j,t}=0$ for $j>i$.
Denote the 
non-fixed elements of the $i$th row of $L^{-1}_t$ as $\bm{l}_{1:i-1,t} = \left(l_{i,1,t},\dots,l_{i,i-1,t}\right)^\top$
for $i\geq 2$, so that the entire $i$th row of $L_t^{-1}$ is $(\bm{l}_{1:i-1,t}^\top,1,\bm{0}_{N-i}^\top)$.
Then each of the $i=1,\ldots,N$ individual equations of the TVP-VAR-SV  can be written as 
\begin{equation}
	y_{i,t}+\bm{y}_{1:i-1,t}^\top\bm{l}_{1:i-1,t} =\left(\bm{y}_{t-1}^\top,\dots,\bm{y}_{t-p}^\top,1\right)\bm{\gamma}_{i,t}+\epsilon_{i,t}
	\label{eq:tvpvarsveqi}
\end{equation}
where  $\bm{y}_{1:i-1,t} = \left(y_{1,t},\dots,y_{i-1,t}\right)^\top$, $\bm{\gamma}_{i,t} = \left(\Gamma_{i,1,t},\dots,\Gamma_{i,p,t},\gamma_{i,0,t}\right)^\top$,
$\Gamma_{i,s,t}$ denotes the $i$th row of  $\Gamma_{s,t}$,
$\gamma_{i,0,t}$ is the $i$th element in $\bm{\gamma}_{0,t}$, and $\epsilon_{i,t}\sim N(0,\exp(h_{i,t}))$. 
Collecting together terms in~\eqref{eq:tvpvarsveqi} and rearranging, the
$i$th equation can be expressed as the regression with time-varying parameters
\[
y_{i,t}=\widetilde\xvec_{i,t}^\top\etavec_{i,t}+\epsilon_{i,t}\,,
\]
where $\widetilde\xvec_{i,t}^\top=\left(\bm{y}_{t-1}^\top,\dots,\bm{y}_{t-p}^\top,1,-\bm{y}_{1:i-1,t}^\top\right)$
and $\etavec_{i,t}^\top = \left(\gammavec_{i,t}^\top,\bm{l}_{1:i-1,t}^\top\right)$. 
The coefficient vector is of dimension $Np+i=J_i/2$ and follows the 
random walk
$\etavec_{i,t}=\etavec_{i,t-1}+\diag{(\sqrt{\vvec_i})}\varepsilonvec_{i,t}$,
with $\varepsilonvec_{i,t}\sim N(\bm{0},I)$ independently. (Here, we use
$\mbox{diag}(\vvec)$ to denote a diagonal matrix with the leading 
diagonal $\vvec$.)

The coefficients $\etavec_{i,t}$ are  
transformed to the ``non-centered'' representation 
$\etavec_{i,t}=\etavec_{i,0}+\mbox{diag}(\sqrt{\vvec_i})\widetilde\etavec_{i,t}$ as a sum of a time-invariant term $\etavec_{i,0}$
and scaled time-varying deviations $\widetilde\etavec_{i,t}$. 
Substituting in this parameterization gives
the regression representation at~\eqref{eq:tvpreg} 
\begin{eqnarray}
	y_{i,t} &= &\widetilde{\xvec}_{i,t}^\top \etavec_{i,0}+\widetilde{\xvec}_{i,t}^\top \mbox{diag}(\widetilde \etavec_{i,t})\sqrt{\vvec_i}+\epsilon_{i,t}\nonumber \\
	&= & \xvec_{i,t}^\top \alphavec_i + \epsilon_{i,t}\label{eq:apptvp1}\,,
\end{eqnarray}
with $\xvec_{i,t}^\top =(\widetilde \xvec_{i,t}^\top,
\widetilde{\xvec}_{i,t}^\top\mbox{diag}(\widetilde \etavec_{i,t}))$ and 
$\alphavec_i^\top=(\etavec_{i,0}^\top,\sqrt{\vvec_i}^\top)$. Note that 
$\xvec_{i,t}$ is a function of the observed time series
data and the time-varying latent variables $\widetilde \etavec_{i,t}$ only. The latter, along with the log-volatilties, are the latent variables used in our
VA in Section~\ref{sec:vb}.

\subsection{Generating the latent variables}\label{app:tvp2}
Another rearrangement of~\eqref{eq:apptvp1} produces the linear
state space model
\begin{eqnarray*}
y_{i,t} &= &\widetilde\xvec_{i,t}^\top \etavec_{i,0}+\widetilde \xvec_{i,t}^\top \mbox{diag}(\sqrt{\vvec_i})\widetilde\etavec_{i,t}+ \epsilon_{i,t}\,,\\
\widetilde \etavec_{i,t} &= & \widetilde \etavec_{i,t-1}+\varepsilonvec_{i,t}\,.
\end{eqnarray*}
Thus, the approach of~\cite{carter1994gibbs} can be used to obtain draws from
the conditional posterior density 
$p(\widetilde \etavec_i|\hvec_i,\thetavec_i,\yvec)$. To generate
the log-volatilities from $p(\hvec_i|\widetilde \etavec_i,\thetavec_i,\yvec)$,
we follow~\cite{kim1998} and use a mixture of seven normals to approximate
the distribution of $\log(\epsilon_{i,t}^2)$ and generate $\hvec_i$ joint with mixture indicators.

%\newpage

%\input{suppmaterial}
\baselineskip=15pt 
\bibliography{Project_bib}
\bibliographystyle{apa}
\newpage
%\begin{table}[h]
%	\caption{Accuracy and speed of variational inference for the UCSV model of U.S. inflation}
%	\begin{center}
%	\begin{tabular}{lccc} \hline \hline
%		Var. Approx. &$\min_{\lambdavec}(\overline{\mbox{KL}}(\lambdavec))$ 
%		&Steps to Stop &Time to Stop \\ \hline
%		Structured Gaussian &0.0559 &4960 &8.17 sec \\
%		Hybrid &0.0003 &1403 &1.73 sec\\ \hline \hline
%	\end{tabular}
%	\end{center}
%\label{tab:AKLD}
%	The two VAs are a structured Gaussian approximation to the 
%	augmented posterior and 
%	our proposal labelled `Hybrid'. The first column reports the minimum average KL divergence (as defined in the text) between each calibrated VA and the exact
%	posterior evaluated using MCMC. The second and third columns
%	report the number of steps and clock time, for the SGA to stop	using an absolute change of $\overline{\mbox{KL}}(\lambdavec)$ 
%	less than 0.0001 as a stopping rule. 
%\end{table}

% Table generated by Excel2LaTeX from sheet 'Sheet1'
\begin{table}[h]
%		\caption{Variational mean and quantiles for
%			$\betavec$ in the large data tobit example.}
	\begin{center}
				\resizebox{18cm}{!}{
	\begin{tabular}{lrrrcrrrcrrr}
		\hline\hline
		                    &                         \multicolumn{3}{c}{Hybrid VA}                         &  &                        \multicolumn{3}{c}{Gaussian VA}                        &  &                      \multicolumn{3}{c}{Gaussian MF VA}                       \\
		                    & \multicolumn{1}{l}{Mean} & \multicolumn{1}{l}{5\%} & \multicolumn{1}{l}{95\%} &  & \multicolumn{1}{l}{Mean} & \multicolumn{1}{l}{5\%} & \multicolumn{1}{l}{95\%} &  & \multicolumn{1}{l}{Mean} & \multicolumn{1}{l}{5\%} & \multicolumn{1}{l}{95\%} \\ \cline{2-4}\cline{6-8}\cline{10-12}
		Intercept           &                  -15.989 &                 -16.076 &                  -15.901 &  &                  -16.682 &                 -17.221 &                  -16.146 &  &                  -16.478 &                 -16.503 &                  -16.454 \\
		\em Lagged Sales    &                          &                         &                          &  &                          &                         &                          &  &                          &  \\ \cline{1-1}
		B1 past D sales     &                    0.183 &                   0.172 &                    0.194 &  &                    0.186 &                   0.177 &                    0.196 &  &                    0.190 &                   0.171 &                    0.210 \\
		B2 past D sales     &                    0.107 &                   0.064 &                    0.150 &  &                    0.111 &                   0.078 &                    0.145 &  &                    0.121 &                   0.087 &                    0.154 \\
		B3 past D sales     &                    0.075 &                   0.039 &                    0.110 &  &                    0.076 &                   0.069 &                    0.084 &  &                    0.076 &                   0.047 &                    0.105 \\
		B1 past R sales     &                    0.161 &                   0.152 &                    0.169 &  &                    0.177 &                   0.172 &                    0.182 &  &                    0.196 &                   0.191 &                    0.200 \\
		B2 past R sales     &                    0.096 &                   0.054 &                    0.137 &  &                    0.100 &                   0.066 &                    0.133 &  &                    0.109 &                   0.077 &                    0.140 \\
		B3 past R sales     &                    0.106 &                   0.098 &                    0.114 &  &                    0.108 &                   0.101 &                    0.114 &  &                    0.109 &                   0.093 &                    0.125 \\
		\em Advertising Variables    &                          &                         &                          &  &                          &                         &                          &  &                          &  \\ \cline{1-1}
		B1 Emails           &                    1.732 &                   1.646 &                    1.818 &  &                    1.821 &                   1.806 &                    1.837 &  &                    1.185 &                   1.164 &                    1.207 \\
		B1 Catal.           &                    1.335 &                   1.055 &                    1.615 &  &                    0.751 &                   0.615 &                    0.884 &  &                    0.906 &                   0.767 &                    1.043 \\
		B1 Paid S.          &                    0.730 &                   0.209 &                    1.231 &  &                    0.472 &                   0.253 &                    0.692 &  &                    0.306 &                   0.096 &                    0.518 \\
		B2 Emails           &                   -0.485 &                  -0.610 &                   -0.353 &  &                   -0.435 &                  -0.485 &                   -0.385 &  &                   -0.154 &                  -0.200 &                   -0.109 \\
		B2 Catal.           &                   -0.809 &                  -1.153 &                   -0.444 &  &                   -0.721 &                  -0.913 &                   -0.530 &  &                   -0.177 &                  -0.344 &                   -0.010 \\
		B2 Paid S.          &                   -0.256 &                  -1.336 &                    0.780 &  &                    0.363 &                  -0.235 &                    0.943 &  &                    0.477 &                  -0.142 &                    1.079 \\
		B3 Emails           &                   -0.519 &                  -0.566 &                   -0.471 &  &                   -0.554 &                  -0.564 &                   -0.544 &  &                   -0.340 &                  -0.349 &                   -0.331 \\
		B3 Catal.           &                   -1.305 &                  -1.649 &                   -0.956 &  &                   -1.086 &                  -1.277 &                   -0.893 &  &                   -0.621 &                  -0.793 &                   -0.448 \\
		B3 Paid S.          &                    0.327 &                  -0.721 &                    1.338 &  &                    0.296 &                  -0.252 &                    0.844 &  &                    0.405 &                  -0.129 &                    0.943 \\
		\em Endogenous Controls &                          &                         &                          &  &                          &                         &                          &  &                          &  \\ \cline{1-1}
		Res Paid S.         &                    0.064 &                   0.000 &                    0.126 &  &                    0.078 &                   0.027 &                    0.128 &  &                    0.104 &                   0.056 &                    0.152 \\
		Organic S. CFs      &                    1.600 &                   1.417 &                    1.784 &  &                    1.616 &                   1.484 &                    1.749 &  &                    1.594 &                   1.467 &                    1.722 \\
		Res Organic S.      &                    0.350 &                   0.273 &                    0.425 &  &                    0.352 &                   0.278 &                    0.426 &  &                    0.373 &                   0.306 &                    0.438 \\
		Res website V.      &                    1.784 &                   1.706 &                    1.864 &  &                    1.799 &                   1.750 &                    1.848 &  &                    1.805 &                   1.755 &                    1.855 \\
		Visits B1           &                    0.185 &                   0.135 &                    0.233 &  &                    0.196 &                   0.151 &                    0.241 &  &                    0.213 &                   0.168 &                    0.259 \\
		\em Other Variables &                          &                         &                          &  &                          &                         &                          &  &                          &  \\ \cline{1-1}
		log price           &                   -0.271 &                  -0.361 &                   -0.178 &  &                   -0.169 &                  -0.333 &                   -0.004 &  &                   -0.231 &                  -0.237 &                   -0.224 \\
		month1              &                   -0.811 &                  -0.944 &                   -0.679 &  &                   -0.816 &                  -0.941 &                   -0.691 &  &                   -0.752 &                  -0.853 &                   -0.652 \\
		month2              &                   -0.781 &                  -0.910 &                   -0.653 &  &                   -0.795 &                  -0.922 &                   -0.670 &  &                   -0.771 &                  -0.866 &                   -0.676 \\
		month3              &                   -0.418 &                  -0.522 &                   -0.310 &  &                   -0.424 &                  -0.544 &                   -0.307 &  &                   -0.402 &                  -0.494 &                   -0.310 \\
		month4              &                   -0.281 &                  -0.398 &                   -0.160 &  &                   -0.291 &                  -0.392 &                   -0.191 &  &                   -0.281 &                  -0.376 &                   -0.185 \\
		month5              &                    0.123 &                   0.014 &                    0.224 &  &                    0.118 &                   0.006 &                    0.230 &  &                    0.118 &                   0.030 &                    0.205 \\
		month7              &                    0.096 &                  -0.069 &                    0.252 &  &                    0.074 &                  -0.097 &                    0.243 &  &                    0.105 &                  -0.030 &                    0.239 \\
		month8              &                    0.056 &                  -0.055 &                    0.161 &  &                    0.052 &                  -0.047 &                    0.151 &  &                    0.063 &                  -0.022 &                    0.147 \\
		month9              &                   -0.206 &                  -0.316 &                   -0.087 &  &                   -0.229 &                  -0.341 &                   -0.115 &  &                   -0.209 &                  -0.300 &                   -0.118 \\
		month10             &                   -0.107 &                  -0.218 &                    0.010 &  &                   -0.126 &                  -0.253 &                    0.002 &  &                   -0.089 &                  -0.185 &                    0.007 \\
		month11             &                    0.172 &                   0.066 &                    0.273 &  &                    0.155 &                   0.039 &                    0.272 &  &                    0.195 &                   0.107 &                    0.282 \\
		month12             &                    1.684 &                   1.575 &                    1.793 &  &                    1.735 &                   1.635 &                    1.834 &  &                    1.744 &                   1.663 &                    1.823 \\
		$\sigma$            &                    7.724 &                   7.673 &                    7.774 &  &                    7.813 &                   7.787 &                    7.840 &  &                    7.841 &                   7.838 &                    7.844 \\ \hline\hline
	\end{tabular}%
}
\end{center}
\caption{Variational mean and quantiles for
	$\betavec$ in the large data tobit example. The variables ($\xvec_{it}$) are defined in the Online Appendix.
	Results are given for our approach (Hybrid VA), the  Gaussian factor VA of~\cite{danaher2020} and mean field VA.
	The estimated error standard deviation $\sigma$ is also reported. }
	\label{tab:tobitbeta}
\end{table}%

% Table generated by Excel2LaTeX from sheet 'Sheet1'
\begin{landscape} 
	\begin{table}[H]
		\caption{Estimate of $V_\alpha$ for the large data tobit example using our proposed VA.}
		\begin{center}
		\resizebox{24cm}{!}{
			\begin{tabular}{llccccccccccccc}
				\hline\hline
				   &           &                                        &  &                                        &                                       &                                       &  &                                       &                                       &                                       &  &                                       &                                       &                                      \\
				   &           &                                        &  &                                                 \multicolumn{3}{c}{B1}                                                 &  &                                                \multicolumn{3}{c}{B2}                                                 &  &                                                \multicolumn{3}{c}{B3}                                                \\
				   &           &               Intercept                &  &                 Emails                 &                Catal.                 &                Paid S.                &  &                Emails                 &                Catal.                 &                Paid S.                &  &                Emails                 &                Catal.                 &               Paid S.                \\ \cline{3-3}\cline{5-7}\cline{9-11}\cline{13-15}
				   &           &                                        &  &                                        &                                       &                                       &  &                                       &                                       &                                       &  &                                       &                                       &                                      \\
				   & Intercept &                 28.481                 &  &                   -                    &                   -                   &                   -                   &  &                   -                   &                   -                   &                   -                   &  &                   -                   &                   -                   &                  -                   \\
				   &           & \footnotesize{\textit{(27.79, 29.38)}} &  &                                        &                                       &                                       &  &                                       &                                       &                                       &  &                                       &                                       &                                      \\
				   &           &                                        &  &                                        &                                       &                                       &  &                                       &                                       &                                       &  &                                       &                                       &                                      \\
				   & Emails    &                 -0.62                  &  &                 2.354                  &                   -                   &                   -                   &  &                   -                   &                   -                   &                   -                   &  &                   -                   &                   -                   &                  -                   \\
				   &           & \footnotesize{\textit{(-0.64, -0.59)}} &  &  \footnotesize{\textit{(2.11, 2.64)}}  &                                       &                                       &  &                                       &                                       &                                       &  &                                       &                                       &                                      \\
				B1 & Catal.    &                 -0.468                 &  &                 0.359                  &                 4.085                 &                   -                   &  &                   -                   &                   -                   &                   -                   &  &                   -                   &                   -                   &                  -                   \\
				   &           & \footnotesize{\textit{(-0.67, -0.27)}} &  &  \footnotesize{\textit{(0.16, 0.54)}}  & \footnotesize{\textit{(1.77, 9.05)}}  &                                       &  &                                       &                                       &                                       &  &                                       &                                       &                                      \\
				   & Paid S.   &                 -0.248                 &  &                 0.168                  &                 0.273                 &                 3.082                 &  &                   -                   &                   -                   &                   -                   &  &                   -                   &                   -                   &                  -                   \\
				   &           & \footnotesize{\textit{(-0.63, 0.11)}}  &  & \footnotesize{\textit{(-0.16, 0.48)}}  & \footnotesize{\textit{(-0.41, 0.77)}} & \footnotesize{\textit{(0.69, 7.99)}}  &  &                                       &                                       &                                       &  &                                       &                                       &                                      \\
				   &           &                                        &  &                                        &                                       &                                       &  &                                       &                                       &                                       &  &                                       &                                       &                                      \\
				   & Emails    &                 0.371                  &  &                 -0.302                 &                -0.291                 &                -0.131                 &  &                 0.963                 &                   -                   &                   -                   &  &                   -                   &                   -                   &                  -                   \\
				   &           &  \footnotesize{\textit{(0.21, 0.53)}}  &  & \footnotesize{\textit{(-0.48, -0.12)}} & \footnotesize{\textit{(-0.65, 0.13)}} & \footnotesize{\textit{(-0.52, 0.34)}} &  & \footnotesize{\textit{(0.61, 1.58)}}  &                                       &                                       &  &                                       &                                       &                                      \\
				B2 & Catal.    &                  0.3                   &  &                 -0.145                 &                 0.035                 &                -0.007                 &  &                 0.051                 &                 4.872                 &                   -                   &  &                   -                   &                   -                   &                  -                   \\
				   &           &  \footnotesize{\textit{(0.12, 0.55)}}  &  &  \footnotesize{\textit{(-0.4, 0.06)}}  & \footnotesize{\textit{(-0.52, 0.52)}} & \footnotesize{\textit{(-0.67, 0.64)}} &  & \footnotesize{\textit{(-0.4, 0.47)}}  & \footnotesize{\textit{(1.25, 11.98)}} &                                       &  &                                       &                                       &                                      \\
				   & Paid S.   &                  0.29                  &  &                 -0.21                  &                -0.086                 &                -0.084                 &  &                 0.084                 &                 0.16                  &                 5.484                 &  &                   -                   &                   -                   &                  -                   \\
				   &           & \footnotesize{\textit{(-0.17, 0.71)}}  &  & \footnotesize{\textit{(-0.55, 0.18)}}  & \footnotesize{\textit{(-0.72, 0.58)}} & \footnotesize{\textit{(-0.71, 0.6)}}  &  & \footnotesize{\textit{(-0.42, 0.49)}} & \footnotesize{\textit{(-0.54, 0.75)}} & \footnotesize{\textit{(1.12, 15.7)}}  &  &                                       &                                       &                                      \\
				   &           &                                        &  &                                        &                                       &                                       &  &                                       &                                       &                                       &  &                                       &                                       &                                      \\
				   & Emails    &                 0.185                  &  &                 -0.162                 &                -0.192                 &                -0.083                 &  &                 0.117                 &                -0.052                 &                 0.054                 &  &                 0.766                 &                   -                   &                  -                   \\
				   &           &  \footnotesize{\textit{(0.08, 0.3)}}   &  & \footnotesize{\textit{(-0.29, -0.05)}} & \footnotesize{\textit{(-0.43, 0.08)}} & \footnotesize{\textit{(-0.39, 0.26)}} &  & \footnotesize{\textit{(-0.09, 0.35)}} & \footnotesize{\textit{(-0.36, 0.27)}} & \footnotesize{\textit{(-0.3, 0.36)}}  &  & \footnotesize{\textit{(0.52, 1.13)}}  &                                       &                                      \\
				B3 & Catal.    &                 0.458                  &  &                 -0.321                 &                 -0.16                 &                -0.122                 &  &                 0.122                 &                 0.268                 &                 0.166                 &  &                 0.054                 &                 3.762                 &                  -                   \\
				   &           &  \footnotesize{\textit{(0.23, 0.72)}}  &  & \footnotesize{\textit{(-0.54, -0.09)}} & \footnotesize{\textit{(-0.67, 0.36)}} & \footnotesize{\textit{(-0.67, 0.46)}} &  & \footnotesize{\textit{(-0.27, 0.44)}} & \footnotesize{\textit{(-0.33, 0.73)}} & \footnotesize{\textit{(-0.44, 0.69)}} &  & \footnotesize{\textit{(-0.24, 0.33)}} & \footnotesize{\textit{(1.37, 8.66)}}  &                                      \\
				   & Paid S.   &                 -0.061                 &  &                 0.083                  &                -0.012                 &                 -0.04                 &  &                -0.074                 &                 0.007                 &                -0.061                 &  &                -0.056                 &                -0.059                 &                3.517                 \\
				   &           & \footnotesize{\textit{(-0.63, 0.49)}}  &  & \footnotesize{\textit{(-0.34, 0.48)}}  & \footnotesize{\textit{(-0.7, 0.69)}}  & \footnotesize{\textit{(-0.7, 0.64)}}  &  & \footnotesize{\textit{(-0.52, 0.41)}} & \footnotesize{\textit{(-0.71, 0.66)}} & \footnotesize{\textit{(-0.69, 0.63)}} &  & \footnotesize{\textit{(-0.42, 0.29)}} & \footnotesize{\textit{(-0.64, 0.55)}} & \footnotesize{\textit{(0.73, 9.87)}} \\ \hline\hline
			\end{tabular}%
		}
	    \end{center}
		\label{tab:REvariance}%
		The diagonal values are estimates of the variances of the random coefficients (i.e. the leading diagonal of $V_\alpha$). 
		The off-diagonal values are estimates of the correlations between the random coefficients (i.e. the correlations
		of the matrix $V_\alpha$). The variational means are reported, along with the 95\% quantiles of the variational
		distribution $q_\lambda$ in parentheses.
	\end{table}%
\end{landscape}

% Table generated by Excel2LaTeX from sheet 'Sheet1'
\begin{table}[h!]
	\caption{Heterogeneity estimates for the large data tobit example} \vspace{-10pt}
	\begin{center}
	\begin{tabular}{lccccc}
		\hline\hline
		               &                   Total (TH)                   &  &                  Focal Brand (FBH)                  &  &                  Cross-Brand (CBH)                  \\ \cline{2-2}\cline{4-4}\cline{6-6}
%		               &                                        &  &                                       &  &                                       \\
		Hybrid VA      &                 28.964                 &  &                 2.201                 &  &                 1.527                 \\
		               & \footnotesize{\textit{(28.31, 29.91)}} &  & \footnotesize{\textit{(2.02, 2.47)}}  &  & \footnotesize{\textit{(1.18, 2.02)}}  \\
%		               &                                        &  &                                       &  &                                       \\
		Gaussian VA    &                 28.002                &  &                 1.919                 &  &                 0.533                 \\
		               & \footnotesize{\textit{(27.73, 28.28)}} &  & \footnotesize{\textit{(1.90, 1.94)}} &  & \footnotesize{\textit{(0.53, 0.54)}} \\
%		               &                                        &  &                                       &  &                                       \\
		Gaussian MF VA &                 25.390                &  &                 0.029                 &  &                 0.076                 \\
		               & \footnotesize{\textit{(24.74, 26.10)}} &  & \footnotesize{\textit{(0.028, 0.029)}}  &  & \footnotesize{\textit{(0.075,0.077)}}  \\ \hline\hline
	\end{tabular}%
	\end{center}
	\label{tab:TH}%
The variational mean of the three heterogeneity measures TH($V_\alpha$), FBH($V_\alpha$) and CBH$(V_\alpha)$ are reported for the three
variational approximations. The variational 95\% posterior probability intervals are reported in parentheses.
\end{table}%

% Table generated by Michael Smith by hand.
\begin{table}[h!]
	\caption{Accuracy and computation times for sub-sampling variational inference applied to the large tobit example}\vspace{-10pt}
	\begin{center}
		\begin{tabular}{lccccc}
			\hline\hline
			&\multicolumn{5}{c}{Number of Sweeps at Step~(b) of Algorithm}\\ \cline{2-6}
$|S(\uvec)|$		 & 1 & 2 & 3 &4 & 5 \\ \hline
			  &\multicolumn{5}{l}{Computation Time (seconds/step)}\\ \cline{2-6}
			1000       &   ---              &  ---          &   0.1923  & 0.2361   &   0.2803\\
			2000        &  ---             &  0.2769  &  0.3622  & 0.4485     & 0.5367    \\
			5000         & 0.4250     &  0.6172  &  0.8349  & 1.0362     & 1.2338\\
			10000       & 0.8409     &  1.2336  &  1.6254  & 2.0188     & 2.4150\\
			20000       & 1.6341    &   2.4299  &  3.2354  & 4.0191     & 4.8145\\ 
	 &\multicolumn{5}{l}{$\mbox{RMSE}(\alphavec,\thetavec)$}\\  \cline{2-6}
					1000      & ---             &    ---         &     0.9026  &  0.9026 &   0.9024\\
					2000       &    ---          & 0.9010   &  0.9007   & 0.9003   & 0.9004\\
					5000       &   0.8992    & 0.8991  &   0.8990   & 0.8989 &   0.8987\\
					10000      &  0.8983   &  0.8985   &  0.8984   & 0.8982  &  0.8980\\
					20000     &   0.8977  &   0.8980   &  0.8980 &   0.8979  &  0.8978 \\
					\hline\hline
		\end{tabular}%
	\end{center}
	\label{tab:tobitss}%
Results are reported for combinations of sub-sampling size $|S(\uvec)|$ and number of 
sweeps of the sampler at step~(b) of Algorithm~\ref{alg:VB}. The case where $|S(\uvec)|=20,000$ corresponds to no sub-sampling. Computation time is measured
in seconds per step of the algorithm coded in MATLAB and executed on a standard laptop. Accuracy is 
measured using the root mean squared error metric discussed in the text. Results denoted as ``---''
are those where the SGA had difficulty converging in 10,000 steps.
\end{table}%
\newpage

\newpage
\baselineskip=20pt
\noindent
\setcounter{page}{1}
\begin{center}
{\bf \Large{Online Appendix for `Efficient variational inference for models with many latent variables'}}
\end{center}

\vspace{10pt}

\setcounter{figure}{0}
\setcounter{table}{0}
\renewcommand{\thetable}{A\arabic{table}}
\renewcommand{\thefigure}{A\arabic{figure}}
\noindent
This Online Appendix has two parts:
\begin{itemize}
	\item[] {\bf Part~A}: Additional details and results for Section~3.
	\item[] {\bf Part~B}: Additional details and results for Section~4.
%	\item[] {\bf Part~C} Details the MATLAB functions provided, and gives examples on how to use them.
\end{itemize}
\newpage

\noindent {\bf \large{Part~A: Additional Details and Results for Section~3}}\\
\ \\
\noindent 
This appendix is split into two sub-sections. Further details on the priors and the required gradients and derivatives for the SGA algorithm are presented in Part~A.1. Additional empirical results in Part~A.2.
\ \\
\textbf{{\large A.1 Priors and gradients}}\\
\textbf{{Priors}}\\
The priors for all the parameters in the model are
$$\tau_{ij}~\sim N(0,1),\ \ \ \chi_{ij}|\nu_{ij}\sim \mathcal{G}^{-1}\left(\frac{1}{2},\frac{1}{\nu_{ij}}\right),\ \ \ \xi_i|\kappa_{i}\sim \mathcal{G}^{-1}\left(\frac{1}{2},\frac{1}{\kappa_{i}}\right)$$ 
$$ \nu_{i1},\dots,\nu_{i2J},\kappa_{i}\sim \mathcal{G}^{-1}\left(\frac{1}{2},1\right),\ \ \ \bar{h}_i\sim N(0,10^2),\ \ \ \frac{\rho_i^h+1}{2}\sim\mathcal{B}(25,5),\ \ \ \sigma_i^2\sim\mathcal{G}\left(\frac{1}{2},\frac{1}{2}\right)$$ 

To conduct variational inference we transform all parameters to the real line as follows:
$$\text{(i) }\chi_{ij} \ \text{is transformed to } \tilde{\chi}_{ij} = \log \chi_{ij}; \ \ \ \ \ \ \ \ \ \ \ \ \ \  \text{(ii) }\xi_{i} \ \text{is transformed to } \tilde{\xi}_{i} = \log \xi_{i};$$
$$\ \text{(iii) }\nu_{ij} \ \text{is transformed to } \tilde{\nu}_{ij} = \log \nu_{ij}; \ \ \ \ \ \ \ \ \ \ \ \ \ \  \ \text{(iv) }\kappa_{i} \ \text{is transformed to } \tilde{\kappa}_{i} = \log \kappa_{i};$$
$$\ \text{(v) }\rho_i^h \ \text{is transformed to } \tilde{\rho}_{i} = \Phi_1^{-1}\left(\frac{\rho_i^h+1}{2}\right); \ \ \ \ \ \ \text{(vi)} \sigma_i^2 \ \text{is transformed to } \tilde{\sigma}_i = \log \sigma_i^2;$$
After the transformations we obtain the following prior density functions (using
the Jacobians of the change of variables)
\begin{align*}
\text{(i)}\hspace{0.3cm}&p(\tau_{ij}) = \phi_1\left(\tau_{ij};0,1\right);\hspace{4.5cm}\text{(ii)}\ p(\tilde{\chi}_{ij}|\nu_{ij}) \propto \left(\frac{1}{\nu_{ij}}\right)^{0.5}\exp\left(-\frac{1}{2}\tilde{\chi}_{ij}-\frac{1}{\nu_{ij}\exp(\tilde{\chi}_{ij})}\right);\\
\text{(iii)}\hspace{0.3cm}&p(\tilde{\xi}_{i}|\kappa_i) \propto \left(\frac{1}{\kappa_{i}}\right)^{0.5}\exp\left(-\frac{1}{2}\tilde{\xi}_{i}-\frac{1}{\kappa_{i}\exp(\tilde{\xi}_{i})}\right);\hspace{0.1cm}\text{(iv)}\ p(\tilde{\nu}_{ij}) \propto \exp\left(-\frac{1}{2}\tilde{\nu}_{ij}-\frac{1}{\exp(\tilde{\nu}_{ij})}\right);\\
\text{(v)}\hspace{0.4cm}&p(\tilde{\kappa}_{i}) \propto \exp\left(-\frac{1}{2}\tilde{\kappa}_{i}-\frac{1}{\exp(\tilde{\kappa}_{i})}\right);\hspace{2.3cm}\text{(vi)}\ p(\bar{h}_{i}) \propto \exp\left(-\frac{1}{2\times 10^2}\bar{h}_i^2\right);\\
\text{(vii)} \hspace{0.3cm}&p(\tilde{\rho}_i)\propto \exp\left(-\frac{\tilde{\rho}_i^2}{2}\right)\Phi_1(\tilde{\rho}_i)^{24}\left(1-\Phi_1(\tilde{\rho}_i)\right)^4;\hspace{0.28cm}\text{(viii)}\ p(\tilde{\sigma}_i) \propto \exp\left(\frac{1}{2}\tilde{\sigma}_i-\frac{e^{\tilde{\sigma}_i}}{2}\right).
\end{align*}
\textbf{{Gradient}}\\
The model-specific gradient vector for the parameters of the TVP-VAR-SV is:
\begin{align*}
\nabla_{\theta_i}\log g(\bm{\theta}_i,\bm{z})=  \left(\right.&\nabla_{\tau_i}\log g(\bm{\theta}_i,\bm{z})^\top,\nabla_{\tilde{\chi}_i}\log g(\bm{\theta}_i,\bm{z})^\top,\nabla_{\tilde{\xi}_i}\log g(\bm{\theta}_i,\bm{z})^\top,\nabla_{\tilde{\nu}_i}\log g(\bm{\theta}_i,\bm{z})^\top,\\
&\left.\nabla_{\tilde{\kappa}_i}\log g(\bm{\theta}_i,\bm{z})^\top,\nabla_{\bar{h}_i}\log g(\bm{\theta}_i,\bm{z})^\top,\nabla_{\tilde{\rho}_i}\log g(\bm{\theta}_i,\bm{z})^\top,\nabla_{\tilde{\sigma}_i}\log g(\bm{\theta}_i,\bm{z})^\top\right)^\top.
\end{align*}
The different terms in this gradient can be computed as
\begin{align*}
\nabla_{\tau_i}\log g(\bm{\theta}_i,\bm{z})           &  = \sqrt{\xi_i}\sqrt{\bm{\chi}_i}\circ\left(\sum_{t=1}^{T}(y_{i,t}-\bm{x}_{i,t}^\top\bm{\alpha}_i)\exp(-h_{i,t})\bm{x}_{i,t}\right)-\bm{\tau}_i\\
\nabla_{\tilde{\chi}_i}\log g(\bm{\theta}_i,\bm{z})   &  =\frac{1}{2}\bm{\tau}_i\circ\left( \sqrt{\xi_i}\sqrt{\bm{\chi}_i}\right)\circ\left(\sum_{t=1}^{T}(y_{i,t}-\bm{x}_{i,t}^\top\bm{\alpha}_i)\exp(-h_{i,t})\bm{x}_{i,t}\right)+\nabla_{\tilde{\chi}_i}\log p(\tilde{\bm{\chi}}_i|\tilde{\bm{\nu}}_i)\\
\nabla_{\tilde{\nu}_i}\log g(\bm{\theta}_i,\bm{z})    &  = \nabla_{\tilde{\nu}_i}\log p(\tilde{\bm{\chi}}_i|\tilde{\bm{\nu}}_i) + \nabla_{\tilde{\nu}_i}\log p(\tilde{\bm{\nu}}_i)\\
\nabla_{\tilde{\xi}_i}\log g(\bm{\theta}_i,\bm{z})    &  = \frac{1}{2}\left(\bm{\tau}_i\circ \sqrt{\xi_i}\sqrt{\bm{\chi}_i}\right)^\top\left(\sum_{t=1}^{T}(y_{i,t}-\bm{x}_{i,t}^\top\bm{\alpha}_i)\exp(-h_{i,t})\bm{x}_{i,t}\right)-\frac{1}{2}+\frac{1}{\kappa_i\xi_i}\\
\nabla_{\tilde{\kappa}_i}\log g(\bm{\theta}_i,\bm{z}) &  =-\frac{1}{2}+\frac{1}{\kappa_i\xi_i}-\frac{1}{2}+\frac{1}{\kappa_i}\\
\nabla_{\bar{h}_i}\log g(\bm{\theta}_i,\bm{z})        &  = -\frac{\bar{h}_i}{10^2}+\frac{h_{i,1}-\bar{h}_i}{s_i^2}-\sum_{t=2}^{T}\frac{(\rho_i^h-1)}{\sigma_i}\left[\frac{h_{i,t}-\bar{h}_i-\rho_i^h(h_{i,t-1}-\bar{h}_i)}{\sigma_i}\right]\\
\nabla_{\tilde{\rho}_i}\log g(\bm{\theta}_i,\bm{z})   &  =\left\{\frac{\rho_i^h}{1-(\rho_i^h)^2}\left[\frac{(h_{i,1}-\bar{h}_i)^2}{s_i^2}-1\right]+\sum_{t=2}^T\frac{h_{i,t-1}-\bar{h}_i}{\sigma_i^2}\left[h_{i,t}-\bar{h}_i-\rho_i^h(h_{i,t-1}-\bar{h}_i)\right]\right\}\times\\
&\hspace{0.9cm}2\phi_1(\tilde{\rho}_i)+\left\{\frac{12}{0.5(\rho_i^h+1)}-\frac{2}{1-0.5(\rho_i^h+1)}\right\}2\phi_1(\tilde{\rho}_i)-\tilde{\rho}_i\\
\nabla_{\tilde{\sigma}_i}\log g(\bm{\theta}_i,\bm{z}) &  =-\frac{1}{2}\sigma_i^2+\frac{(h_{i,1}-\bar{h}_i)^2}{2s_i^2}+\sum_{t=2}^T\left\{-\frac{1}{2}+\frac{\left[h_{i,t}-\bar{h}_i-\rho_i^h(h_{i,t-1}-\bar{h}_i)\right]^2}{2\sigma_i^2}\right\}
\end{align*}
where $s_i^2 = \frac{\sigma_i^2}{1-(\rho_i^h)^2}$, $\nabla_{\tilde{\chi}_i}\log p(\tilde{\bm{\chi}}_i|\tilde{\bm{\nu}}_i) = \left(-\frac{1}{2}+\frac{1}{\nu_{i1}\chi_{i1}},\dots,-\frac{1}{2}+\frac{1}{\nu_{iJ_i}\chi_{iJ_i}}\right)^\top$,\\ $\nabla_{\tilde{\nu}_i}\log p(\tilde{\bm{\chi}}_i|\tilde{\bm{\nu}}_i) = \left(-\frac{1}{2}+\frac{1}{\nu_{i1}\chi_{i1}},\dots,-\frac{1}{2}+\frac{1}{\nu_{iJ_i}\chi_{iJ_i}}\right)^\top$,  $\nabla_{\tilde{\nu}_i}\log p(\tilde{\bm{\nu}}_i) = \left(-\frac{1}{2}+\frac{1}{\nu_{i1}},\dots,-\frac{1}{2}+\frac{1}{\nu_{iJ_i}}\right)^\top$.
\newpage

\noindent {\bf \large{A.2 Supplemental figures}}\\
\begin{figure}[H]
	\begin{center}
		\includegraphics[scale = 0.6]{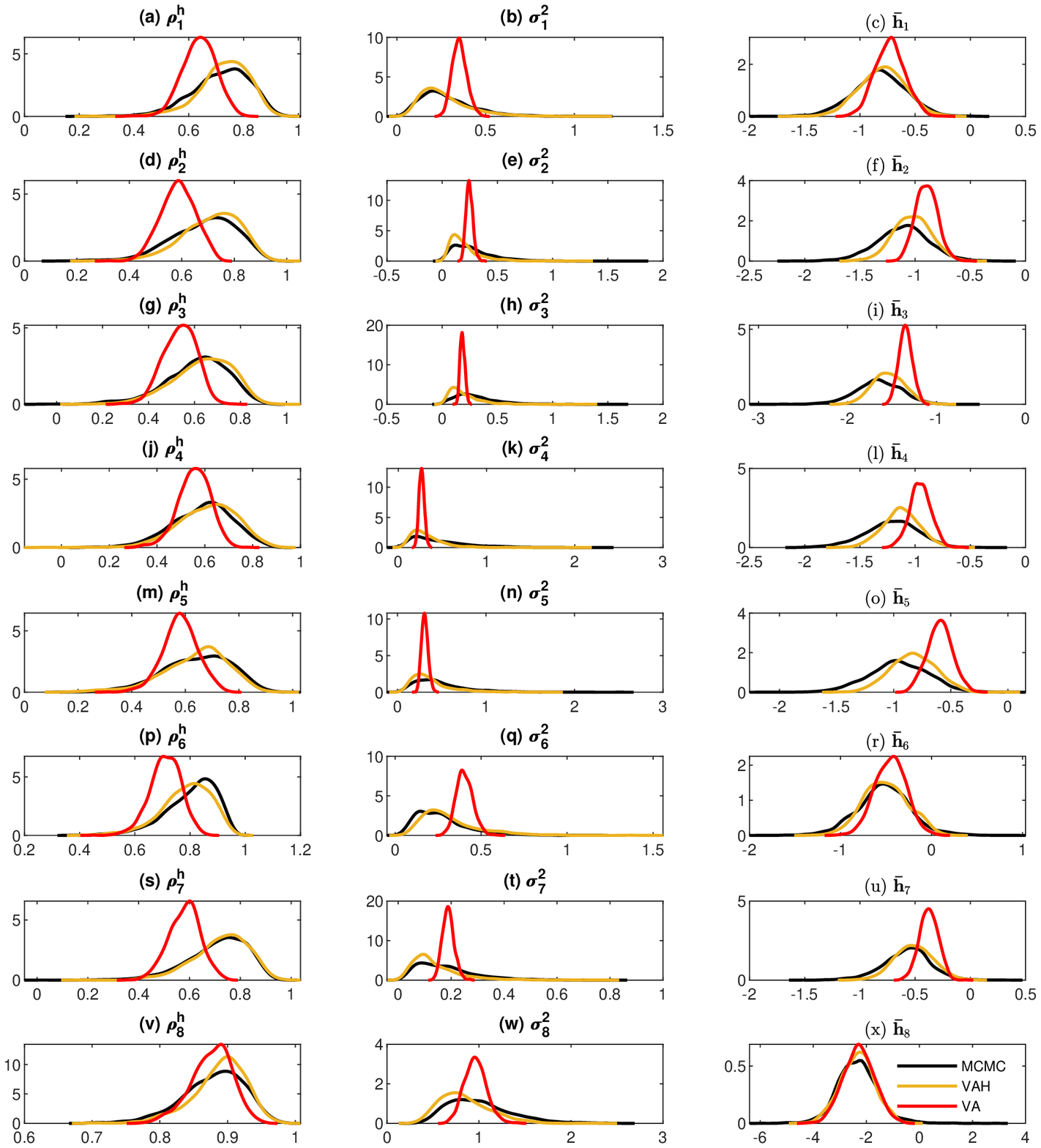}
	\end{center}
\caption{Marginal posterior density estimates of $\rho_i^h,\sigma_i^2,\bar{h}_i$ 
	for the	TVP-VAR-SV model. In each panel, exact posterior estimates (computed by MCMC) are plotted in black, 
	the hybrid VA in yellow,
	and the structured Gaussian VA in red. Each row corresponds to a different 
	equation for $i=1,\ldots,8$.}
	\label{fig:Parameter_posteriors_gibbs}
\end{figure}

\begin{figure}[H]	
	\caption{Comparison of posterior estimates of time-varying autoregressive coefficients for the PCECC96 equation}
	\begin{center}
		\includegraphics[scale = 0.6]{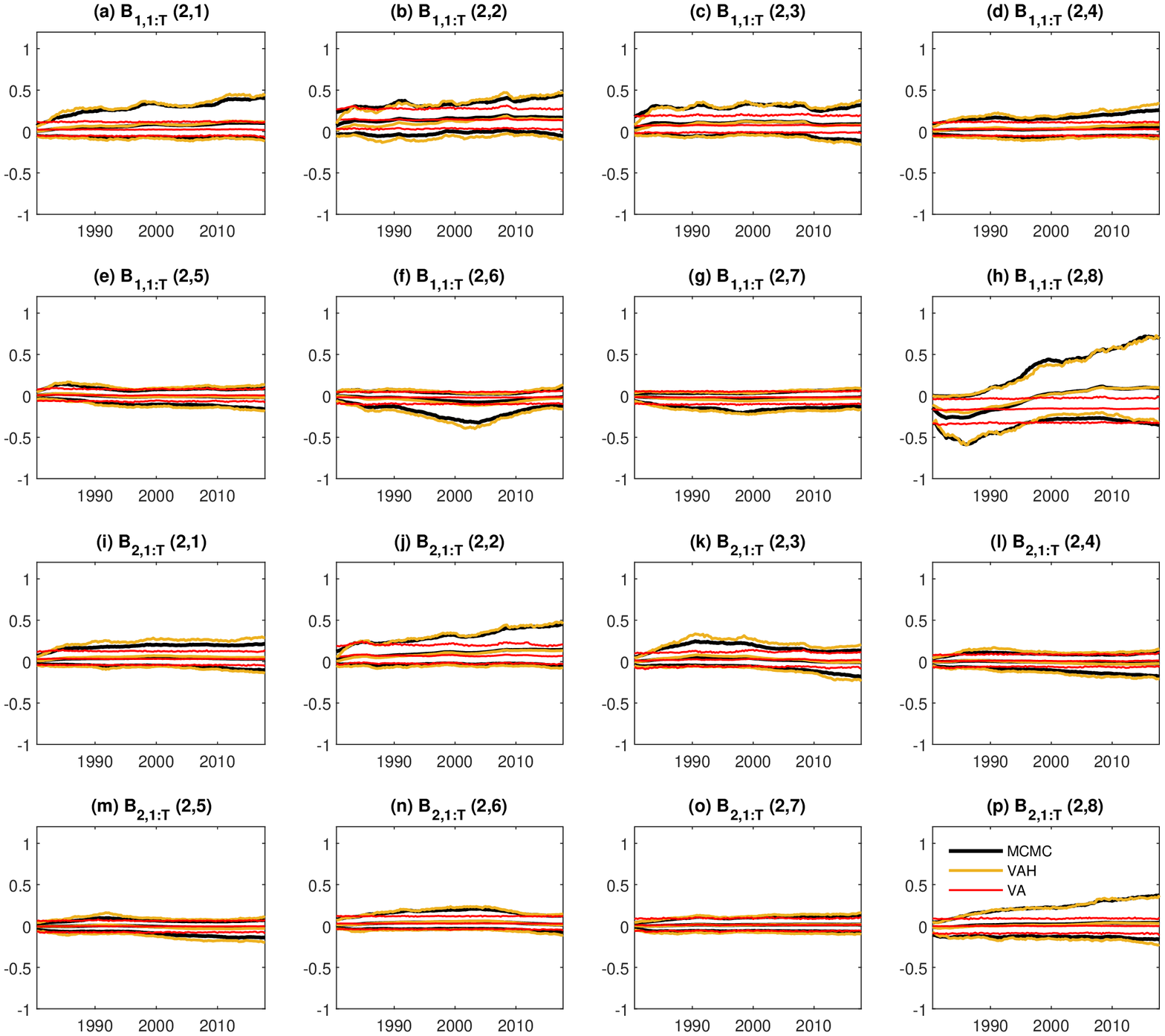}
	\end{center}
Each panel plots the posterior mean and 90\% intervals of 
an element in a time-varying autoregressive coefficient matrix $B_{s,t}$ against time $t=1,\ldots,T$. The top eight panels are the 
elements in row 2 of $B_{1,t}$, and the bottom eight panels
are the elements in row 2 of $B_{2,t}$; these are the coefficients
for the PCECC96 equation. Exact posterior estimates (computed by MCMC) are plotted in black, those for the hybrid VA in yellow,
and that for the structured Gaussian VA in red.
	\label{fig:Parameter_posteriors_betas_2}
\end{figure}

\begin{figure}[H]	
	\caption{Comparison of posterior estimates of time-varying autoregressive coefficients for the FPIx equation.}
	\begin{center}
		\includegraphics[scale=0.6]{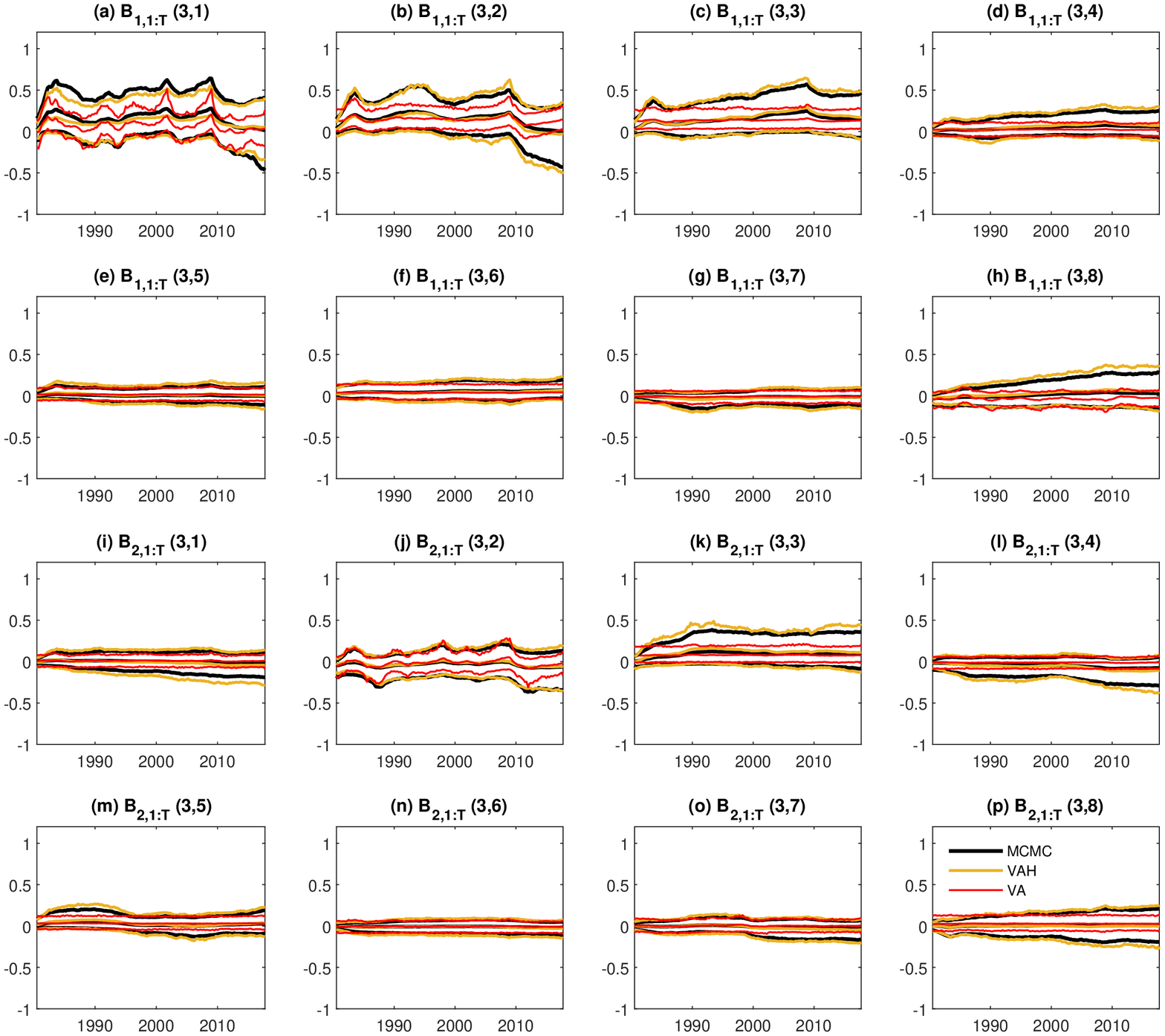}
	\end{center}
Each panel plots the posterior mean and 90\% intervals of 
an element in a time-varying autoregressive coefficient matrix $B_{s,t}$ against time $t=1,\ldots,T$. The top eight panels are the 
elements in row 3 of $B_{1,t}$, and the bottom eight panels
are the elements in row 3 of $B_{2,t}$; these are the coefficients
for the FPIx equation. Exact posterior estimates (computed by MCMC) are plotted in black, those for the hybrid VA in yellow,
and that for the structured Gaussian VA in red.	\label{fig:Parameter_posteriors_betas_3}
\end{figure}

\begin{figure}[H]	
	\caption{Comparison of posterior estimates of time-varying autoregressive coefficients for the CE16OV equation.}
	\begin{center}
		\includegraphics[scale=0.6]{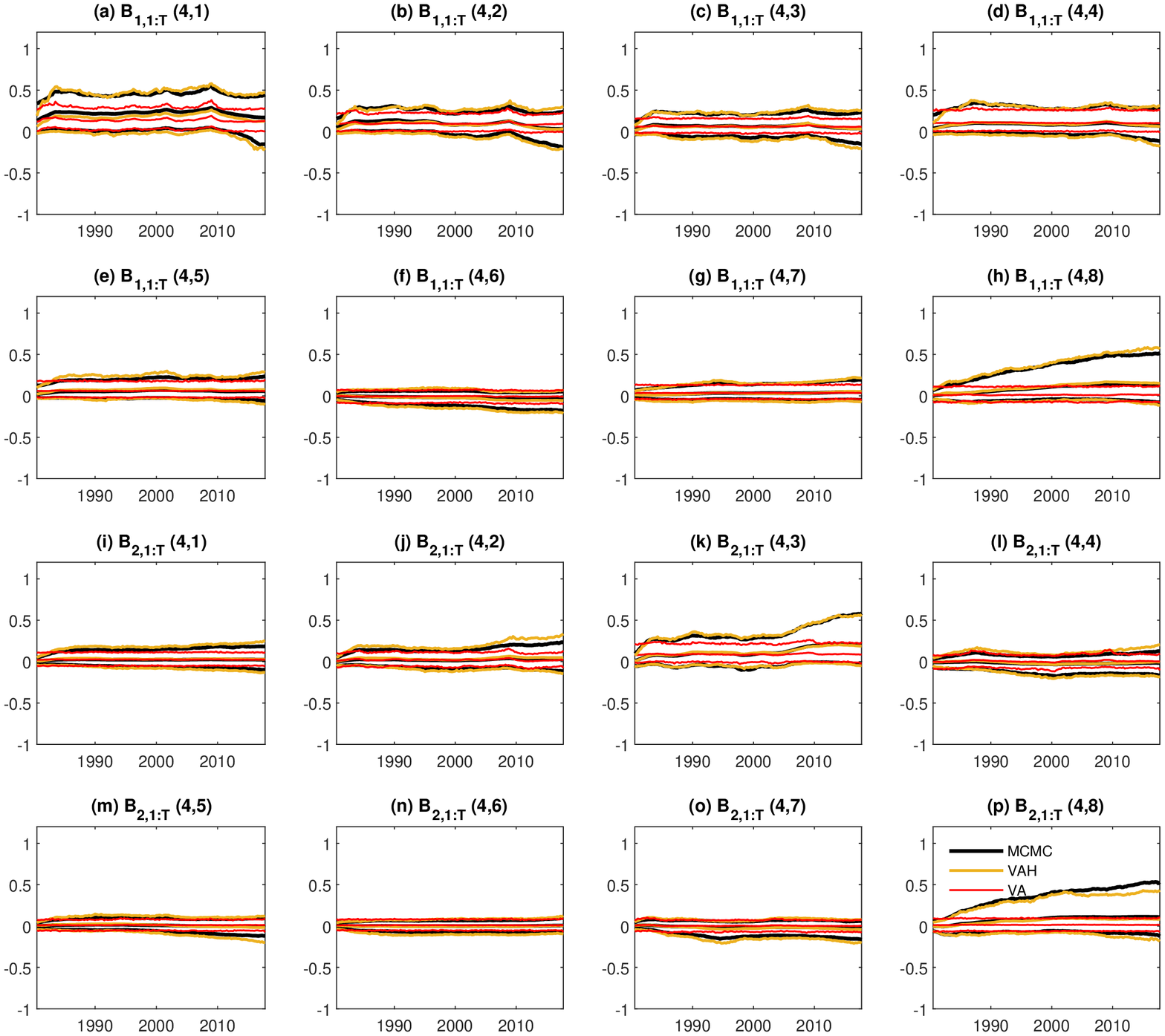}
	\end{center}
Each panel plots the posterior mean and 90\% intervals of 
an element in a time-varying autoregressive coefficient matrix $B_{s,t}$ against time $t=1,\ldots,T$. The top eight panels are the 
elements in row 4 of $B_{1,t}$, and the bottom eight panels
are the elements in row 4 of $B_{2,t}$; these are the coefficients
for the CE160V equation. Exact posterior estimates (computed by MCMC) are plotted in black, those for the hybrid VA in yellow,
and that for the structured Gaussian VA in red.	\label{fig:Parameter_posteriors_betas_4}
\end{figure}

\begin{figure}[H]	
	\caption{Comparison of posterior estimates of time-varying autoregressive coefficients for the CES0600000008 equation.}
	\begin{center}
		\includegraphics[scale=0.6]{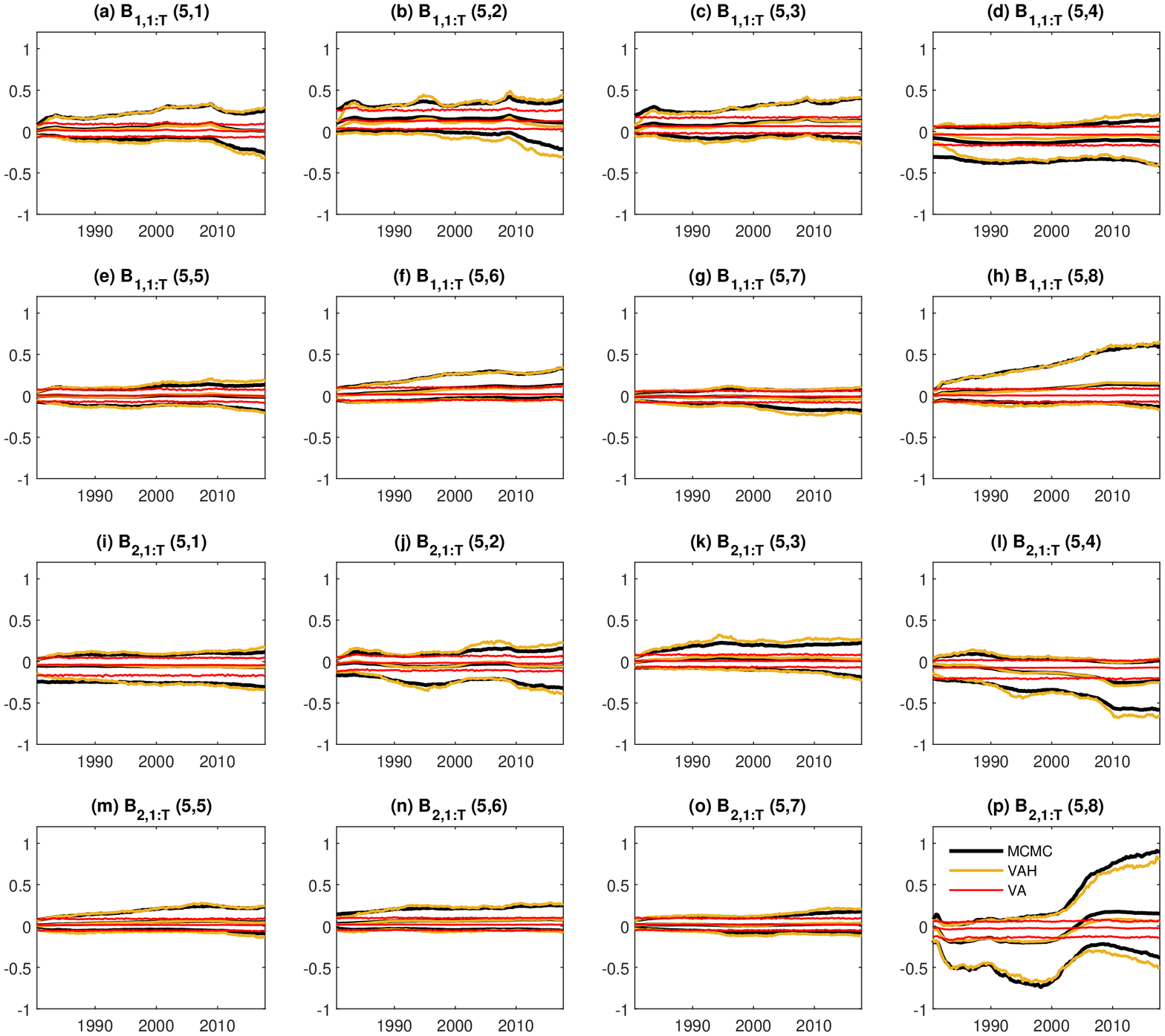}
	\end{center}
Each panel plots the posterior mean and 90\% intervals of 
an element in a time-varying autoregressive coefficient matrix $B_{s,t}$ against time $t=1,\ldots,T$. The top eight panels are the 
elements in row 5 of $B_{1,t}$, and the bottom eight panels
are the elements in row 5 of $B_{2,t}$; these are the coefficients
for the CES0600000008 equation. Exact posterior estimates (computed by MCMC) are plotted in black, those for the hybrid VA in yellow,
and that for the structured Gaussian VA in red.	\label{fig:Parameter_posteriors_betas_5}
\end{figure}

\begin{figure}[H]	
	\caption{Comparison of posterior estimates of time-varying autoregressive coefficients for the GDPCTPI equation.}
	\begin{center}
		\includegraphics[scale=0.6]{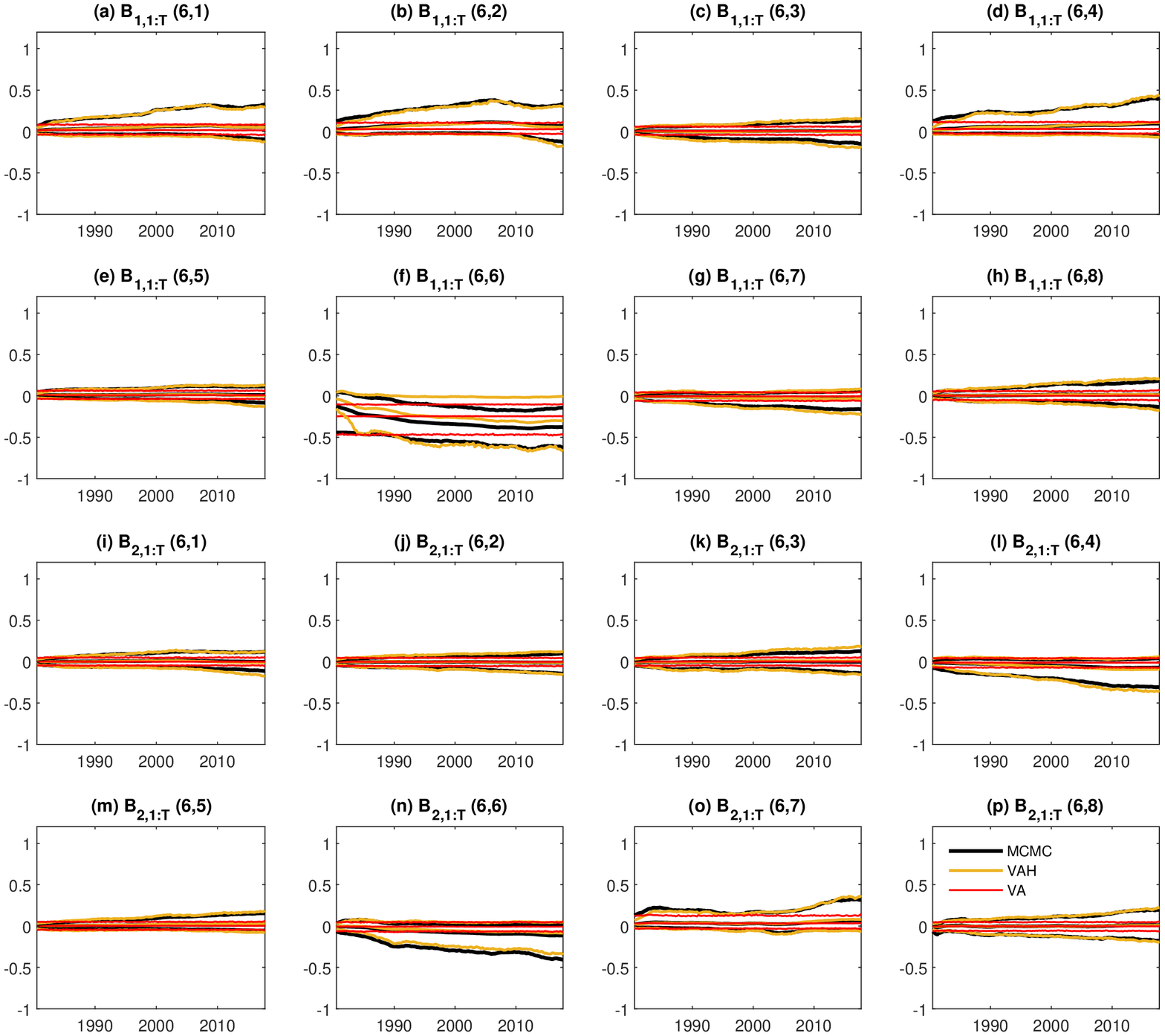}
	\end{center}
Each panel plots the posterior mean and 90\% intervals of 
an element in a time-varying autoregressive coefficient matrix $B_{s,t}$ against time $t=1,\ldots,T$. The top eight panels are the 
elements in row 6 of $B_{1,t}$, and the bottom eight panels
are the elements in row 6 of $B_{2,t}$; these are the coefficients
for the GDPCTPI equation. Exact posterior estimates (computed by MCMC) are plotted in black, those for the hybrid VA in yellow,
and that for the structured Gaussian VA in red.	\label{fig:Parameter_posteriors_betas_6}
\end{figure}

\begin{figure}[H]	
	\caption{Comparison of posterior estimates of time-varying autoregressive coefficients for the CES0600000008 equation.}
	\begin{center}
		\includegraphics[scale=0.6]{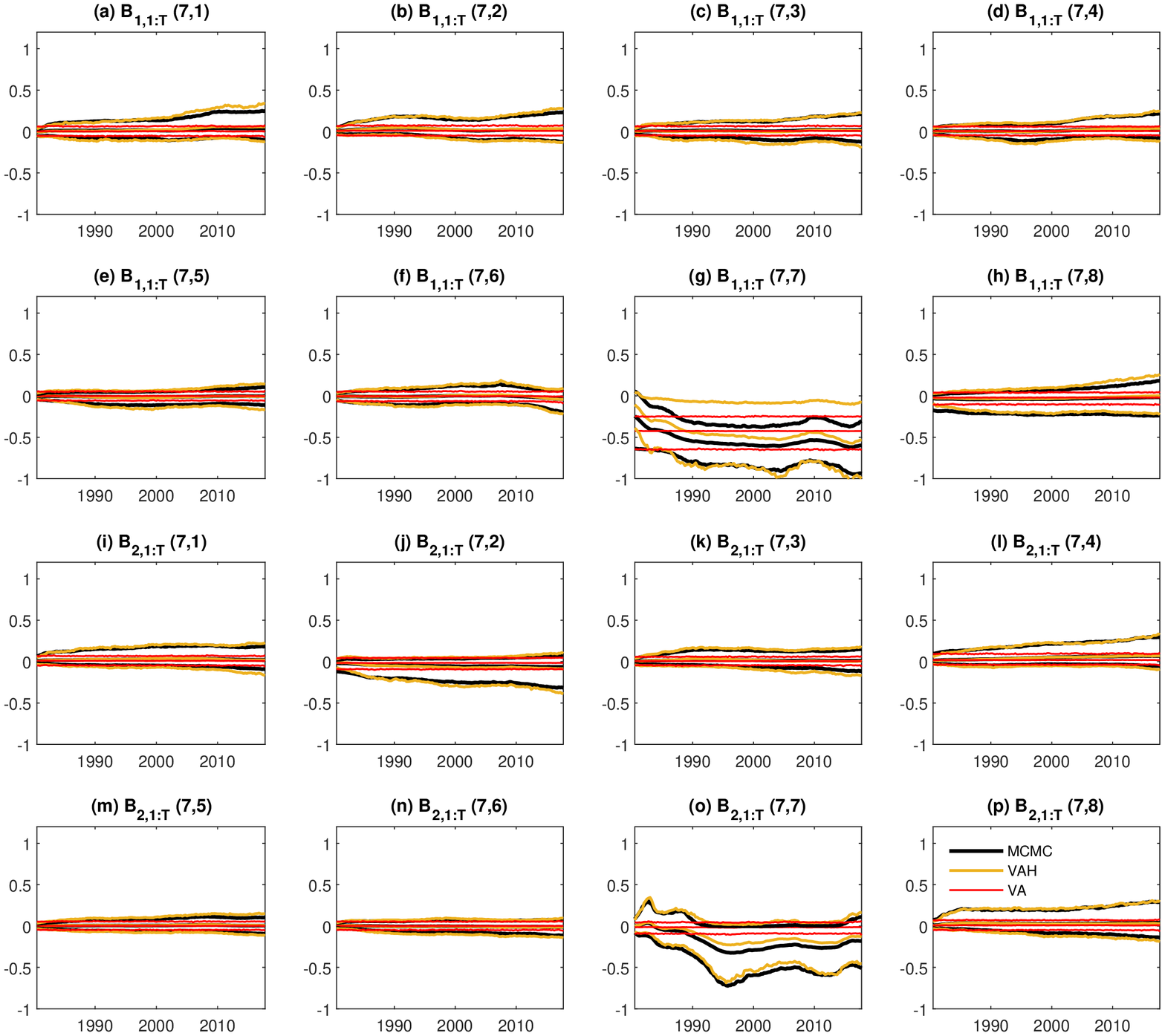}
	\end{center}
Each panel plots the posterior mean and 90\% intervals of 
an element in a time-varying autoregressive coefficient matrix $B_{s,t}$ against time $t=1,\ldots,T$. The top eight panels are the 
elements in row 7 of $B_{1,t}$, and the bottom eight panels
are the elements in row 7 of $B_{2,t}$; these are the coefficients
for the CES0600000008 equation. Exact posterior estimates (computed by MCMC) are plotted in black, those for the hybrid VA in yellow,
and that for the structured Gaussian VA in red.	\label{fig:Parameter_posteriors_betas_7}
\end{figure}

\begin{figure}[H]	
	\caption{Comparison of posterior estimates of time-varying autoregressive coefficients for the FEDFUNDS equation.}
	\begin{center}
		\includegraphics[scale=0.6]{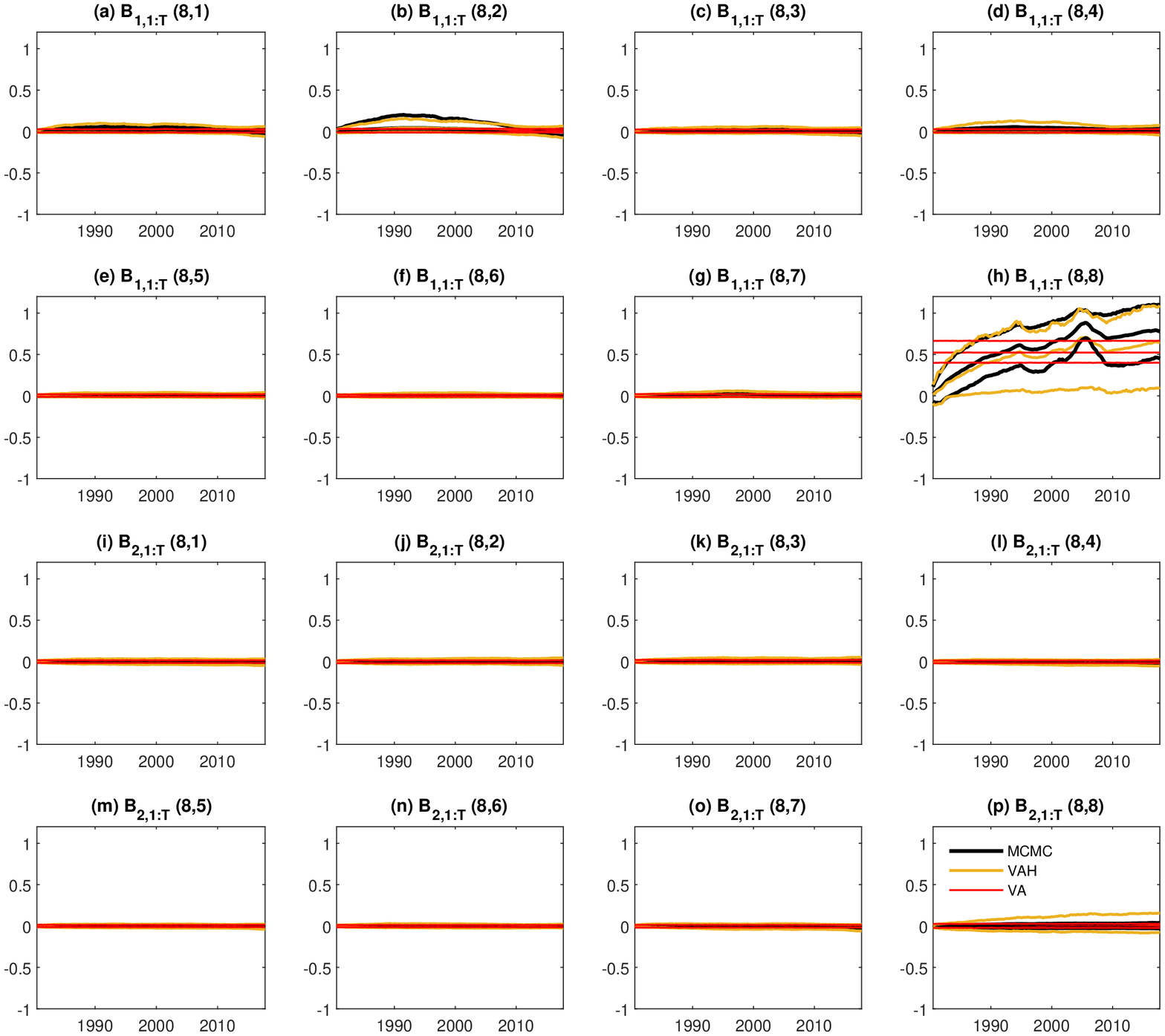}
	\end{center}
Each panel plots the posterior mean and 90\% intervals of 
an element in a time-varying autoregressive coefficient matrix $B_{s,t}$ against time $t=1,\ldots,T$. The top eight panels are the 
elements in row 8 of $B_{1,t}$, and the bottom eight panels
are the elements in row 8 of $B_{2,t}$; these are the coefficients
for the FEDFUNDS equation. Exact posterior estimates (computed by MCMC) are plotted in black, those for the hybrid VA in yellow,
and that for the structured Gaussian VA in red.	\label{fig:Parameter_posteriors_betas_8}
\end{figure}

\begin{figure}[H]	
	\begin{center}
			\caption{Comparison of posterior estimates of the top left quadrant of $L_t^{-1}$.}
		\includegraphics[scale=0.7]{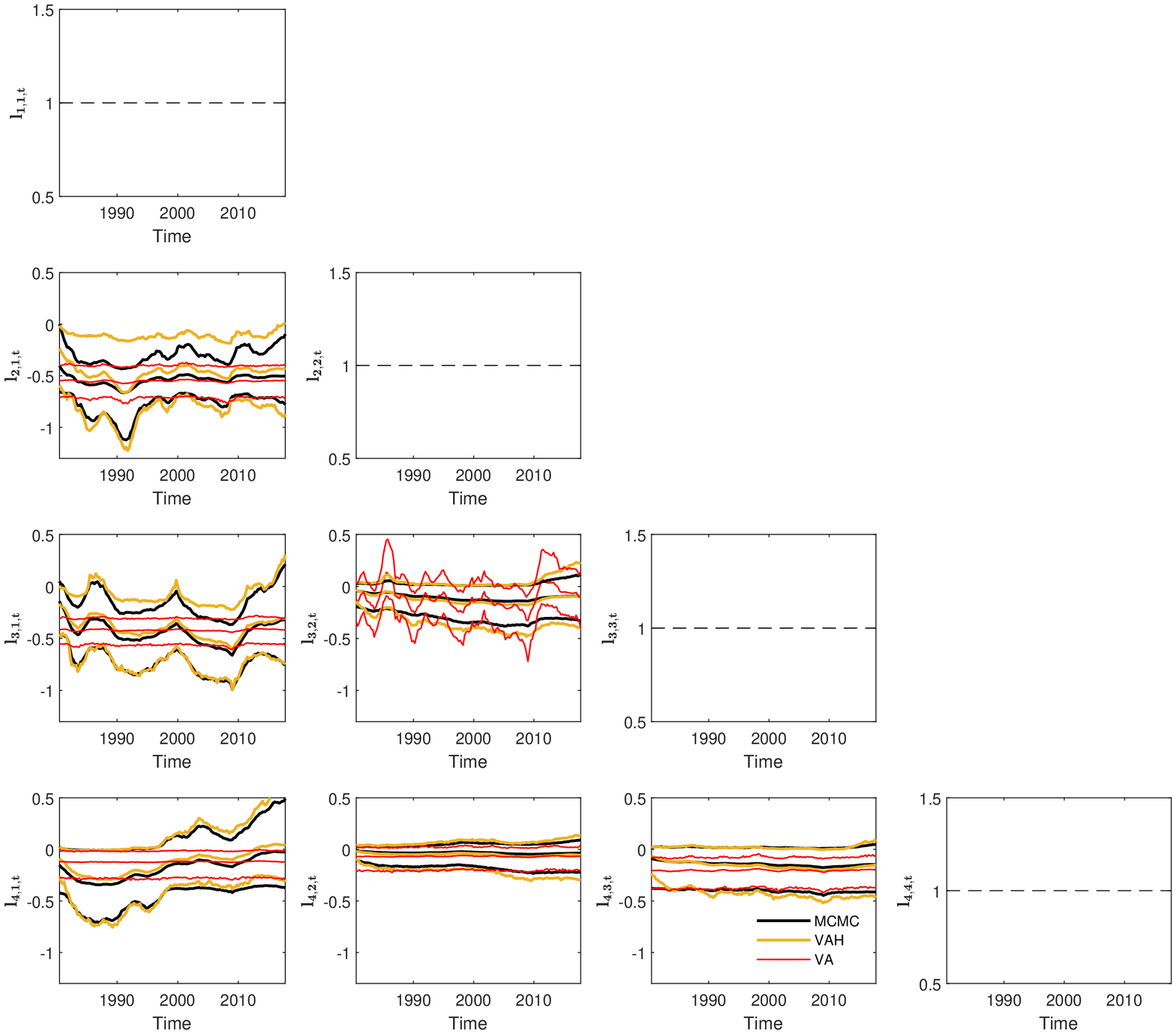}
	\end{center}
	The posterior mean and posterior 90\% intervals are provided for the top left elements of the matrix $L_t^{-1}$ of the TV-VAR-SV model, for $t=1,\ldots,T$. Exact posterior estimates (computed by MCMC) are plotted in black, the hybrid VA in yellow,
	and the structured Gaussian VA in red.	
	\label{fig:Parameter_posteriors_Lmat1}
\end{figure}

\begin{figure}[H]	
	\begin{center}
			\caption{Comparison of posterior estimates of the bottom left quadrant of $L_t^{-1}$.}
		\includegraphics[scale=0.7]{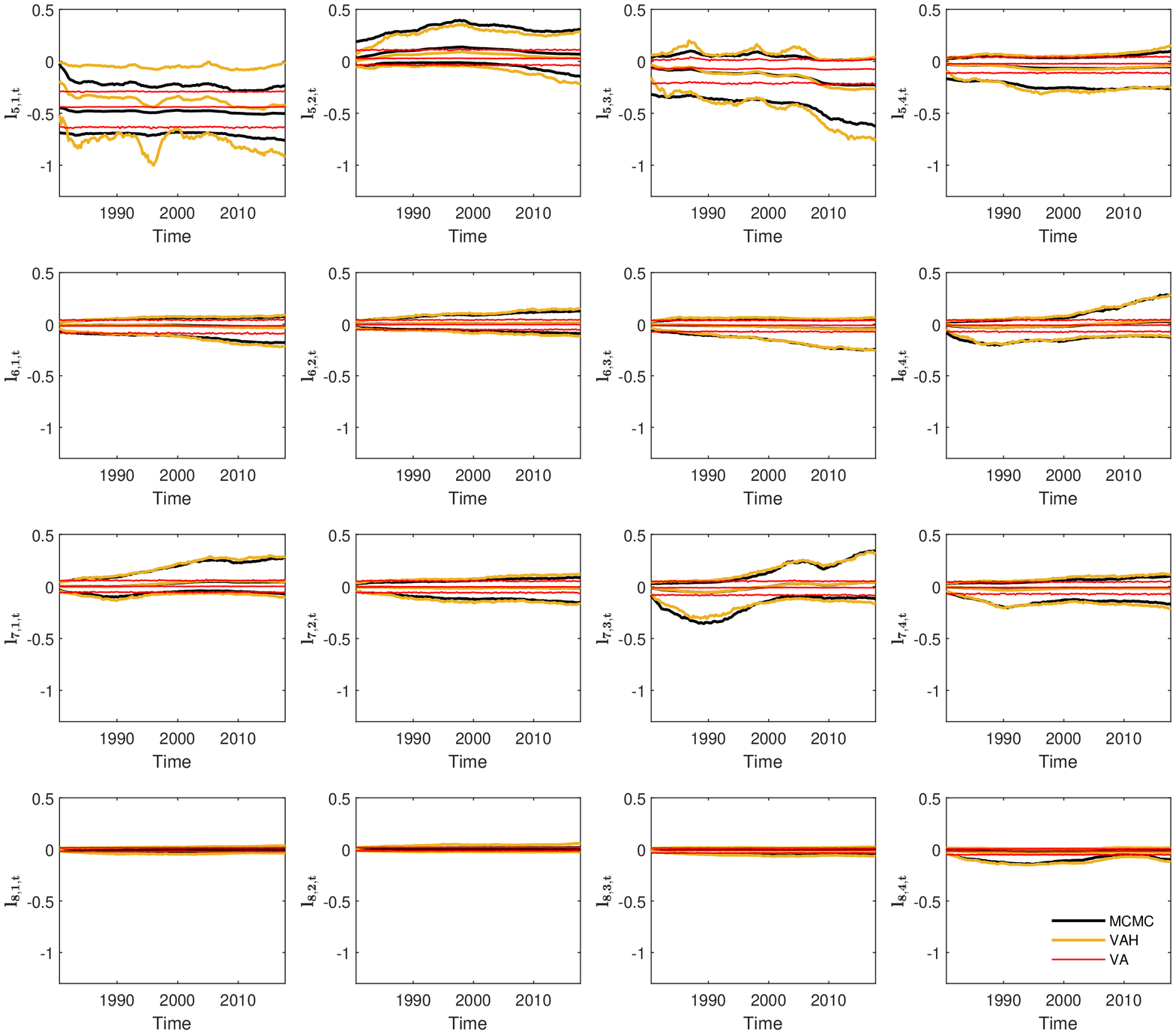}
	\end{center}
	The posterior mean and posterior 90\% intervals are provided for the bottom left elements of the matrix $L_t^{-1}$ of the  TV-VAR-SV model, for $t=1,\ldots,T$. Exact posterior estimates (computed by MCMC) are plotted in black, the hybrid VA in yellow,
	and the structured Gaussian VA in red.		\label{fig:Parameter_posteriors_Lmat2}
\end{figure}

\begin{figure}[H]	
	\begin{center}
			\caption{Comparison of posterior estimates of the bottom right quadrant of $L_t^{-1}$.}
		\includegraphics[scale=0.7]{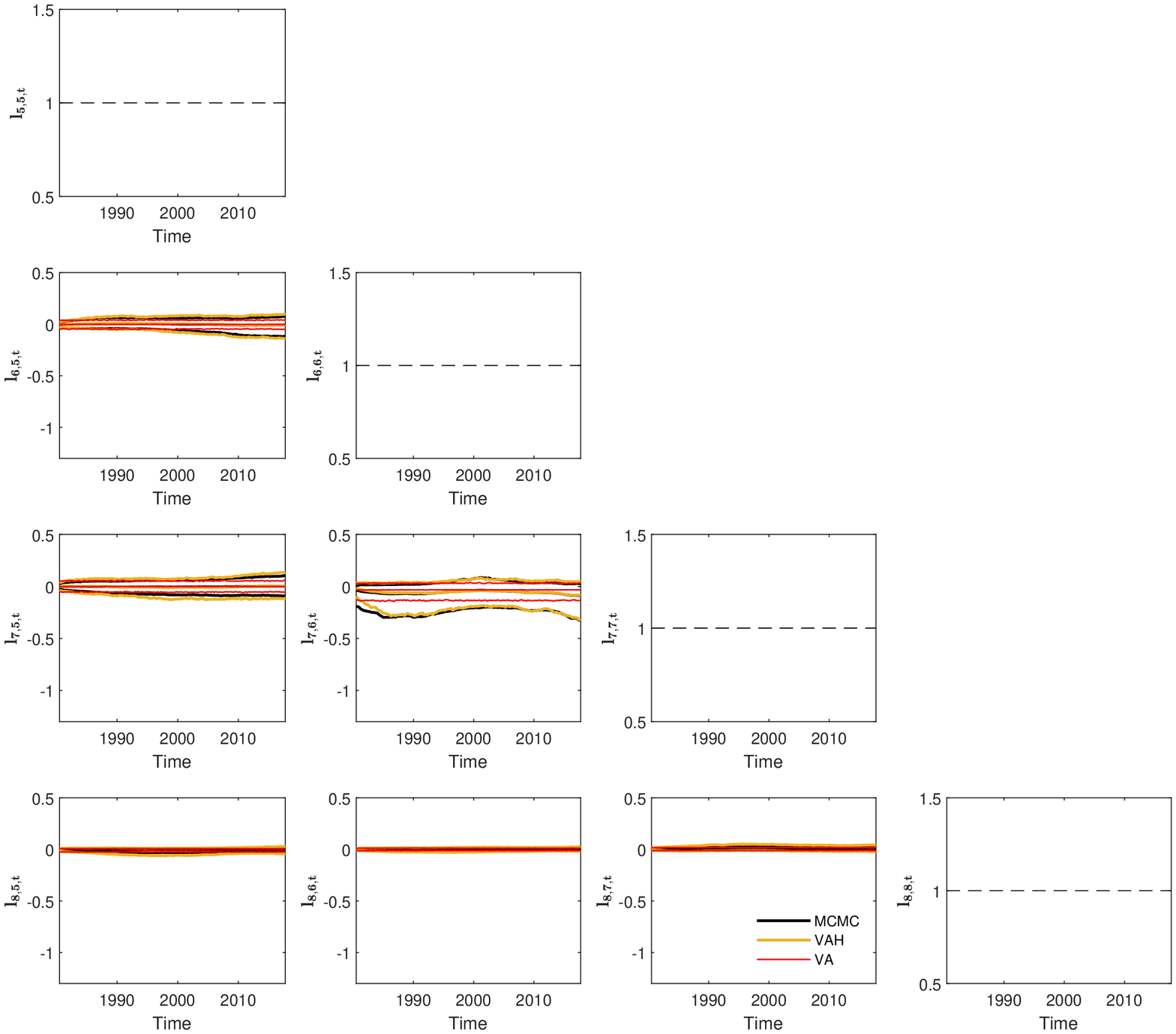}
	\end{center}
	The posterior mean and posterior 90\% intervals are provided for the bottom right elements of the matrix $L_t^{-1}$ of the TV-VAR-SV model, for $t=1,\ldots,T$. 
	Exact posterior estimates (computed by MCMC) are plotted in black, the hybrid VA in yellow,
	and the structured Gaussian VA in red.		\label{fig:Parameter_posteriors_Lmat3}
\end{figure}

\newpage
\noindent {\bf \large{Part~B: Additional Details and Results for Section~4}}\\
\ \\
\noindent 
This appendix is split into four sub-sections. Part~B.1 provides further details on the parameterization and choice of priors. 
In Part~B.2 all the required gradients and derivatives for the SGA algorithm are derived. Part~B.3 provides details on MCMC estimation of the tobit model, how to draw $\zvec$ in step~(b) of Algorithm~1
and details on how long the MCMC sampler took to run for both the small and large
tobit examples. Finally, Part~B.4 provides additional results for the empirical application.\\
\textbf{{\large B.1 Parameterization and Priors}}\\ 
The parameters of the mixed effects tobit model are $\thetavec=(\betavec^\top,\mbox{vech}(L)^\top,\omegavec^\top,\sigma^2)^\top$. All the elements in $\bm{\theta}$ need to be transformed to the real line. To do this, we introduce $c = \log(1/\sigma)$, $\xi_i = \log(\omega_i)$, $\bm\xi = \left(\xi_1,\dots,\xi_r\right)^\top$, $\kappa_j = \log(L_{j,j})$, $\bm\kappa = \left(\kappa_1,\dots,\kappa_{k_\alpha}\right)^\top$, and $\bm{l}$ which denotes the elements in $\mbox{vech}(L)$ with the diagonal elements of $L$ excepted. We then re-express this parameter vector as $\thetavec=(\betavec^\top,\xivec^\top,c,\bm{\kappa}^\top,\bm{l}^\top)^\top$. 

The prior is defined as $p\left(\bm{\theta}\right) = p\left(\bm{\beta}\right)p\left(\bm{\xi}\right)p\left(c\right)p\left(\bm{\kappa}\right)p\left(\bm{l}\right)$, with
\begin{align*}
\text{(i)}\hspace{0.3cm} &p\left(\bm{\beta}\right) = \phi_{p}\left(\bm{\beta};\bm{0},\Sigma_\beta\right),\hspace{1cm}\text{(ii)}\ p\left(\bm{\xi}\right) = \prod_{i=1}^{r}\exp\left(\xi_i\right)^{-1}\exp\left(-\frac{1}{\exp\left(\xi_i\right)}\right),\hspace{1cm}\text{(iii)}\ p(c)\propto 1\\
\text{(iv)} \hspace{0.1cm} &p\left(\bm{\kappa}\right) = \prod_{j=1}^{k_\alpha}2\phi_1\left(\exp(\kappa_j);0,\sigma_l^2\right)\exp(\kappa_j),\hspace{1.1cm}\text{(v)}\ p\left(\bm{l}\right) = \prod_{i=2}^{r}\prod_{j=1}^{\text{min}(i-1,k_\alpha)}\phi_1\left(L_{i,j};0,\sigma_l^2\right),\hspace{1cm}
\end{align*}
Here, $p(c)$ was constructed by considering the prior $p(\sigma^2)\propto\frac{1}{\sigma^2}$ and deriving the corresponding prior on $c$. The prior $p\left(\bm{\xi}\right)$ was constructed by using the prior $p\left(\bm{\omega}\right) = \prod_{i=}^{r}p\left(\omega_{i}\right), \text{ with } p_\omega\left(\omega_{i}\right) = \text{Inv-Gamma}\left(\omega_{i},1,1\right)$, and then using the Jacobian of the transformation to derive the corresponding prior on $\bm{\xi}$.
Finally, $p\left(\bm{\kappa}\right)$ was constructed by considering the truncated normal prior $2\phi_1\left(L_{j,j};0,\sigma_l^2\right)$ $I\left(L_{j,j}>0\right)$  on $L_{j,j}$ and deriving the corresponding prior for $\kappa_j$ via the Jacobian of the transformation.
\newpage

\noindent
\textbf{{\large B.2 Computing gradients}}\\
Given the expressions for the augmented posterior and prior densities above, the function $\log g(\bm\theta,\bm{z})$ can be written as:
\begin{align}\nonumber
\log g(\bm\theta,\bm{z}) = & \log\left\{\left[\prod_{i=1,t=1}^{N,T}\phi_1\left(y^{*}_{i,t};\eta_{i,t},\sigma^2\right)\right]\left[\prod_{i=1}^N\phi_{r}\left(\bm{\alpha}_i;\bm{0},V_\alpha\right)\right]
p\left(\bm{\beta}\right)p\left(\bm{\xi}\right)p\left(c\right)p\left(\bm{\kappa}\right)p\left(\bm{l}\right)\right\}\\\nonumber
= & -\frac{n\log(2\pi)}{2}+nc-\frac{e^{2c}}{2}\sum_{i,t}\left(y_{i,t}^*-\eta_{i,t}\right)^2-\frac{rN\log(2\pi)}{2}-\frac{N}{2}\log(|V_\alpha|)-\\
&\frac{1}{2}\sum_{i}\bm{\alpha}_j^\top V_\alpha^{-1}\bm{\alpha}_j-\frac{p\log(2\pi)}{2}-\frac{1}{2}\log(|\Sigma_\beta|)-\frac{1}{2}\bm{\beta}^\top \Sigma_\beta^{-1}\bm{\beta}-\frac{1}{2}rk_\alpha\log(2\pi)-\nonumber\\
&\frac{rk_\alpha-(k_\alpha-1)k_\alpha}{2}\log(\sigma_l^2)+k_\alpha\log(2)-\sum_{i=1}^r\sum_{j=1}^{k_\alpha}\frac{l_{i,j}^2}{2\sigma_l^2}+\sum_{j=1}^{k_\alpha}\kappa_j-\sum_{i=1}^r\left(\xi_i+\frac{1}{\exp(\xi_i)}\right)\nonumber\\
=&  \text{const}+nc -\frac{e^{2c}}{2}\sum_{i,t}\left(y_{i,t}^*-\eta_{i,t}\right)^2-\frac{N}{2}\log(|V_\alpha|)-\frac{1}{2}\sum_{i}\left(\bm{\alpha}_i^\top \otimes \bm{\alpha}_j^\top\right)\text{vec}\left(V_\alpha^{-1}\right)-\nonumber\\
&\frac{1}{2}\bm{\beta}^\top \Sigma_\beta^{-1}\bm{\beta}-\sum_{i=1}^r\sum_{j=1}^{k_\alpha}\frac{l_{i,j}^2}{2\sigma_l^2}+\sum_{j=1}^{k_\alpha}\kappa_j-\sum_{i=1}^r\left(\xi_i+\frac{1}{\exp(\xi_i)}\right)\nonumber\\
& = \text{const} + f(\yvec^\star,\bm{\alpha},\bm{\theta}) + \sum_{j=1}^{k_\alpha}\kappa_j\nonumber
\end{align}
where $\yvec^\star$ is the $Nr$ vector of the elements $y_{i,t}^\star$. 
The required gradients with respect to $c$ and $\bm{\beta}$ can be computed as:
$$\nabla_c \log g(\bm\theta,\bm{z}) = n-e^{2c}\sum_{i,t}\left(y_{i,t}^*-\eta_{i,t}\right)^2$$
$$\nabla_\beta \log g(\bm\theta,\bm{z})= e^{2c}\sum_{i,t}\left(y_{i,t}^*-\eta_{i,t}\right)\bm{x}_{i,t}^\top-\bm{\beta}^\top\Sigma_{\beta}^{-1}$$
To construct the gradients with respect to $\bm{l}$ and $\bm{\kappa}$, we must first compute
\begin{align}
\nabla_{L} f(\bm{y}^*,\bm{\alpha},\bm{\theta}) =&
-\frac{N}{2}\frac{1}{|V_{\alpha}|}\frac{\partial|V_{\alpha}|}{\partial V_{\alpha}}\frac{\partial V_{\alpha}}{\partial L}-\frac{1}{2}\sum_{i}\left(\bm{\alpha}_i^\top\otimes\bm{\alpha}_i^\top\right)\frac{\partial V_{\alpha}^{-1}}{\partial V_{\alpha}}\frac{\partial V_{\alpha}}{\partial L}-\frac{1}{\sigma_l^2}\text{vec}\left(L\right)^\top\nonumber\\
=&-\frac{N}{2|V_{\alpha}|}\text{vec}\left(|V_{\alpha}|V_{\alpha}^{-1}\right)^\top\frac{\partial V_{\alpha}}{\partial L}-\frac{1}{2}\sum_{i}\left(\bm{\alpha}_i^\top\otimes\bm{\alpha}_i^\top\right)\frac{\partial V_{\alpha}^{-1}}{\partial V_{\alpha}}\frac{\partial V_{\alpha}}{\partial L}-\frac{1}{\sigma_l^2}\text{vec}\left(L\right)^\top\nonumber
\end{align}
The elements of the gradients $\nabla_{l} f(\bm{y}^*,\bm{\alpha},\bm{\theta})$ and $\nabla_{\kappa} f(\bm{y}^*,\bm{\alpha},\bm{\theta})$ can then be constructed using 
\begin{align*}
\nabla_{L_{j,i}} \log g(\bm\theta,\bm{z}) =& \nabla_{L_{j,i}} f(\bm{y}^*,\bm{\alpha},\bm{\theta})\\
\nabla_{\kappa_{j}} \log g(\bm\theta,\bm{z})=& \nabla_{L_{j,j}} f(\bm{y}^*,\bm{\alpha},\bm{\theta})\exp(\kappa_j)+1\nonumber
\end{align*}
Finally, the gradient with respect to $\bm\xi$ can be constructed as:
\begin{align}
\nabla_{\xi} \log g(\bm\theta,\bm{z})=&
-\frac{N}{2}\frac{1}{|V_{\alpha}|}\frac{\partial|V_{\alpha}|}{\partial V_{\alpha}}\frac{\partial V_{\alpha}}{\partial \bm{\omega}}\frac{\partial \bm{\omega}}{\partial \bm{\xi}}-\frac{1}{2}\sum_{i}\left(\bm{\alpha}_i^\top\otimes\bm{\alpha}_i^\top\right)\frac{\partial V_{\alpha}^{-1}}{\partial V_{\alpha}}\frac{\partial V_{\alpha}}{\partial \bm{\omega}}\frac{\partial \bm{\omega}}{\partial \bm{\xi}}-\left(\bm{1}_r-\bm\omega^{-1}\right)\nonumber\\
=&-\frac{N}{2|V_{\alpha}|}\text{vec}\left(|V_{\alpha}|V_{\alpha}^{-1}\right)^\top\frac{\partial V_{\alpha}}{\partial \bm{\omega}}\frac{\partial \bm{\omega}}{\partial \bm{\xi}}-\frac{1}{2}\sum_{i}\left(\bm{\alpha}_i^\top\otimes\bm{\alpha}_i^\top\right)\frac{\partial V_{\alpha}^{-1}}{\partial V_{\alpha}}\frac{\partial V_{\alpha}}{\partial \bm{\omega}}\frac{\partial \bm{\omega}}{\partial \bm{\xi}}-\left(\bm{1}_r-\bm\omega^{-1}\right)\nonumber
\end{align}

All the expression above can be computed by noting that  $\frac{\partial V_{\alpha}}{\partial L} = \left(I_{r^2}+K_{r,r}\right)\left(L\otimes I_r\right)$ and $\frac{\partial V_{\alpha}^{-1}}{\partial V_{\alpha}} = -\left(V_{\alpha}^{-1}\otimes V_{\alpha}^{-1}\right)$. We can further simplify $\frac{\partial V_{\alpha}^{-1}}{\partial V_{\alpha}}\frac{\partial V_{\alpha}}{\partial L} =-\left(I_{r^2}+K_{r,r}\right)\left(V_{\alpha}^{-1}L\otimes V_{\alpha}^{-1}\right)$. For the gradeints with respect to $\bm{\xi}$ we can use $\frac{\partial V_{\alpha}}{\partial \bm{\omega}} = I_{r^2}P$, where $P$ is the matrix of ones and zeros that extract columns 1, $r+2$, $2r+3$,...$r^2$, and $\frac{\partial \bm{\omega}}{\partial \bm{\xi}}=\Omega$.
\ \\
\ \\
\textbf{{\large B.3 Exact Bayesian inference and drawing $\zvec$}}\\ 
For exact Bayesian inference on the augmented posterior we employ the following MCMC sampling scheme:\\
\ \\
$\underline{\text{Sampling Scheme}}$\\
\ \ \hspace{2cm} Step 1: Generate from $\bm{\alpha}|\yvec_{\tiny U}^\star,\bm{\theta},\bm{y}$.\\
\ \ \hspace{2cm} Step 2: Generate from $\bm{y}^\star_{{\tiny U}}|\bm{\alpha},\bm{\theta},\bm{y}$.\\
\ \ \hspace{2cm} Step 3: Generate from $\bm{\theta}|\yvec_{\tiny U}^\star,\bm{\alpha},\bm{y}$.\\
\ \\
To perform Step 1, note that $p(\bm{\alpha}|\yvec_{\tiny U}^\star,\bm{\theta},\bm{y})=\prod_{i=1}^Np\left(\bm{\alpha}_i|\bm{y},\yvec_{\tiny U}^\star,\bm{\beta}\right)$, where each density in the product is an $r$-dimensional Gaussian density, so that
$p\left(\bm{\alpha}_i|\bm{y},\yvec_{\tiny U}^\star,\bm{\beta}\right)=\phi_r\left(\bm{\alpha}_i;A_i^{-1}M_i^\top,A_i^{-1}\right)$ with $M_i = \frac{1}{\sigma^2}\sum_{t=1}^{T}\left(y_{i,t}^*-\bm{x}_{i,t}^\top\bm{\beta}\right)\bm{w}_{i,t}^\top$ and $A_i = V_{\alpha}^{-1}+\frac{1}{\sigma^2}\sum_{t=1}^{T}\bm{w}_{i,t}\bm{w}_{i,t}^\top$.
In Step 2 we draw from $p(\yvec_{\tiny U}^\star|\bm{\alpha},\bm{\theta},\bm{y})=\prod_{i,t|y_{i,t}=0}p(y_{i,t}^\star|y_{i,t},\bm{\alpha},\bm{\theta})$, with
\begin{equation*}
p\left(y_{i,t}^*|y_{i,t},\bm{\alpha},\bm{\theta}\right)=
\frac{\phi_1\left(y^{*}_{i,t};\eta_{i,t},\sigma^2\right)}{\Phi_1\left(0;\eta_{i,t},\sigma^2\right)}I\left(y^{*}_{i,t}\le 0\right) \,.
\end{equation*}
In Step 3, generation is conducted via random walk Metropolis-Hastings. At the beginning of each iteration, the elements of $\bm{\theta}$ are randomly assigned to groups of $10$ elements. The groups are then sampled, one group conditional on the other, with the $10$-dimensional proposal density equal to the product of $10$ independent univariate normals. The variances of the univariate normals are set adaptively to target acceptance rates between $10\%$ and $20\%$.

Steps~1 and 2 in this sampling scheme are also used to generate from
$\zvec$ in step~(b) of Algorithm~1
used to calibrate the Hybrid VA. The latent vector $\zvec$ is initialized at its
last value in the previous step of the SGA algorithm. If $\thetavec$ was constant,
this would define a Gibb sampler for $\zvec$ over the SGA steps. However, $\thetavec$ is not constant over steps, so our approach only provides an approximate
draw from $p(\zvec|\thetavec,\yvec)$. Nevertheless, once the SGA algorithm
converges the value of $\thetavec$ does not change much over SGA step, so that
our approach to generate $\zvec$ at step~(b) is an approximate Gibbs sampler
in this sense. Our empirical results suggest that this works well.

Sampling schemes for this tobit model can be slow to mix when there are large
numbers of latent variables, as in our application. Computational details
for the two data sets are given below. 
\begin{itemize}
	\item[] $\underline{\text{Small Tobit Example}}$:
For the small tobit example we
ran the sampling scheme above for 2 million sweeps, taking a total of 27.081 hours.
While this is a large number of draws, we computed the ``effective sample size'' (ESS)
of these draws
for each element in $\thetavec$, and found that the ESS was between 1,011.3 and 12,413 (with a mean ESS of 3,591 across all elements of $\thetavec$) so that such a large sample is necessary here. 
\item[] $\underline{\text{Large Tobit Example}}$:
We did not run MCMC for the large tobit example, but estimate the run time using different number of sweeps of the sampling scheme. Runs of lengths 0.5, 1 and 2 million sweeps are estimated to take 13.6, 27.2 and 54.4 days, respectively, using the 
same computing environment as our SGA algorithm (i.e. in MATLAB on a contemporary 
laptop).
\end{itemize}
\ \\
\newpage
\noindent {\bf \large{B.4 Supplemental figures and tables}}\\

\begin{figure}[H]
	\begin{center}
			\caption{Posterior moments of $\thetavec$ for the small tobit example.}
		\includegraphics[scale=0.8]{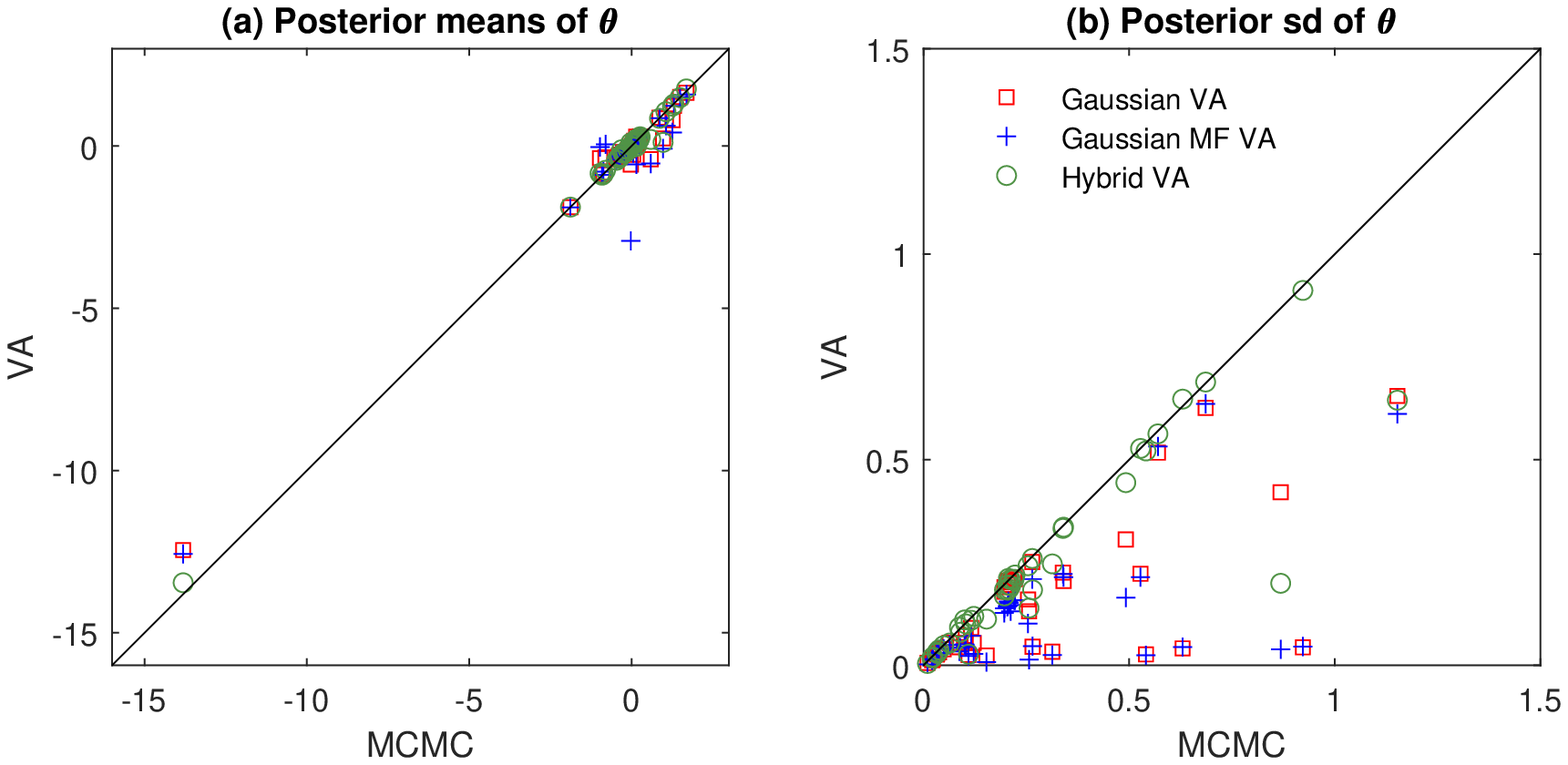}
	\end{center}
	Panel~(a) plots the exact posterior mean of $\thetavec$ (computed using MCMC)
	against the mean of the calibrated VA $q_{\hat{\lambda}}^0(\thetavec)$. Each
	point in the scatter corresponds to a parameter in $\thetavec$. Results are give
	for the Gaussian VA (red box),
	Gaussian mean field VA (blue cross) and our proposed hybrid VA (green circles).
	Panel~(b) is an equivalent plot for the posterior standard deviation. 
	VAs with more accurate moments will have scatters that fall closer to the 45 degree line. 
	\label{fig:poteriormoments}
\end{figure}

\begin{figure}[H]
	\begin{center}
	\caption{Accuracy of the latent $\yvec_{\tiny U}^\star$ estimates
	for the tobit small data example.}
		\includegraphics[scale = 1]{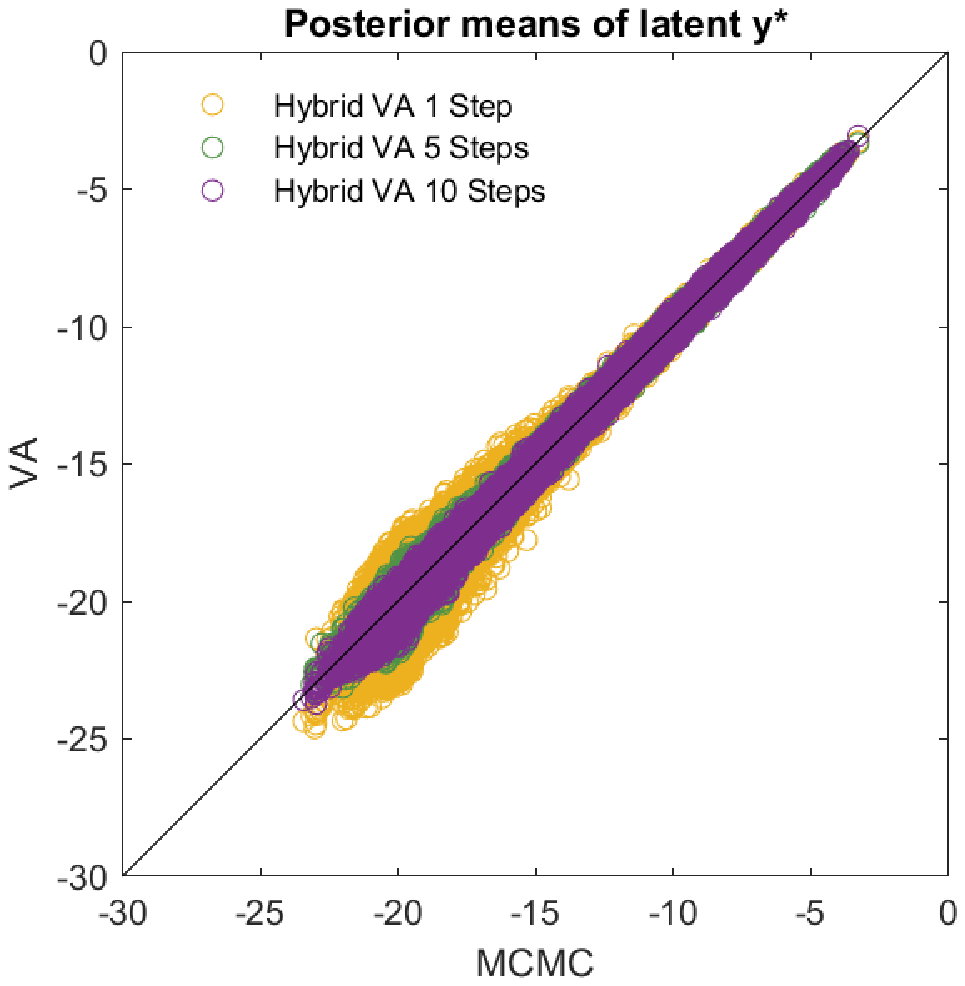}
	\end{center}
	Scatter-plots of VB mean estimates of the latent $\yvec_{\tiny U}^\star$
	values against their true posterior means computed
using MCMC. Accurate estimates fall on the 45 degree line. 
Results are given for the our proposed VA using 1, 5 and 10 sweeps
of a Gibbs sampler at step~(b) of Algorithm~1.
	\label{fig:ystar_posteriors_gibbs_tobit}
\end{figure}

\begin{figure}[H]
  \caption{Tobit small data example: marginal posterior densities of coefficients (group 1)}
	\begin{center}
		\includegraphics[width=1\textwidth]{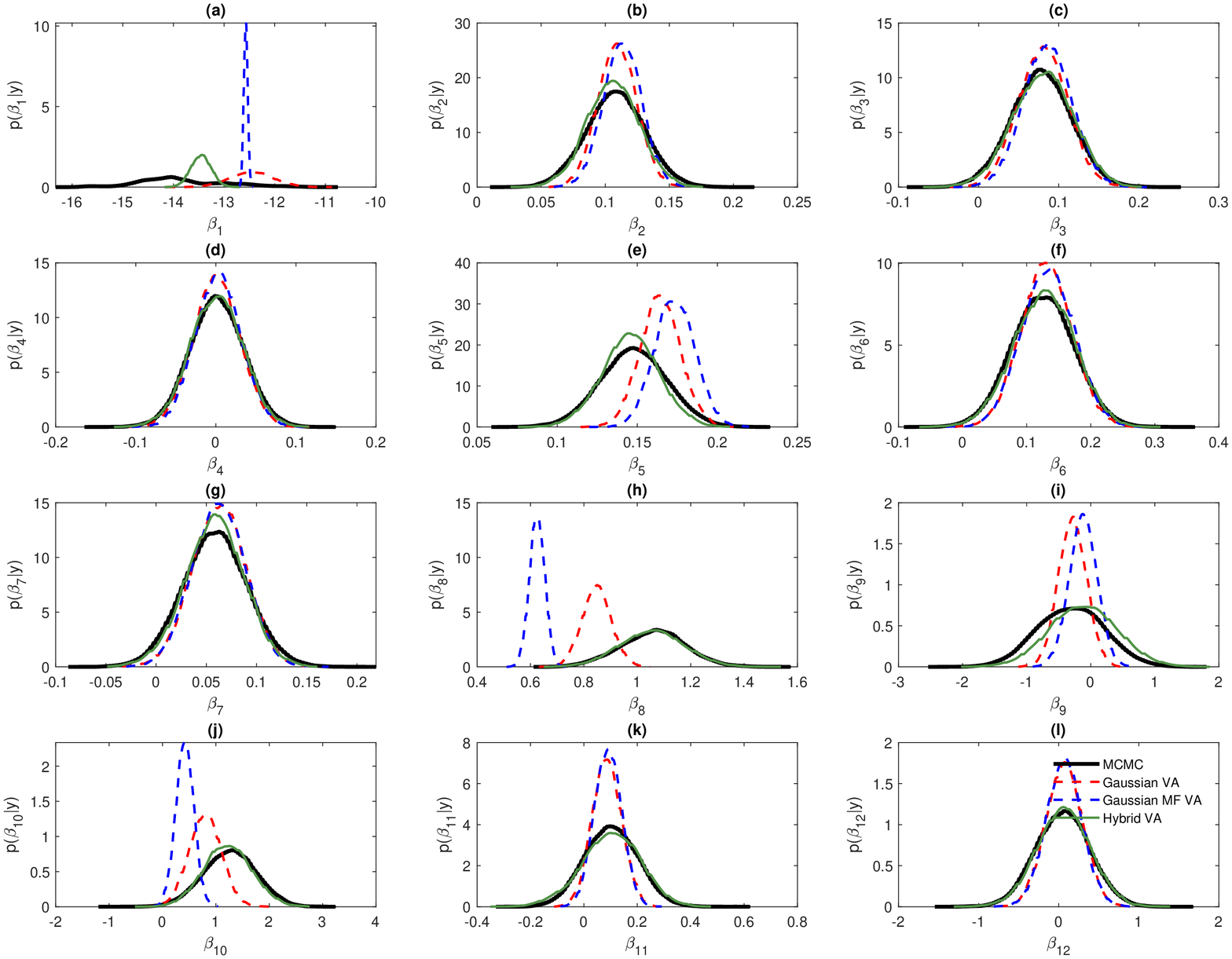}
	\end{center}
	
	\label{fig:poteriordens1}
\end{figure}

\begin{figure}[H]
			\caption{Tobit small data example: marginal posterior densities of coefficients (group 2)}
	\begin{center}
		\includegraphics[width=1\textwidth]{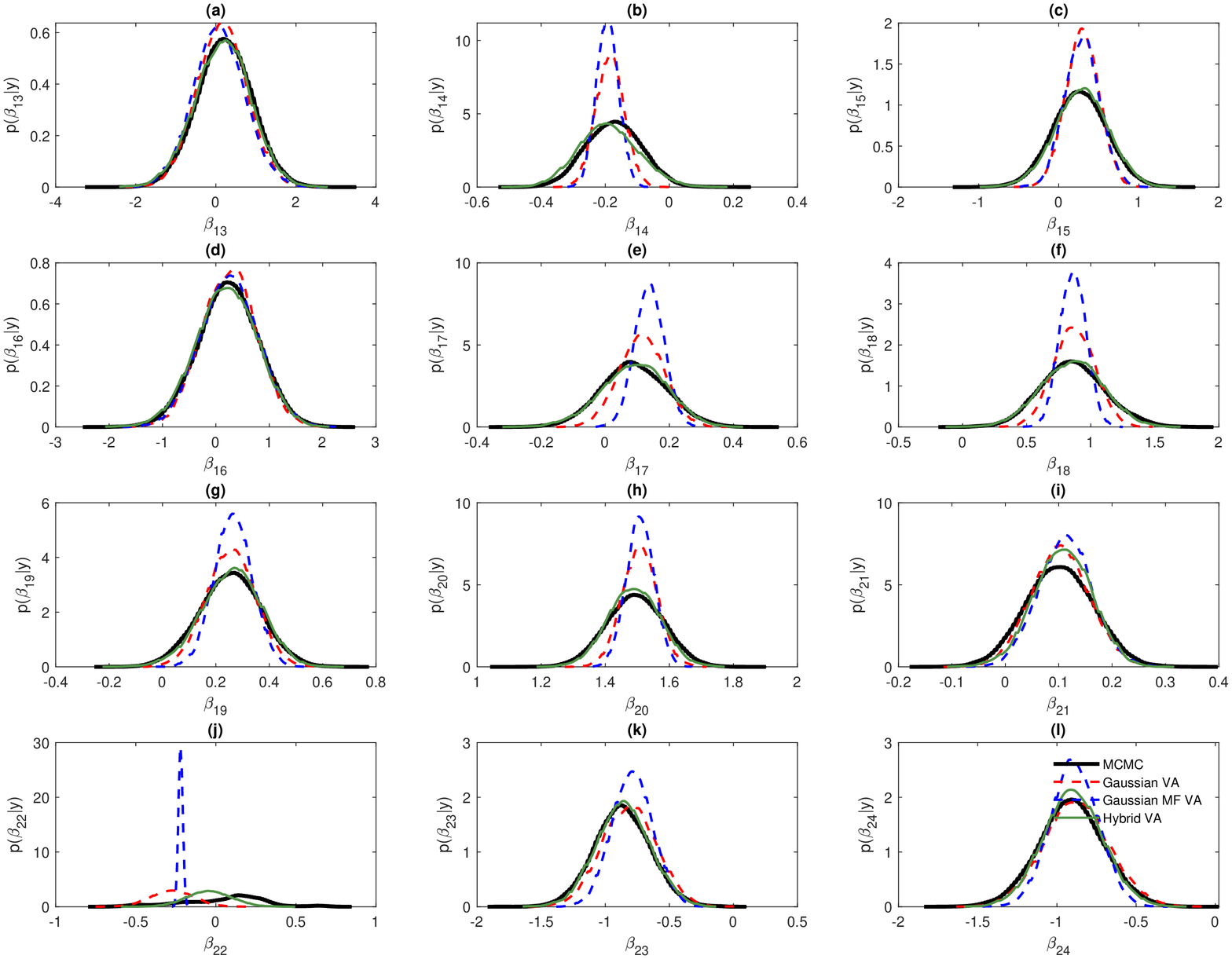}
	\end{center}
	\label{fig:poteriordens2}
\end{figure}

\begin{figure}[H]
			\caption{Tobit small data example: marginal posterior densities of coefficients (group 3)}
	\begin{center}
		\includegraphics[width=1\textwidth]{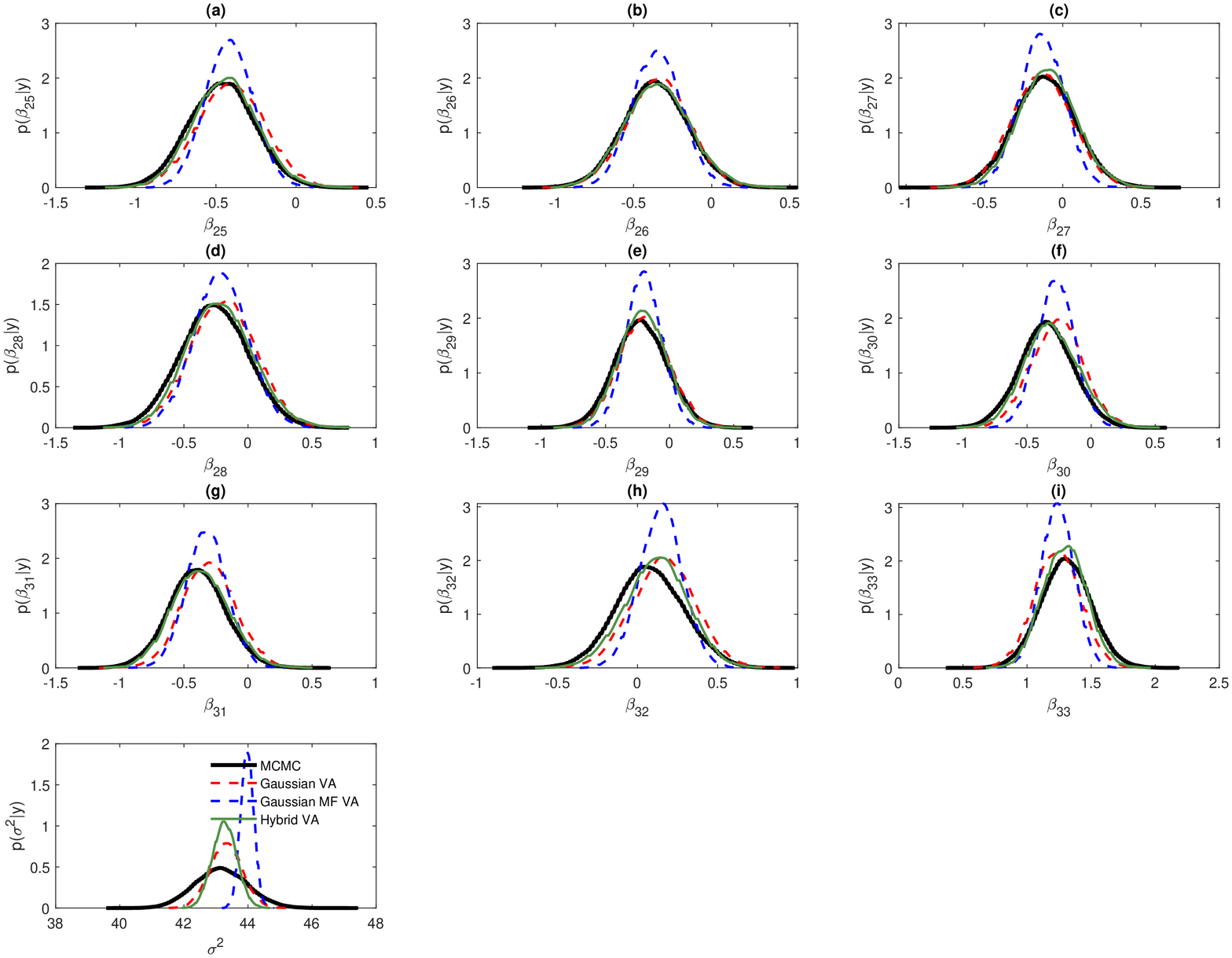}
	\end{center}
	\label{fig:poteriordens3}
\end{figure}

\begin{landscape} 
% Table generated by Excel2LaTeX from sheet 'Sheet1'
\begin{table}[htbp]
		\caption{Estimate of $V_\alpha$ for the tobit small data example for MCMC,  Hybrid VA and Gaussian factor VA.}
		\begin{center}
	\begin{tabular}{lccccclcccc}
		\hline\hline
		          &                                        &                                       &                                       &                                      &  &           &                                        &                                        &                                        &                                      \\
		          &                                                          \multicolumn{4}{c}{\textbf{Panel A: MCMC}}                                                           &  &           &                                                        \multicolumn{4}{c}{\textbf{Panel B: Gaussian VA}}                                                        \\
		          &                                        &                                       &                                       &                                      &  &           &                                        &                                        &                                        &                                      \\
		          &              {Intercept}               &               {Emails}                &               {Catal.}                &              {Paid S.}               &  &           &              {Intercept}               &                {Emails}                &                {Catal.}                &              {Paid S.}               \\ \cline{2-5}\cline{7-11}
		          &                                        &                                       &                                       &                                      &  &           &                                        &                                        &                                        &                                      \\
		Intercept &                 34.633                 &                                       &                                       &                                      &  & Intercept &                 28.083                 &                                        &                                        &                                      \\
		          & \footnotesize{\textit{(29.44, 40.47)}} &                                       &                                       &                                      &  &           & \footnotesize{\textit{(26.01, 30.47)}} &                                        &                                        &                                      \\
		Emails    &                 -0.57                  &                 1.676                 &                                       &                                      &  & Emails    &                 -0.41                  &                 0.688                  &                                        &                                      \\
		          & \footnotesize{\textit{(-0.68, -0.43)}} & \footnotesize{\textit{(1.04, 2.41)}}  &                                       &                                      &  &           & \footnotesize{\textit{(-0.46, -0.36)}} &  \footnotesize{\textit{(0.63, 0.75)}}  &                                        &                                      \\
		Catal.    &                 0.074                  &                -0.054                 &                 2.971                 &                                      &  & Catal.    &                 0.322                  &                 -0.14                  &                 0.748                  &                                      \\
		          & \footnotesize{\textit{(-0.59, 0.73)}}  & \footnotesize{\textit{(-0.5, 0.41)}}  & \footnotesize{\textit{(0.43, 9.25)}}  &                                      &  &           &  \footnotesize{\textit{(0.27, 0.38)}}  & \footnotesize{\textit{(-0.17, -0.11)}} &  \footnotesize{\textit{(0.69, 0.81)}}  &                                      \\
		Paid S.   &                 -0.59                  &                 0.398                 &                -0.051                 &                 2.45                 &  & Paid S.   &                 -0.341                 &                 0.148                  &                 -0.116                 &                1.158                 \\
		          & \footnotesize{\textit{(-0.83, -0.25)}} & \footnotesize{\textit{(0.16, 0.63)}}  & \footnotesize{\textit{(-0.55, 0.46)}} & \footnotesize{\textit{(0.91, 5.13)}} &  &           & \footnotesize{\textit{(-0.39, -0.29)}} &  \footnotesize{\textit{(0.12, 0.18)}}  & \footnotesize{\textit{(-0.14, -0.09)}} & \footnotesize{\textit{(1.07, 1.25)}} \\
		          &                                        &                                       &                                       &                                      &  &           &                                        &                                        &                                        &                                      \\ \cline{1-5}\cline{7-11}
		          &                                        &                                       &                                       &                                      &  &           &                                        &                                        &                                        &                                      \\
		          &                                                     \multicolumn{4}{c}{\textbf{Panel C: Gaussian MF VA}}                                                      &  &           &                                                         \multicolumn{4}{c}{\textbf{Panel D: Hybrid VA}}                                                         \\
		          &                                        &                                       &                                       &                                      &  &           &                                        &                                        &                                        &                                      \\
		          &              {Intercept}               &               {Emails}                &               {Catal.}                &              {Paid S.}               &  &           &              {Intercept}               &                {Emails}                &                {Catal.}                &              {Paid S.}               \\ \cline{2-5}\cline{7-11}
		          &                                        &                                       &                                       &                                      &  &           &                                        &                                        &                                        &                                      \\
		Intercept &                 24.977                 &                                       &                                       &                                      &  & Intercept &                 35.137                 &                                        &                                        &                                      \\
		          & \footnotesize{\textit{(22.63, 27.76)}} &                                       &                                       &                                      &  &           & \footnotesize{\textit{(31.55, 39.18)}} &                                        &                                        &                                      \\
		Emails    &                 0.215                  &                 0.057                 &                                       &                                      &  & Emails    &                 -0.566                 &                 1.696                  &                                        &                                      \\
		          &  \footnotesize{\textit{(0.15, 0.2)}}   & \footnotesize{\textit{(0.05, 0.06)}}  &                                       &                                      &  &           & \footnotesize{\textit{(-0.68, -0.44)}} &  \footnotesize{\textit{(1.16, 2.32)}}  &                                        &                                      \\
		Catal.    &                 0.095                  &                 0.021                 &                 0.589                 &                                      &  & Catal.    &                 -0.035                 &                 0.019                  &                 2.178                  &                                      \\
		          &  \footnotesize{\textit{(0.04, 0.16)}}  & \footnotesize{\textit{(0.01, 0.04)}}  & \footnotesize{\textit{(0.54, 0.64)}}  &                                      &  &           &  \footnotesize{\textit{(-0.8, 0.68)}}  & \footnotesize{\textit{(-0.41, 0.47)}}  &  \footnotesize{\textit{(0.33, 8.21)}}  &                                      \\
		Paid S.   &                 -0.04                  &                -0.009                 &                -0.004                 &                0.569                 &  & Paid S.   &                 -0.575                 &                 0.339                  &                  0.02                  &                2.284                 \\
		          &  \footnotesize{\textit{(-0.1, 0.02)}}  & \footnotesize{\textit{(-0.02, 0.01)}} &  \footnotesize{\textit{(-0.01, 0)}}   & \footnotesize{\textit{(0.52, 0.62)}} &  &           & \footnotesize{\textit{(-0.84, -0.25)}} &  \footnotesize{\textit{(0.14, 0.52)}}  & \footnotesize{\textit{(-0.45, 0.52)}}  & \footnotesize{\textit{(0.81, 5.22)}} \\
		          &                                        &                                       &                                       &                                      &  &           &                                        &                                        &                                        &                                      \\ \hline\hline
	\end{tabular}
		\end{center}
	\label{tab:REvariance_allMethods}%
The diagonal values are estimates of the variances of the random coefficients (i.e. the leading diagonal of $V_\alpha$). 
The off-diagonal values are estimates of the correlations between the random coefficients (i.e. the correlations
of the matrix $V_\alpha$). The variational means are reported, along with the 95\% quantiles of the variational
	distribution $q_\lambda$ in parentheses. Panels A to D correspond to MCMC, Gaussian factor VA, mean field Gaussian VA and our proposed Hybrid VA, respectively.
\end{table}%

\end{landscape}

% Table generated by Excel2LaTeX from sheet 'Sheet1'
\begin{landscape} 
	\begin{table}[hp]
		\caption{Estimate of $V_\alpha$ for the tobit full data example using the Gaussian factor VA.}
\begin{center}
		\resizebox{24cm}{!}{
			\begin{tabular}{llccccccccccccc}
				\hline\hline
				   &           &                                         &  &                                        &                                       &                                        &  &                                      &                                       &                                       &  &                                      &                                       &                                      \\
				   &           &                                         &  &                                                 \multicolumn{3}{c}{B1}                                                  &  &                                                \multicolumn{3}{c}{B2}                                                &  &                                               \multicolumn{3}{c}{B3}                                                \\
				   &           &                Intercept                &  &                 Emails                 &                Catal.                 &                Paid S.                 &  &                Emails                &                Catal.                 &                Paid S.                &  &                Emails                &                Catal.                 &               Paid S.                \\ \cline{3-3}\cline{5-7}\cline{9-11}\cline{13-15}
				   &           &                                         &  &                                        &                                       &                                        &  &                                      &                                       &                                       &  &                                      &                                       &                                      \\
				   & Intercept &             {27.174       }             &  &                                        &                                       &                                        &  &                                      &                                       &                                       &  &                                      &                                       &                                      \\
				   &           & \footnotesize{\textit{(26.89, 27.46)}}  &  &                                        &                                       &                                        &  &                                      &                                       &                                       &  &                                      &                                       &                                      \\
				   &           &                                         &  &                                        &                                       &                                        &  &                                      &                                       &                                       &  &                                      &                                       &                                      \\
				   & Emails    &            {-0.626        }             &  &            {2.31          }            &                                       &                                        &  &                                      &                                       &                                       &  &                                      &                                       &                                      \\
				   &           & \footnotesize{\textit{(-0.63, -0.62)}}  &  & \footnotesize{\textit{(2.29, 2.34)  }} &                                       &                                        &  &                                      &                                       &                                       &  &                                      &                                       &                                      \\
				B1 & Catal.    &            {0.015         }             &  &            {-0.009        }            &            {0.366        }            &                                        &  &                                      &                                       &                                       &  &                                      &                                       &                                      \\
				   &           & \footnotesize{\textit{(0.01, 0.02)  }}  &  & \footnotesize{\textit{(-0.01, 0)    }} & \footnotesize{\textit{(0.36, 0.37) }} &                                        &  &                                      &                                       &                                       &  &                                      &                                       &                                      \\
				   & Paid S.   &            {-0.075        }             &  &            {0.048         }            &            {-0.001       }            &            {1.782         }            &  &                                      &                                       &                                       &  &                                      &                                       &                                      \\
				   &           & \footnotesize{\textit{(-0.08, -0.07)}}  &  & \footnotesize{\textit{(0.04, 0.05)  }} & \footnotesize{\textit{(-0.01, 0)   }} & \footnotesize{\textit{(1.76, 1.81)  }} &  &                                      &                                       &                                       &  &                                      &                                       &                                      \\
				   &           &                                         &  &                                        &                                       &                                        &  &                                      &                                       &                                       &  &                                      &                                       &                                      \\
				   & Emails    &            {0.621         }             &  &            {-0.4          }            &            {0.01         }            &            {-0.048        }            &  &            {0.294       }            &                                       &                                       &  &                                      &                                       &                                      \\
				   &           & \footnotesize{\textit{(0.62, 0.62)  } } &  & \footnotesize{\textit{(-0.4, -0.4)  }} & \footnotesize{\textit{(0, 0.02)    }} & \footnotesize{\textit{(-0.05, -0.04)}} &  & \footnotesize{\textit{(0.29, 0.3) }} &                                       &                                       &  &                                      &                                       &                                      \\
				B2 & Catal.    &            {0.459         }             &  &            {-0.295        }            &            {0.007        }            &            {-0.035        }            &  &            {0.293       }            &            {1.592        }            &                                       &  &                                      &                                       &                                      \\
				   &           & \footnotesize{\textit{(0.45, 0.46)  }}  &  & \footnotesize{\textit{(-0.3, -0.29) }} & \footnotesize{\textit{(0, 0.01)    }} & \footnotesize{\textit{(-0.04, -0.03)}} &  & \footnotesize{\textit{(0.29, 0.3) }} & \footnotesize{\textit{(1.58, 1.61) }} &                                       &  &                                      &                                       &                                      \\
				   & Paid S.   &            {0.034         }             &  &            {-0.022        }            &            {0            }            &            {-0.003        }            &  &            {0.022       }            &            {0.016        }            &            {2.332        }            &  &                                      &                                       &                                      \\
				   &           & \footnotesize{\textit{(0.03, 0.04)  }}  &  & \footnotesize{\textit{(-0.03, -0.02)}} & \footnotesize{\textit{(-0.01, 0.01)}} & \footnotesize{\textit{(-0.01, 0)    }} &  & \footnotesize{\textit{(0.02, 0.03)}} & \footnotesize{\textit{(0.01, 0.02) }} & \footnotesize{\textit{(2.31, 2.36) }} &  &                                      &                                       &                                      \\
				   &           &                                         &  &                                        &                                       &                                        &  &                                      &                                       &                                       &  &                                      &                                       &                                      \\
				   & Emails    &            {0.704         }             &  &            {-0.453        }            &            {0.011        }            &            {-0.054        }            &  &            {0.449       }            &            {0.332        }            &            {0.025        }            &  &            {0.187       }            &                                       &                                      \\
				   &           & \footnotesize{\textit{(0.7, 0.71)   }}  &  & \footnotesize{\textit{(-0.46, -0.45)}} & \footnotesize{\textit{(0.01, 0.02) }} & \footnotesize{\textit{(-0.06, -0.05)}} &  & \footnotesize{\textit{(0.45, 0.45)}} & \footnotesize{\textit{(0.33, 0.34) }} & \footnotesize{\textit{(0.02, 0.03) }} &  & \footnotesize{\textit{(0.19, 0.19)}} &                                       &                                      \\
				B3 & Catal.    &            {0.43          }             &  &            {-0.277        }            &            {0.007        }            &            {-0.033        }            &  &            {0.274       }            &            {0.203        }            &            {0.015        }            &  &            {0.311       }            &            {1.446        }            &                                      \\
				   &           & \footnotesize{\textit{(0.42, 0.44)  }}  &  & \footnotesize{\textit{(-0.28, -0.27)}} & \footnotesize{\textit{(0, 0.01)    }} & \footnotesize{\textit{(-0.04, -0.03)}} &  & \footnotesize{\textit{(0.27, 0.28)}} & \footnotesize{\textit{(0.2, 0.21)  }} & \footnotesize{\textit{(0.01, 0.02) }} &  & \footnotesize{\textit{(0.31, 0.32)}} & \footnotesize{\textit{(1.42, 1.47) }} &                                      \\
				   & Paid S.   &            {0             }             &  &            {0             }            &            {0            }            &            {0             }            &  &            {0           }            &            {0            }            &            {0            }            &  &            {0           }            &            {0            }            &            {2.346       }            \\
				   &           & \footnotesize{\textit{(0, 0.01)     }}  &  & \footnotesize{\textit{(0, 0)        }} & \footnotesize{\textit{(-0.01, 0.01)}} & \footnotesize{\textit{(-0.01, 0.01) }} &  & \footnotesize{\textit{(0, 0)      }} & \footnotesize{\textit{(-0.01, 0.01)}} & \footnotesize{\textit{(-0.01, 0.01)}} &  &    \footnotesize{\textit{(0, 0)}}    & \footnotesize{\textit{(-0.01, 0.01)}} & \footnotesize{\textit{(2.33, 2.37)}} \\ \hline\hline
			\end{tabular}%
		}
	\end{center}
		\label{tab:REvarianceJMRVA}%
The diagonal values are estimates of the variances of the random coefficients (i.e. the leading diagonal of $V_\alpha$). 
The off-diagonal values are estimates of the correlations between the random coefficients (i.e. the correlations
of the matrix $V_\alpha$). The variational means are reported, along with the 95\% quantiles of the variational
distribution $q_\lambda$ in parentheses.
	\end{table}%
\end{landscape}

% Table generated by Excel2LaTeX from sheet 'Sheet1'
\begin{table}[hp]
	\centering
	\begin{tabular}{llccc}
		\hline\hline
		                   &                                                  &\multicolumn{3}{c}{\textbf{Random Coefficients}}  \\
		\textbf{Label}     & \textbf{Brief description}                       &          \textbf{Small}          &  &  \textbf{Full}                  \\
		                   &                                                  &          \textbf{E.g.}                                  & &       \textbf{E.g.}                  \\
		Intercept          & Scalar set to $1$ for all observations           &                 \checkmark                 &  &                 \checkmark                 \\
		\multicolumn{4}{l}{\underline{\em Lagged Sales Variables}} \\ 
		B1 past D sales    & Total online sales of B1 in previous 4 weeks   &                                            &  &                                            \\
		B2 past D sales    & Total online sales of B2 in previous 4 weeks   &                                            &  &                                            \\
		B3 past D sales    & Total online sales of B3 in previous 4 weeks   &                                            &  &                                            \\
		B1 past R sales    & Total in-store sales of B1 in previous 4 weeks   &                                            &  &                                            \\
		B2 past R sales    & Total in-store sales of B2 in previous 4 weeks   &                                            &  &                                            \\
		B3 past R sales    & Total in-store sales of B3 in previous 4 weeks   &                                            &  &                                            \\
		\multicolumn{4}{l}{\underline{\em Advertising Variables (in Adstock Form)}} \\  
		B1 Emails          & Number of brand B1 ad emails received  &                 \checkmark                 &  &                 \checkmark                 \\
		B1 Catal.          & Number of brand B1 ad catalogs received  &                 \checkmark                 &  &                 \checkmark                 \\
		B1 Paid S.         & Number of brand B1 paid search click-throughs &                 \checkmark                 &  &                 \checkmark                 \\
		B2 Emails          & Number of brand B2 ad emails received        &                                            &  &                 \checkmark                 \\
		B2 Catal.          & Number of brand B2 ad catalogs received          &                                            &  &                 \checkmark                 \\
		B2 Paid S.         & Number of brand B2 paid search click-throughs    &                                            &  &                 \checkmark                 \\
		B3 Emails          & Number of brand B3 ad emails received            &                                            &  &                 \checkmark                 \\
		B3 Catal.          & Number of brand B3 ad catalogs received         &                                            &  &                 \checkmark                 \\
		B3 Paid S.         & Number of brand B3 paid search click-throughs    &                                            &  &                 \checkmark                 \\
		\multicolumn{4}{l}{\underline{\em Variables included to control for endogeniety}} \\ 
		Res Paid S.        & Residuals from the paid search control function           &                                            &  &                                            \\
		Organic S. CFs     & No. clicks on organic search links for B1 ($\log(\mbox{clicks}+1)$)          &                                            &  &                                            \\
		Res Organic S.     & Residuals from the organic search control function          &                                            &  &                                            \\
		Res website V.     & Residuals from the website visit control function &                                            &  &                                            \\
		Visits B1 		& Number of visits to B1 website ($\log(\mbox{Visits}+1)$)                                                &                                            &  &                                            \\		\multicolumn{4}{l}{\underline{\em Other variables}} \\ 
		log price          & Log of B1 price index 				&                                            &  &                                            \\
		month1             & January dummy variable               &                                            &  &                                            \\
		month2             & February dummy variable                &                                            &  &                                            \\
		month3             & March dummy variable                &                                            &  &                                            \\
		month4             & April dummy variable               &                                            &  &                                            \\
		month5             & May dummy variable               &                                            &  &                                            \\
		month7             & July dummy variable               &                                            &  &                                            \\
		month8             & August dummy variable               &                                            &  &                                            \\
		month9             & September dummy variable               &                                            &  &                                            \\
		month10            & October dummy variable               &                                            &  &                                            \\
		month11            & November dummy variable               &                                            &  &                                            \\
		month12            & December dummy variable               &                                            &  &                                            \\ \hline\hline
	\end{tabular}%
	\label{tab:descriptions}%
	\caption{Brief description of the covariates in the mixed tobit model, 
	with a full description given in~\cite{danaher2020}. All covariates are included as fixed effects.
	Individual-level (zero mean) random coefficients are also considered for the intercept and advertising variables
	as indicated in the last two columns for the small and full data examples, respectively. Advertising variables
	are entered in `AdStock' form, which is exponentially smoothed with a short estimated lag.}
\end{table}%

\end{document}